\documentclass[draftcls, onecolumn]{IEEEtran}
\usepackage{amsmath}
\usepackage{amssymb}
\usepackage{mathtools}
\usepackage[pdftex]{graphicx}
\usepackage{bm}

\usepackage{algpseudocode}
\usepackage{algorithm}

\usepackage{comment} 

\usepackage{enumerate}
\usepackage{cite}

\newtheorem{definition}{Definition}
\newtheorem{theorem}{Theorem}
\newtheorem{corollary}{Corollary}
\newtheorem{lemma}{Lemma}
\newtheorem{property}{Property}
\newtheorem{proposition}{Proposition}
\usepackage{arydshln}

\title{
Algorithm Families for Computing Information-Theoretic Forms of Strong Converse Exponents in Channel Coding and Lossy Source Coding}
\author{
Yutaka Jitsumatsu~\IEEEmembership{Member,~IEEE}
\thanks{Dept. of Information and Communications Engineering, 
Tokyo Institute of Technology 
  }
and  Yasutada Oohama~\IEEEmembership{Member,~IEEE}
\thanks{
Dept. of Network and Computer Engineering, 
The University of Electro-Communications}
\\
}

\begin{document}

\maketitle

\begin{abstract}
The error exponent of a discrete memoryless channel is expressed in two forms. One is Gallager's expression with a slope parameter $\rho\in[0,1]$ and the other is Csisz\'ar and K\"orner's information-theoretic representation expressed using the mutual information and the relative entropy. They differ in appearance, and existing proof methods for their agreement are not elementary, requiring an evaluation of the Karush-Kuhn-Tacker (KKT) condition that an optimal distribution must satisfy. Similarly, there are two expressions of the correct decoding probability exponent or the strong converse exponent. They are Arimoto's expression with slope parameter $\rho\in(-1,0]$ and
Dueck and K\"orner's information-theoretic expression. 
The purpose of this paper is to clarify the relationship between two different ways of representing exponents, namely, representation using slope parameters and representation using information-theoretic quantities, from the viewpoint of algorithms for computing exponents.
Arimoto's algorithm for computing the exponents is based on representation using slope parameters. On the other hand, the authors' and Tridenski and Zamir's algorithms compute the strong converse exponent based on Dueck and K\"orner's information-theoretic expression. A family of algorithms that includes the above two algorithms as special cases was recently proposed by Tridenski et al. We clarify that the convergence of Tridenski and Zamir's algorithm proves the agreement of Arimoto's  and Dueck and K\"orner's strong converse exponents.
We discuss another family of algorithms and, using the surrogate objective function used therein, prove that the two representations of the error exponent coincide. This proof method is elementary in the sense that it does not require evaluation of the KKT condition.
In the second half of the paper, we discuss the error and correct decoding probability exponents of lossy source coding of discrete memoryless sources and the algorithms for computing them. This paper defines a family of algorithms for computing the source coding strong converse exponent. A member of the algorithm family has a nice property that the convergence of the algorithm implies the agreement of the two kinds of representation of the strong converse exponents.

\end{abstract}

\section{Introduction}
Fifty years have passed since the Arimoto-Blahut algorithm was published~\cite{Arimoto1972, Blahut1972}. 
This elegant algorithm efficiently finds the optimal distribution that achieves the capacity of a communication channel and rate distortion function of an information source. It is an alternating optimization algorithm and its connection to the Expectation-Maximization (EM) algorithm has attracted interest~\cite{Csiszar-Tusnady1983}. 
The extension of the Arimoto-Blahut algorithm to other coding problems~\cite{Nagaoka1998, Vontobel, DupuisISIT2004, Cheng2005,YasuiISIT2007,Yasui2008,YasuiISIT2010,Naiss_extention_of_Blahut-Arimoto} is still an important research topic.
One such extension is the algorithm found by Arimoto for computing the reliability functions~\cite{Arimoto1976}. 
Here, the reliability functions are the functions that express the exponents of error and correct decoding probabilities in channel coding and lossy source coding. 
A family of algorithms for computing the correct decoding probability exponent in channel coding 
was recently proposed by Tridenski et al.\cite{Tridenski2020arXiv}.
This paper reveals several properties of the Tridenski et al.'s family of algorithms that help us to understand the relation between the different expressions of the exponent function. 

The objective of this research is to establish all algorithms for computing the optimal distribution for optimization and minimax problems that give error and strong converse exponents in channel coding and lossy source coding. 
To achieve this goal, we compare the optimization and minimax problems that give the exponents shown in Table~\ref{table:existing_algorithsm} and investigate the relationship among them.
The left half of Table~\ref{table:existing_algorithsm} shows the research results that established the error exponent and the strong converse exponent for channel coding as well as the research results that proposed algorithms for their computation. 
Table~\ref{table:existing_algorithsm} also shows the results of this paper. 
Arimoto's algorithm~\cite{Arimoto1976} is an algorithm for computing Gallager's random coding error exponent and Arimoto's strong converse exponent. 
The authors proposed an algorithm for computing the Dueck and K\"orner's exponent~\cite{YutakaISIT2015}, which is later extended to the case of channel coding under cost constraint~\cite{Jitsumatsu_Oohama_IT_Trans2020}.
Subsequently, Tridenski and Zamir~\cite{Tridenski2018arXiv} proposed another algorithm that was based on their strong converse exponent expression\footnote{
Tridenski and Zamir proposed two algorithms:
One computes directly the strong converse exponent for a fixed $R \ge 0$~\cite{Tridenski2018ISIT} and the other is an algorithm for fixed parameter $\rho$~\cite{Tridenski2018arXiv, Tridenski2020ITtran}.
The latter is discussed in this paper.}. 
Then, Tridenski et al. proposed a family of algorithms that is parameterized by four non-negative parameters~\cite{Tridenski2020arXiv} and includes~\cite{YutakaISIT2015} and \cite{Tridenski2018arXiv} as special cases. 
Algorithm~\ref{new_algorithm_B} in Table~\ref{table:existing_algorithsm} is also included in algorithm family of Tridenski et al. but its importance is studied in detail for the first time. 

\begin{table*}
    \caption{Error and strong converse exponents in channel coding and lossy source coding and their computation algorithms}
    \begin{center}
    \renewcommand{\arraystretch}{1.6}
    \setlength{\arrayrulewidth}{1pt}
    \setlength{\doublerulesep}{0pt}
    \fontsize{8pt}{8pt}\selectfont
    \begin{tabular}{||l||l|l||l|l||}\hline
    & \multicolumn{2}{|c|}{Channel coding} & \multicolumn{2}{|c|}{Lossy source coding} \\\hline
    Error exponent & Gallager (1965) \cite{Gallager1965} & Csisz\'ar \& K\"orner & Blahut (1974)\cite{Blahut1974} & Marton$^*$ (1974)\cite{Marton1974} \\[-2ex] 
    expressions & & (1981) \cite{Csiszar-KornerBook} & 
    {\fontsize{7pt}{0pt}\selectfont [Remark]} & 
    {\fontsize{7pt}{0pt}\selectfont [Remark]}\\[-3ex] & & & 
    {\fontsize{7pt}{0pt}\selectfont Suboptimal in general} & {\fontsize{7pt}{0pt}\selectfont May not be continuous}\\\hdashline[1pt/1.2pt]
    Algorithms & Arimoto (1976)\cite{Arimoto1976} & Open & Arimoto (1976)\cite{Arimoto1976} & Open\\\hline 
    Strong converse & Arimoto$^*$ (1973) \cite{Arimoto1973}& Dueck \& K\"orner$^*$ & \multicolumn{1}{|l|}{Theorem~\ref{theorem5}} & Csisz\'ar \& K\"orner$^*$ \\[-2ex]
    exponent expressions & & (1979)\cite{Dueck_Korner1979}  &  & (1981) \cite{Csiszar-KornerBook}\\\hdashline[1pt/1.2pt]
    Algorithms & Arimoto (1976) \cite{Arimoto1976} & O.\& J. (2015) & Theorem~\ref{theorem5} shows & J. \& O. (2016)\cite{YutakaISIT2016b}\\[-2ex]
    & & without cost constraint\cite{YutakaISIT2015} & that Arimoto (1976)\cite{Arimoto1976} & \\[-2ex]
    & &                         & is applicable & \\[-2ex]
    & & J. \& O. (2020) & & \\[-2ex]
    & & extension to the channel & & \\[-2ex]
    & & under cost constraint\cite{Jitsumatsu_Oohama_IT_Trans2020} & & \\
    & & Tridenski \& Zamir (2018)\cite{Tridenski2018arXiv} & & Algorithm~\ref{algorithm_GCK1}  \\ 
    & & Algorithm~\ref{new_algorithm_B}  & & Algorithm~\ref{algorithm_GCK2}\\
    & & [Parametric family of algorithms] & & [Parametric family of algorithms] \\[-2ex]
    & & Tridenski et al. (2020)\cite{Tridenski2020arXiv} & & Algorithm~\ref{alg:family_lossy_source_coding}\\
    \hline
    \multicolumn{5}{l}{\fontsize{7pt}{0pt}
    \selectfont The * symbol after the name indicates that the function has been proven optimal }
    \\
    \end{tabular}
    \end{center}
    \label{table:existing_algorithsm}
\end{table*}

One of the characteristics of the Arimoto-Blahut algorithm is that it yields an alternative expression. 
Because this property is used repeatedly in this paper, we illustrate it by an example here. 
The standard expression for the channel capacity of a discrete memoryless channel (DMC) is $C(W)=\max_{P} I(P, W)$, 
where $P=P(x)$ is an input distribution, $W=W(y|x)$ is a transition probability of the communication channel, and $I(P, W)$ is the mutual information.
Arimoto-Blahut algorithm uses a surrogate objective function $\Psi(P, Q|W)=\sum_x \sum_y P(x)W(y|x)\log ( {P(x)}/{Q(x|y)})$ for alternating optimization and solves the double maximization problem $\max_{P} \max_{Q} \Psi(P, Q|W)$ by alternately optimizing $P(x)$ and $Q(x|y)$. 
Because $\max_{Q} \Psi(P,Q|W) = I(P,W)$ and 
$\max_{P} \Psi( P, Q |W) = \log \sum_x \prod_y Q(x|y)^{W(y|x)}$ hold, 
the channel capacity has another expression, $\tilde C(W) = \max_{Q} \log \sum_x \prod_y Q(x|y)^{W(y|x)}$, in addition to the standard expression. 
The match of the two expressions, $C(W) = \tilde C(W)$ for any transition probability $W$, is proved by the convergence of the Arimoto-Blahut algorithm. 
Such characteristics of Arimoto-Blahut type algorithms plays an important role in this paper.

After defining the mathematical expressions of exponent functions, the first half of this paper describes the algorithms for computing the reliability functions in the channel coding and the second half describes the algorithms for computing the reliability functions in lossy source coding.
For DMCs, 
the random coding error exponent is the exponent of the decoding error probability averaged over an ensemble of randomly generated codes. Two forms of its representation are known.
One is the form of Gallager~\cite{Gallager1965} and the other is the form of Csisz\'ar and K\"orner~\cite{Csiszar-KornerBook}. 
The former is expressed using Gallager's $E_0$-function, while the latter is defined using information theoretic quantities, i.e., mutual information and Kullback-Leibler (KL) divergence.
They match at all rates of $R\geq 0$, although they look very different.
On the other hand, the correct decoding probability exponent, also known as the strong converse exponent, of channel coding expresses the exponential upper bound of the correct decoding. 
Like the error exponent, the channel coding strong converse exponent is also known in two forms.
One is Arimoto's exponent~\cite{Arimoto1973} that is expressed by using the $E_0$-function and the other is Dueck and K\"orner's exponent~\cite{Dueck_Korner1979} that is expressed by using information theoretic quantities. 
These two strong converse exponents are also known to coincide for any rate $R\ge 0$.

%
%

\begin{figure*}
\centering
\includegraphics[width=0.95\textwidth]{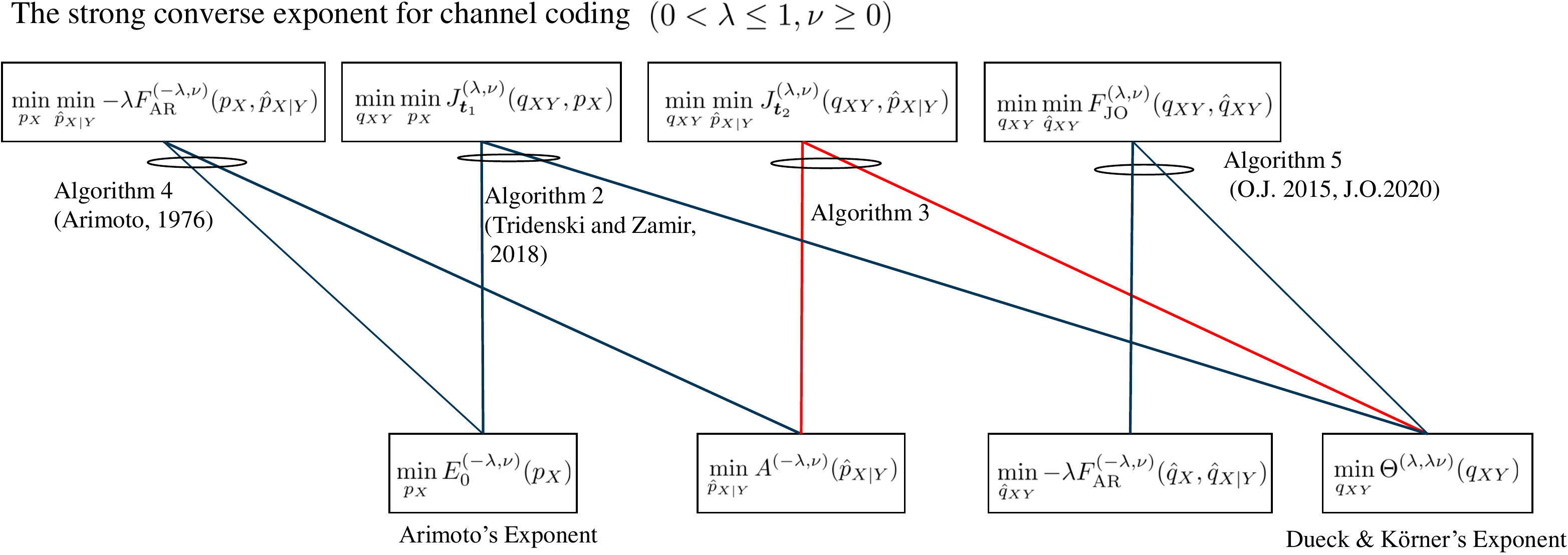}
\caption{
Relation between the four algorithms discussed in this paper and the expressions of the strong converse exponents in channel coding
}
\label{fig.1}
\end{figure*}

Fig.~\ref{fig.1} shows the relationship between Arimoto's algorithm~\cite{Arimoto1976},
the authors' previous algorithm~\cite{YutakaISIT2015, Jitsumatsu_Oohama_IT_Trans2020},
Tridenski and Zamir's algorithm~\cite{Tridenski2018arXiv, Tridenski2020ITtran}, and Algorithm~\ref{new_algorithm_B}.
The box in the leftmost on the bottom row labeled 
$
\min_{p_X} E_0^{(-\lambda, \nu)}(p_X) 
$
is the optimization problem that appears in Arimoto's strong converse exponent. 
The box in the rightmost on the bottom row labeled 
$ \min_{ q_{XY} } \Theta^{(\lambda, \lambda \nu)}(q_{XY})$ is the optimization problem that expresses
Dueck and K\"orner's exponent. 
The meaning of each function and variable is explained in Section II.
The double minimization problems in the Arimoto algorithm and the authors' previous algorithm are written to the left of these two optimization problems in the upper row, respectively.
In a double minimization problem, fixing one variable and minimizing with respect to the other variable yields a single minimization problem.
Line segments connecting a double minimization form and a single minimization form in Fig.~\ref{fig.1} indicate that the match of two expressions can be proved by optimizing one variable in the double minimization form.  
The Arimoto algorithm and the authors' algorithm have alternative expressions $ \min_{\hat p_{X|Y}} A^{(-\lambda, \nu)} ( \hat p_{X|Y} )  $ and $ \min_{\hat q_{XY}} -\lambda F^{(-\lambda, \nu)}_{\rm AR} (\hat q_{X}, \hat q_{X|Y} )  $, respectively.
Tridenski and Zamir's algirithm, on the other hand, is different from either of these two algorithms.
In Tridenski and Zamir's algorithm, the joint distribution $q_{XY}$ and input distribution $p_X$ are alternately updated. 
In this paper, we point out that the objective function of Arimoto's exponent and that of Dueck and K\"orner's exponent appear alternately during the update. 
Thus, we can say that Tridenski and Zamir's algorithm is an algorithm that can simultaneously compute Arimoto's exponent and Dueck and K\"orner's exponent.
This fact is important because the convergence theorem of Tridenski and Zamir's algorithm immediately implies that the two strong converse exponents coincide.
Among the four algorithms in Fig.~\ref{fig.1}, only Tridenski and Zamir's algorithm can prove the match of the two exponents.
This paper is the first to prove the match of two exponents by the convergence of an algorithm.

In~\cite{Tridenski2020arXiv}, Tridenski et al. proposed a parameterized algorithms, in which
the surrogate objective functions that are shown by the three right boxes 
in the upper row in Fig.~\ref{fig.1} are unified as a parameterized objective function. 
Tridenski et al.'s parameterised algorithm includes Algorithm\ref{new_algorithm_A}\cite{Tridenski2018arXiv}, Algorithm~\ref{new_algorithm_B}, and Algorithm~\ref{alg:previous_work} as special cases.
The three optimization problems that are shown by the three left boxes 
in the bottom row in Fig.~\ref{fig.1} are also unified as a parameterized optimization problems.
This implies that we obtain a family of expressions that are alternative to the 
Dueck and K\"orner's expression of the correct decoding probability exponent. 
The family of algorithms give a unified point of view for the computing algorithms. 
On the other hand, their special cases defined as Algorithms \ref{new_algorithm_A},~\ref{new_algorithm_B}, and~\ref{alg:previous_work} have their own attractive properties. We also discuss the
relationship among these special cases.

The algorithm for computing Dueck and K\"orner's epxonent cannot be applied to compute Csisz\'ar and K\"orner's exponent because of the following reason:  
Dueck and K\"orner's strong converse exponent is defined as the minimum of an objective function 
with respect to the joint probability distribution.
The objective function of Dueck and K\"orner's exponent is convex, which enabled us to derive algorithms for finding the global minimum. 
On the other hand, Csisz\'ar and K\"orner's error exponent is defined as the saddle point of a minimax problem and the structure of the problem is different.
To the best of the author's knowledge, no algorithm for finding the saddle point of the minimax problem of Csisz\'ar and K\"orner's function has been found. 
This issue remains unresolved. 
However, by using $J_{\bm{t}_2}^{(\lambda, \nu)}(q_{XY}, \hat p_{X|Y})$ used in Algorithm~\ref{new_algorithm_B}, we are able to give a new proof that Gallager's exponent coincides with that of Csisz\'ar and K\"orner as a byproduct.
This is another result of this paper.
The conventional proof method~\cite[Problem 10.24]{Csiszar-KornerBook} evaluates the upper and lower bounds of the saddle point using the KKT conditions satisfied by the saddle point of the minimax problem that determines Csisz\'ar and K\"orner's exponent.
The advantage of the proof method in this paper is that it is more elementary, since the saddle point can be evaluated by an equality rather than by the KKT condition, i.e. by a set of inequalities.

In the second half of this paper, we discuss the computation algorithms for the strong converse exponent in lossy source coding of discrete memoryless sources (DMSs). 
The right half of Table~\ref{table:existing_algorithsm} shows the research results that established the error and the strong converse exponents for lossy source coding as well as the results that established their computation algorithms. 
Blahut's exponent~\cite{Blahut1974} is expressed using a function similar to Gallager's $E_0$-function with a slope parameter $\rho>0$ and is suboptimal in general. 
The optimal error exponent was established by Marton~\cite{Marton1974} which is expressed using information-theoretic quantities. 
For the strong converse exponent of lossy source coding, Csisz\'ar and K\"orner's exponent is the only known expression. 
The first algorithm for computing Csisz\'ar and K\"orner's exponent is due to the authors~\cite{YutakaISIT2016b}.

\begin{figure*}
    \centering
    \includegraphics[width=0.95\textwidth]{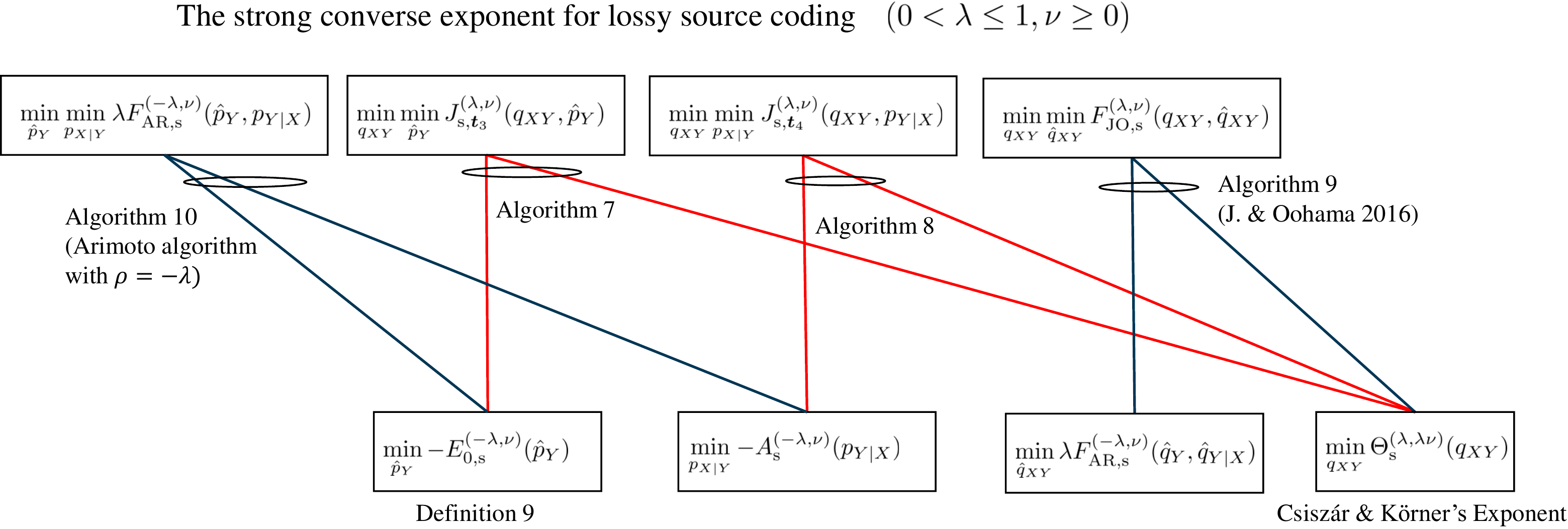}
\caption{
Relation between the four algorithms discussed in this paper and the expressions of the strong converse exponents in lossy source coding 
}
    \label{fig:st_conv_source_coding}
\end{figure*}

In this paper, we define a new family of algorithms for computing the strong converse exponent for lossy source coding, 
based on the idea similar to the algorithm family of Tridenski et al. 
This new algorithm family includes our previous algorithm~\cite{YutakaISIT2016b} as a special case.
We then discuss two other special cases, which are called Algorithms \ref{algorithm_GCK1} and \ref{algorithm_GCK2}. 
The relation between 
these three cases of the new algorithm family and the Arimoto's algorithm is shown in Fig.~\ref{fig:st_conv_source_coding}. 
The structure of the relationship between these algorithms corresponds perfectly to Fig.~\ref{fig.1}.
The new algorithm family uses a parameterized objective function 
which includes the objective functions shown in the three right boxes 
in the upper row in Fig.~\ref{fig:st_conv_source_coding}.  
We introduce in Definition~\ref{def:GCK_check} a new expression of the strong converse exponent that is similar to Blahut exponent, except that the range of the slope parameter $\rho$ is $[-1,0]$. 
The rightmost $\min_{q_{XY} } \Theta_{\rm s}^{(\lambda, \lambda \nu)} (q_{XY})
$ in the bottom row is an optimization problem expressing Csisz\'ar and K\"orner's strong converse exponent. 
The exponent that we introduce in Definition~\ref{def:GCK_check} is 
the leftmost $\min_{\hat p_Y} -E_{0,\rm s}^{(-\lambda, \nu)} (\hat p_Y)$
in the bottom row in Fig.~\ref{fig:st_conv_source_coding}. 
In Algorithm~\ref{algorithm_GCK1}, the objective function in Blahut's exponent, but with a negative $\rho$, and that of Csisz\'ar and K\"orner's exponent appear alternately.  
The convergence of the algorithm proves that for all rate $R\geq 0$,
the exponent of Definition~\ref{def:GCK_check} coincides with Csisz\'ar and K\"orner's exponent. 
This paper also shows in Theorem~\ref{theorem5} that Arimoto algorithm originally proposed for positive $\rho$ works correctly for the parameter $\rho\in [-1,0]$ as well.

This paper is organized as follows:
Section II defines symbols and notations used in this paper and reviews the error and the strong converse exponents in channel coding and source coding. 
We discuss channel coding in Sections III and IV and source coding in Sections V and VI.  
Algorithms for computing the strong converse exponent in channel coding is discussed in 
Section III. 
New iterative algorithms are defined and the convergence of the algorithms to an optimal distribution is proved. 
We then compare the proposed algorithm with Arimoto's algorithm, Tridenski and Zamir's algorithm, and our previously proposed algorithm.
In Section IV, we give a new proof for the fact that Csisz\'ar and K\"orner's error exponent matches with Gallager's exponent.
In Section V, computation algorithms for the strong converse exponent in lossy source coding are proposed. 
In Section VI, we prove that Arimoto's algorithm for computing error exponent of lossy source coding is applicable for computation of strong converse exponent. 
We also show that Ar{\i}kan and Merhav's guessing exponent~\cite{ArikanMerhav1998}
can also be computed by Arimoto's algorithm for lossy source coding.

\section{Definitions of exponent functions}
As a preliminary to state the main result, 
this section reviews the definition of the error and the strong converse exponent functions in channel coding and lossy source coding listed in Table~\ref{table:existing_algorithsm}. 
Each of these exponent functions is defined as an optimization problem with respect to a probability distribution.  


\subsection{Exponent functions in channel coding} 
We consider a DMC with finite input and output alphabet $\mathcal{X}$ and $\mathcal{Y}$
that is subject to an input cost constraint. 
Let $W(y|x)$, $y\in \mathcal{Y}$, $x\in \mathcal{X}$, be the transition probability of the DMC.
Let $c(x)\geq 0$ be a cost function for sending $x\in \mathcal{X}$ and 
the average cost of sending codeword $x_1 x_2 \cdots x_n$ is $c(x_1, \ldots, c_n) = (1/n) \sum_{i=1}^n c(i)$. 
Let $R\geq 0 $ be a coding rate  and 
$\Gamma\geq 0$ be a maximum allowable cost.  
We define four functions. 
\begin{definition}
For $R\geq 0$ and $\Gamma\geq 0$,
Gallager's random coding error exponent is defined by 
\begin{align}
E_{\rm r} (R, \Gamma|W) 
= 
\max_{ \rho \in [0,1] }
\inf_{ \nu \geq 0 }
\max_{p_X \in\mathcal{P(X)} }
\left\{
-\rho R + \rho \nu \Gamma
+ E_0^{(\rho, \nu)}(p_X|W) 
\right\},
\label{E_r}
\end{align}
\end{definition}
where 
$\mathcal{P(X)}$ is a set of probability distributions
on the input alphabet and 
\begin{align}
&E_0^{(\rho, \nu)}(p_X|W) \notag\\\
&= -\log \sum_{y \in \mathcal{Y}} \left[
\sum_{x \in \mathcal{X}}  p_X(x) \{ W(y|x) {\rm e}^{ \rho \nu c(x)} \}^{\frac{1}{1+\rho}} 
\right] ^{1+\rho}
\label{E_0}
\end{align}
is called Gallager's $E_0$ function for cost constraint channels.
We use the natural logarithm. 

\begin{definition}
For $R\geq 0$ and $\Gamma\geq 0$,
Arimoto's strong converse exponent is defined by 
\begin{align}
G_{\rm AR} (R, \Gamma| W) 
= 
\sup_{ \rho \in (-1,0] }
\sup_{ \nu \geq 0}
\min_{p_X \in\mathcal{P(X)}} 
\{
-\rho R + \rho \nu \Gamma 
+ E_0^{(\rho, \nu )}(p_X|W) 
\}. 
\label{G_AR}
\end{align}
\end{definition}

\begin{definition}
Csisz\'ar and K\"orner's error exponent is defined by 
\begin{align}
E_{\rm CK} (R,\Gamma |W) = 
\max_{
\substack{
q_X \in\mathcal{P(X)}: \\
\mathrm{E}_q[c(X)] \leq \Gamma }
} 
\min_{ q_{Y|X} \in \mathcal{ P(Y|X)} }  
\{
D(q_{Y|X} || W | q_X )
+ 
| I(q_X, q_{Y|X} ) - R |^+  \}, 
\label{E_CK}
\end{align}
where $\mathcal{P(Y|X)}$ is a set of conditional probability distributions 
of $X\in \mathcal{X}$ given $Y \in \mathcal{Y}$, 
$D(q_{Y|X} || W | q_X)$ is the conditional relative entropy defined by
\begin{align*}
    D(q_{Y|X} || W | q_X)
    = \sum_{ x\in \mathcal{X} } q_X(x) \sum_{y\in \mathcal{Y} }
      q_{Y|X} (y|x) \log \frac{q_{Y|X} (y|x) }{W(y|x)}
\end{align*}
and $I(q_X, q_{Y|X})$ is the mutual information. We follow the convention
that $0\log \frac{0}{0} =0$, 
$0\log\frac{0}{p}=0$ and $p\log \frac{p}{0}=+\infty$ for $p>0$.
\end{definition}

\begin{definition}
Dueck and K\"orner's strong converse exponent is defined as  
\begin{align}
G_{\rm DK} (R, \Gamma|W) = 
    \min_{
    \substack{
    q_{XY} \in\mathcal{P(X\times Y)}: \\
    \mathrm{E}_q[c(X)] \leq \Gamma }
    } 
\{
D(q_{Y|X} || W | q_X )
+ 
| R- I(q_X, q_{Y|X} ) |^+
\}, 
\label{G_DK}
\end{align}
where 
$\mathcal{P(X\times Y)}$ is a set of probability distributions on 
$\mathcal{X\times Y}$.

Tridenski and Zamir gave a new expression of the optimal strong converse
exponent~\cite[Eq.(32)]{TridenskiISIT2017}. 
The following expression is an extension of Tridenski and Zamir's exponent
to DMCs under cost constraint. 
\begin{definition}
Tridenski and Zamir's strong converse exponent is defined by
\begin{align}
    G_\mathrm{TZ}(R, \Gamma|W)
    &=
    \min_{
    \substack{
    q_{XY} \in\mathcal{P(X\times Y)}: \\
    \mathrm{E}_q[c(X)] \leq \Gamma }
    } 
    \min_{p_{X} \in\mathcal{P(X)}} 
    \{
        D(q_{XY} || q_X \circ W ) + |R-D( q_{XY} || p_X \times q_{Y} ) |^+ 
    \}, 
    \label{G_TZ}
\end{align}
\end{definition}
where 
$q_X \circ W$ and $ p_X \times q_{Y}$ are joint distributions defined by 
$q_X(x) W(y|x)$ and $p_X(x) q_Y(y)$. 
\end{definition}

This exponent coincides with 
$G_{\rm DK}(R, \Gamma|W)$ as well as $G_{\rm AR}(R, \Gamma|W)$. 
We will prove the match of $G_{\rm TZ}(R, \Gamma|W)$ and $G_{\rm DK}(R, \Gamma|W)$
in Lemma~\ref{lemma:GDK_GTZ} in Section~\ref{section:new_algorithms}\footnote{
The outline of the proof was given in~\cite[Footnote 13]{Tridenski2017arXiv}.}.

\subsection{Exponent functions in Lossy source coding}
\label{section:def_exponent_source_coding}
Consider a DMS with source alphabet $\mathcal{X}$ and probability distribution $P_X \in \mathcal{P(X)}$. 
Let $\mathcal{Y}$ be a reproduction alphabet and $d(x,y)\geq 0$ be a distortion measure.
We consider the case that the distortion with a block length $n$ is measured by  
$d(x^n, y^n) = (1/n) \sum_{i=1}^n d(x_i, y_i)$.
Let $R$ be the coding rate and $\Delta$ is the maximum allowable distortion.
We assume that both $\mathcal{X}$ and $\mathcal{Y}$ are finite sets and allow $\mathcal{Y}$ to be different form $\mathcal{X}$
and that for every $x\in \mathcal{X}$ there exist at least one $y\in \mathcal{Y}$ satisfying $d(x,y) = 0$. %
We define three exponent functions.
\begin{definition}
Blahut's error exponent~\cite{Blahut1974} is defined by 
\begin{align}
  E_{\rm B}(R, \Delta|P_X)
  =\sup_{\rho \geq 0} \inf_{ \nu \geq0}
  \max_{ \hat p_{Y} }
  \left[ 
  \rho R + \rho \nu \Delta 
  -\log \sum_{x} P_X(x)
  \left\{
  \sum_{y}
  \hat p_Y(y){\rm e}^{-\nu  d(x,y)}
  \right\}^{-\rho }
  \right] 
  \label{E_B}. 
\end{align}
\end{definition}
We define the third term in the parenthesis of (\ref{E_B}) as 
\begin{align}
E_{0, \textrm{s} }^{(\rho, \nu )} (\hat p_Y | P_X) 
= \log\sum_x P_X(x) 
   \left\{
    \sum_y
    \hat p_Y(y) {\rm e}^{- \nu  d(x, y) }
    \right\}^{-\rho } ,
    \label{E_S0}
\end{align}
which plays a role similar to $E_0$-function. 

The error exponent of the lossy coding problem was established by
Marton~\cite{Marton1974}. 
\begin{definition}
Marton's error exponent for lossy source coding was defined by 
\begin{align}
    E_{\rm M} ( R, \Delta | P_X)  
    = \min_{ \genfrac{}{}{0pt}{}{ q_X\in \mathcal{P(X)}: }{ R(\Delta | q_X) \geq R } } 
    D(q_X||P_X), 
    \label{E_M}
\end{align}
\end{definition}
where $$ R(\Delta|q_X) = \min_{
\genfrac{}{}{0pt}{}{q_{Y|X} \in \mathcal{P(Y|X)}:}{\mathrm{E}_{q_{XY}} [ d(X,Y) ] \leq \Delta }
} I(q_X, q_{Y|X})$$
is a rate distortion function.  
Marton proved that exponents of upper and lower bounds of 
the error probability for an optimal pair of encoder and decoder
are both given by $E_{\rm M} ( R, \Delta | P_X)$. That is,  
Marton's exponent is optimal. 

The exponential strong converse theorem for lossy source coding was established by
Csisz\'ar and K\"orner~\cite{Csiszar-KornerBook}.
\begin{definition}
\label{def:G_CK}
Csisz\'ar and K\"orner's strong converse exponent for lossy source coding is defined by 
\begin{align}
    G_{\rm CK} ( R, \Delta | P_X)  
    = \min_{q_X \in \mathcal{P(X)}} 
    \{ D(q_X||P_X)  + | R(\Delta | q_X) - R |^+ \}. 
\end{align}
\end{definition}

$G_{\rm CK}( R, \Delta | P_X) $ was proven to be optimal~\cite{Csiszar-KornerBook}.
The strong converse exponent for lossy source coding in Blahut style has not been known. We define 
\begin{definition}
    \label{def:GCK_check}
For $R\geq 0$ and $\Delta\geq 0$, we define 
\begin{align}
    {G}_{\rm JO} (R, \Delta|P_X) 
    & = 
    \sup_{ \rho \in [-1,0) } \sup_{\nu \ge 0} \min_{\hat p_Y \in \mathcal{P(Y)} } 
    \{
      \rho R + \rho \nu \Delta - E_{0,\rm s}^{(\rho, \nu)}(\hat p_Y|P_X) 
    \} . 
    \label{def:eq.GCK_check}
\end{align}
\end{definition}
We will show in Section~\ref{section_sourcecoding_Arimotoalgorithm} that 
${G}_{\rm JO} (R, \Delta|P_X)$ coincides with 
$G_{\rm CK} ( R, \Delta | P_X)$.

Marton mentioned in~\cite[Section III]{Marton1974} that the continuity of $E_{\rm M}(R, \Delta |P)$ with respect to $R$ had not been established. 
Note that the rate distortion function $R(\Delta|q_X)$ is not convex in $q_X$ for a fixed $\Delta\geq 0$.
Therefore, for a fixed $R$, the feasible set in (\ref{E_M}) is not convex. 
This implies that $E_{\rm M}(R,\Delta|P_X)$ may jump at some $R$.  
Then, in 1990, Ahlswede gave an example that 
Marton's exponent is not continuous in $R$, using a distortion measure with a special structure~\cite{Ahlswede1990}. 
This implies Eq.(\ref{E_M}) is a non-convex optimization problem 
and Blahut's and Marton's exponents do not match in general. 
The difficulty lies in that we can select any distortion measure $d(x,y)\geq 0$ on 
$x\in \mathcal{X}, y\in \mathcal{Y}$. 
We normally expect $d(x,y)$ is a function similar to the distance between $x$ and $y$ and,
fortunately, it is guaranteed that 
if $d(x,y)$ belongs to an important class of distortion measure that 
can be expressed as a function that only depends on the difference
of $x$ and $y$, then $E_{\rm M}(R|P_X)$ is a convex function of $R$\cite{ ArikanMerhav1998}, which immediately implies that $E_{\rm M}(R|P_X) = E_{\rm B}(R|P_X)$.
Hamming distortion is included in this class.  
See also \cite{BergerTEXTBOOK} and \cite[Exercise 9.5]{Csiszar-KornerBook}.

Hereafter, when it is obvious from the context, we omit the symbol for the set of probability distribution such as $\mathcal{P(X)}$ and $\mathcal{P(Y|X)}$ and write only the probability distribution under $\max$ and $\min$ symbols.

\section{Algorithms for the channel-coding strong converse exponent}
The algorithms for computing the channel coding strong converse exponent are discussed in this section. 
Arimoto's algorithm~\cite{Arimoto1976} was the first for computing the strong converse exponent, which is based on Arimoto's expression (\ref{G_AR}). 
Usually we are interested not only in the value of the exponent function, i.e. the maximum or the minimum value of the objective function, but also in the probability distribution that attains the optimal value. 
The optimal distribution is not necessarily unique. 
In Arimoto's algorithm as well as other algorithms described in this paper, 
the probability distribution converges to one of the optimal distributions 
when the optimal distribution is not unique.

About 40 years later, the authors~\cite{YutakaISIT2015} proposed a new algorithm for computing the strong converse exponent based on Dueck and K\"orner's expression (\ref{G_DK}). 
Subsequently, Tridenski and Zamir~\cite{Tridenski2018arXiv} proposed another algorithm that was based on their strong converse exponent expression (\ref{G_TZ}). 
Then, Tridenski et al.~\cite{Tridenski2020arXiv} proposed a family of algorithms which includes~\cite{YutakaISIT2015} and \cite{Tridenski2018arXiv} as special cases. 

The main purpose of this section is to illustrate the overall picture of Fig.~\ref{fig.1}. 
We begin with the algorithm family of Tridenski et al.~\cite{Tridenski2020arXiv}
and then consider important special cases.
Section~\ref{section:new_algorithms} explains Tridenski and Zamir's algorithm. 
We will show the outstanding property of this algorithm that the objective function takes alternately the forms of the objective functions appeared in Arimoto's and Dueck and K\"orner's exponent.
Thus, the convergence of the algorithms to the global minimum directly implies the match of the two exponents. This paper is the first to point out such a desirable property of Tridenski and Zamir's exponent. 
Then, in Section~\ref{section:new_algorithm2}, we describe the properties of another special case denoted by Algorithm~\ref{new_algorithm_B}.
The importance of this special case is clarified for the first time in this paper. 
The surrogate objective function $J_{\bm{t}_2}^{(\lambda)}(q_{XY}, \hat p_{X|Y})$ used for Algorithm~\ref{new_algorithm_B} plays an important role for the error exponent, too. 
This topic is described in Section~\ref{section:Er_and_E_CK_match}. 
We then compare Tridenski and Zamir's algorithm and Algorithm~\ref{new_algorithm_B} with Arimoto's and our previously proposed algorithm.

Before proceeding to Section~\ref{section:new_algorithms}, we review the basic properties of Dueck and K\"orner's exponent, which are commonly utilized in 
all the algorithms in Fig.~\ref{fig.1} except for Arimoto's algorithm.
Put $\Gamma_{\min} = \min_{x\in \mathcal{X}} c(x)$.
Let 
$C(\Gamma|W) = \displaystyle 
\sup_{
\substack{
p_X \in \mathcal{P(X)}: \\
\mathrm{E}_{p_X}[c(X)] \leq \Gamma
}
} 
I(p_X, W)$ be the channel capacity under input constraint $(c,\Gamma)$. 
Dueack and K\"orner's exponent satisfies the following property.

\begin{property}
\begin{enumerate}[(a)]
    \item For a fixed $\Gamma \geq \Gamma_{\min}$, 
    $ G_{\rm DK}(R, \Gamma |W)$ is monotone non-decreasing function of $R$. 
    For a fixed $R\geq 0$, 
    $ G_{\rm DK}(R, \Gamma |W)$ is monotone non-increasing function of $\Gamma$. 
    \item $G(R,\Gamma|W)$ is a convex function of $(R,\Gamma)$. 
    \item For $0 \leq R \leq C(\Gamma| W)$, $G_{\rm DK}(R, \Gamma |W)=0$ and 
     for $R > C(\Gamma|W)$, $G_{\rm DK}(R, \Gamma |W)$ is strictly positive.  
\end{enumerate}
\end{property}
{\it Proof:} 
See~\cite{Oohama_str_conv_theorem_DMCswithcost} for the proof.

Then, we give a parametric expression of the
exponent function. To this aim, we define 
the following functions: 
\begin{definition}
For any fixed $\lambda \in [0,1]$, we define 
\begin{align}
    G_{\rm DK}^{(\lambda)}(\Gamma|W)
    = \min_{\substack{q_{XY}:}{\mathrm{E}_{q_X} [ c(X) ] \leq \Gamma} }
    \{ D(q_{Y|X} || W | q_X ) - \lambda I(q_X, q_{Y|X} ) 
    \}.
\end{align}
\end{definition}
\begin{definition}
For fixed $\lambda \in [0,1]$ and
$\mu\ge 0$, we define  
\begin{align}
\Theta^{(\lambda, \mu )}(q_{XY}|W)
&= D(q_{Y|X} || W | q_X) -\lambda I(q_X, q_{Y|X} ) + \mu \mathrm{E}_{q_X}[c(X)]
\notag\\
&=
\mathrm{E}_{ q_{XY} } \left[
\log \frac{ q_{Y|X}^{1 - \lambda}(Y|X)
 q_{Y}^{\lambda}(Y) 
}
{ W(Y|X) {\rm e}^{-\mu c(X)} }
\right], 
\label{Theta}\\
\Theta^{(\lambda, \mu)}(W) &= \min_{q_{XY}} \Theta^{(\lambda, \mu)}(q_{XY}|W). 
\label{Optimal_Theta}
\end{align}
\end{definition}
It was shown~\cite[Appendix C]{Jitsumatsu_Oohama_IT_Trans2020} that, 
for a fixed $\lambda\in[0,1]$ and $\mu\geq 0$, the function  
$\Theta^{(\lambda, \mu)}(q_{XY}|W)$ is 
convex in $q_{XY}$. 
The following lemma states that the Dueck and K\"orner's exponent
is obtained by evaluating the minimum of $\Theta^{(\lambda, \mu )}(q_{XY}|W)$.
\begin{lemma}[\cite{Jitsumatsu_Oohama_IT_Trans2020}]
For any fiex $\lambda\in[0,1]$, $\Gamma\geq 0$, and $W\in \mathcal{P(Y|X)}$, we have
\begin{align}
    G_{\rm DK}^{(\lambda)} ( \Gamma | W) 
    = \sup_{\mu\ge 0} \{ - \mu \Gamma + \Theta^{(\lambda,\mu)} \}.
    \label{G_DK_lambda}
\end{align}
For any $R\geq 0$, $\Gamma\geq 0$ and $W\in \mathcal{P(Y|X)}$, we have 
\begin{align}
G_{\rm DK}(R, \Gamma|W)
&=
\max_{0\le \lambda \le 1} \{ \lambda R + G_{\rm DK}^{(\lambda)}(\Gamma |W) \} \label{eq.G_DK_Legendre1} \\
&= 
\max_{0\le \lambda \le 1}
\sup_{\mu \geq 0}
\left\{
\lambda R -\mu \Gamma 
+ \Theta^{(\lambda, \mu)}(W) 
\right\}.
\label{eq.G_DK_Legendre}
\end{align}
\end{lemma}

Computation of the exponent functions involves the Legendre-Fenchel transformation (LFT)~\cite{Rockafellar}. The LFT of a function $F(x)$ of an $n$-dimensional vector $x$ is defined by 
\begin{align}
    F^*(y) = \sup_{x\in \mathbb{R}^n} \{ x^T y - F(x) \},
    \label{def:LFT}
\end{align}
where $x^T$ is the transpose of $x$.
We use the LFT to derive a parametric expression of the exponent functions, 
by considering supporting lines to the curve
of exponent functions. This is the same as 
the computation of the rate distortion function and the channel capacity under cost-constraint, which is obtained by considering the supporting lines of the rate-distortion function and the capacity-cost function~\cite[Chapter 8]{Csiszar-KornerBook}.

\begin{figure*}
    \centering
    \includegraphics[scale=0.6]{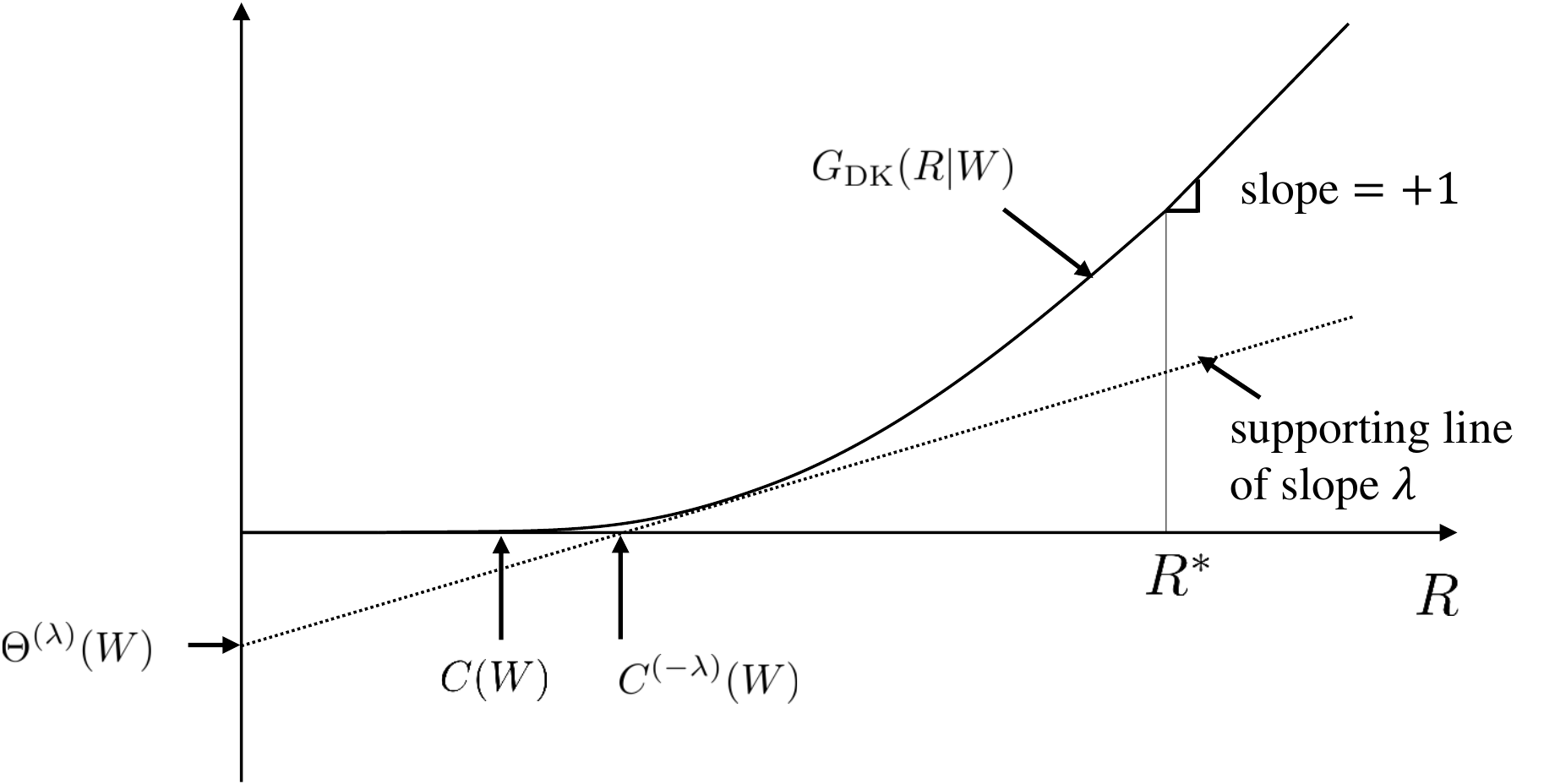}
    \caption{The strong converse exponent function $G_{\rm DK}(R,\Gamma|W)$ for a fixed $\Gamma$ and the generalized cutoff rate in channel coding}
    \label{fig:strong_converse_exponent_channel}
\end{figure*}

By comparing Eq.(\ref{G_DK_lambda}) with (\ref{def:LFT}),
we see that for fixed $\lambda \in [0,1]$, 
$G_{\rm DK}^{(\lambda)}(\Gamma|W)$ as a function of $y=\Gamma$ 
is the one-dimensional LFT of $ -\Theta^{(\lambda, \mu)}(W)$ as a function of 
$x=-\mu \in (-\infty,0]$. Then, 
by comparing Eq.(\ref{eq.G_DK_Legendre1}) with (\ref{def:LFT}),
we see that for a fixed $\Gamma$, 
$G_{\rm DK}(R, \Gamma |W)$ is the one-dimensional LFT of 
$-G_{\rm DK}^{(\lambda)}(\Gamma|W)$ as a function of $\lambda \in [0,1]$.
This implies that $G_{\rm DK}(R, \Gamma |W)$ is computed by the 
two-dimensional LFT of $-\Theta^{(\lambda,\mu)}(W)$ 
as a function of $ x  = (\lambda, -\mu) \in [0,1]\times(-\infty,0]$. 
The numerical computation of the LFT can be performed efficiently.  
See, for example, \cite{Lucet}.  
Thus, the target for the algorithm is to compute an optimal $q_{XY}^* = q_{XY}^*(\lambda, \mu)$ that attains
$\Theta^{(\lambda,\mu)}(W)$ for any given $\lambda\in [0,1]$ and $\mu\geq 0$. 
Fig.~\ref{fig:strong_converse_exponent_channel} shows a rough sketch of the error and the strong converse exponents in channel coding. 
In Fig.~\ref{fig:strong_converse_exponent_channel}, for a fixed $\Gamma\geq 0$, 
$G_{\rm DK}(R, \Gamma |W)$ as a function of $R$ is depicted by a solid curve, 
and its supporting line of slope $\lambda$, 
expressed by $\lambda R + G_{\rm DK}^{(\lambda)}(\Gamma|W)$, is depicted by dotted line. 
The $R$-axis intercept of this line is called the generalized cutoff rate~\cite{Csiszar1998}, denoted 
by $C^{(-\lambda)}(\Gamma|W)$ in this figure. 
If $R\leq C(\Gamma|W)$, $G_{\rm DK}(R,\Gamma|W) =0$, while 
if $R \ge R^*$, $G_{\rm DK}(R,\Gamma|W)
= R + G_{\rm DK}^{(1)}(\Gamma|W)
$, where  $R^* = \left[ \frac{\partial}{\partial \lambda}
\{ G_{\rm DK}^{(\lambda)}(\Gamma|W) \} 
\right]_{\lambda \to 1-}$. See~\cite[Lemma 10]{Jitsumatsu_Oohama_IT_Trans2020} for detail.

For given $\lambda \in (0,1)$, $\mu\geq 0$ and $W\in\mathcal{P(Y|X)}$, we do not have general formula for the explicit expression of the optimal joint distribution that attains (\ref{Optimal_Theta}). 
For $\lambda = 0$ and $1$, one can easily see that $\Theta^{(0,\mu)}(W) = \min_{q_{XY}}  \{ D(q_{Y|X} || W |q_X) + \mu \mathrm{E}_{q_X} [c(X)] = \mu \min_{x} c(x) =\mu \Gamma_{\rm min}$ holds and $\Theta^{(1, \mu)}(W)$ is evaluated as follows: 
\begin{align}
\Theta^{(1, \mu)}(W) 
&=\min_{ q_{XY} }
\Theta^{(1, \mu)}(q_{XY}|W) 
 = 
\min_{ q_{XY} } 
{\rm E}_{q_{XY} }
\left[ 
\log \frac{q_{Y}(Y)}{ W(Y|X) {\rm e}^{-\mu c(X) }}
\right] \notag\\
& = 
\min_{ q_{XY} } 
\sum_{y}
q_Y(y) 
\left\{ 
\log q_Y(y) 
-
\sum_{x}
q_{X|Y}(x|y) \log {W(y|x) {\rm e}^{-\mu c(x) } }
\right\}
\notag\\
& 
\stackrel{\rm (a)}
=
\min_{ q_{Y} } 
\sum_{y}
q_Y(y) 
[ 
\log q_Y(y) - \log \max_{x} W(y|x) {\rm e}^{-\mu c(x) }
]
\notag\\
&
\stackrel{\rm (b)}
= -\log \sum_{y} \max_{x} W(y|x) {\rm e}^{-\mu c(x) }, 
\end{align}
where (a) holds with equality if and only if $q_{X|Y}(x'|y) =1$
for $(x',y)$ such that $x' = \arg \max_{x} W(y|x)$ for all $y\in \mathcal{Y}$
and (b) holds with equality if and only if 
$q_Y(y) = \displaystyle 
\frac{\max_{x} W(y|x) {\rm e}^{-\mu c(x) } } { \sum_{y' \in \mathcal{Y} } \max_{x} W(y'|x) {\rm e}^{-\mu c(x) } }$.

We now perform a change of variables: let $\mu = \lambda \nu$, and instead of the parameter pair $(\lambda, \mu)$, let $(\lambda, \nu)$ be the variable parameter pair.
By doing so, in the limit of $\lambda \to 0$, one of the new algorithms is reduced to the Arimoto-Blahut algorithm that computes the channel capacity under cost constraints. 

In the following subsections, we describe algorithms shown in Fig.~\ref{fig.1}.
We begin with the algorithm family of Tridenski et al.~\ref{section:generalized_algorithm_TSZ}. 
The algorithm family includes Tridenski and Zamir's algorithm~\cite{Tridenski2018arXiv} as a special case,
which is explained in Section~\ref{section:new_algorithms}. 
Then, another special case, Algorithm~\ref{new_algorithm_B}, is explained in Section~\ref{section:new_algorithm2}. 
Comparison with Arimoto's and the authors' algorithms are given in Sections \ref{section:arimoto_channel_coding} and \ref{section:our_previous_algorithm}.

\subsection{Algorithm family of Tridenski et al.}
\label{section:generalized_algorithm_TSZ}


In~\cite{Tridenski2020arXiv}, Tridenski et al. proposed a family of computation algorithms that includes two previously proposed algorithms as spacial cases\footnote{ 
As in the case of~\cite{Tridenski2018arXiv, Tridenski2018ISIT}, there are two types of the algorithm families.
One is for a fixed rate $R$ and the other is for a fixed slope parameter $\lambda$. The latter is discussed here. }. 
One is our algorithm~\cite{Jitsumatsu_Oohama_IT_Trans2020, YutakaISIT2015} and the other is Tridenski and Zamir's algorithm~\cite{Tridenski2018arXiv}.
The objective function for the algorithm family is~\cite[Eq.(24)]{Tridenski2020arXiv} 
\begin{align}
    J_{\bm{t}}^{(\lambda, \nu)}(q_{XY}, p_{XY} |W)
    = \Theta^{(\lambda, \lambda \nu )} (q_{XY}|W) + (1-\lambda) D_{\bm{t}}(q_{XY}, p_{XY}), 
    \label{def:F_TSZ}
\end{align}
where $\bm{t}=(t_1, t_2, t_3, t_4)$ is a vector of four non-negative coefficients and 
\begin{align}
    D_{\bm{t}}(q_{XY}, p_{XY})
    &= 
    t_1 D(q_X || p_X) + t_2 D(q_{Y|X} || p_{Y|X} | q_X )\notag\\
    &\quad + t_3 D(q_Y || p_Y) + t_4 D(q_{X|Y} || p_{X|Y} | q_Y ).
    \label{def:D_t}
\end{align}

From the definition, we have the following lemma.
\begin{lemma}
\label{lemma5}
For any fixed $\lambda \in [0,1)$, $\nu\geq 0$, $\bm{t}\in \mathcal{T}$, 
and any fixed $q_{XY}\in \mathcal{P(X\times Y)}$, 
$J_{\bm{t}}^{(\lambda, \nu)}(q_{XY}, p_{XY} |W)$ is minimized by $p_X = q_X$ and 
its minimum value is
\begin{align}
    J_{\bm{t}}^{(\lambda, \nu)}(q_{XY}, q_{XY} |W) = \Theta^{(\lambda, \lambda\nu)}(q_{XY}|W).
\end{align}
This implies that 
\begin{align}
& \min_{q_{XY} } \min_{p_{XY}} J_{\bm{t}}^{(\lambda, \nu)} ( q_{XY}, p_{XY}|W) \notag \\
&= \min_{q_{XY} }  J_{\bm{t}}^{(\lambda, \nu)} ( q_{XY}, q_{XY} |W) \notag \\
&= \min_{q_{XY} }  \Theta^{(\lambda, \lambda \nu)} ( q_{XY} |W) = \Theta^{(\lambda, \lambda \nu)}(W).
    \label{eq.in.lemma5}
\end{align}
\end{lemma}
\textit{Proof:}
This lemma follows from Eqs.~(\ref{def:F_TSZ}) and the condition for $D_{\bm{t}}( q_{XY}, p_{XY} )=0$.
\hfill \IEEEQED

Vontobel et al.~\cite{Vontobel} pointed out that to derive an Arimoto-Blahut type algorithm for a given optimization problem, say $\min_{P} F(P)$, it is required to find a surrogate objective function $\Psi(P,Q)$ that satisfies the following three conditions:
\begin{itemize}
    \item[a)] The surrogate function takes the same value as $F(P)$ at $Q=P$, i.e., $\Psi(P,P)=F(P)$.
    \item[b)] $\Psi(P,Q)$ is never below $F(P)$, i.e., $\Psi(P,W)\geq F(P)$ for all $Q$.
    \item[c)] Minimizing $\Psi(P,Q)$ over $P$ can be done in a computationally efficient way.
\end{itemize}
Assume a $\Psi$ satisfies the three conditions. Then, by the iterative algorithm $P^{[i+1]} = \arg\min_{P} \Psi(P, P^{[i]})$ with a given initial value $P^{[0]}$, the value of the objective function $F(P^{[i]})$ is monotone non-increasing. To prove convergence of the algorithm to its optimal value, a separate proof for each objective function $F(P)$ is needed.
For the current optimization problem, we replace $F$, $\Psi$, $P$ and $Q$, respectively with 
$\Theta^{(\lambda, \lambda\nu)}(\cdot |W)$,
$J_{\bm{t}} ^{(\lambda, \nu)}(\cdot |W)$, $q_{XY}$, and $p_{XY}$.
It is easily confirmed that the condition a) and b) are satisfied in this case. 
Regarding the condition c), Tridenski et al. showed the condition on $t_1, t_2, t_3$ and $t_4$ 
so that optimal $q_{XY} $  that minimizes $J_{\bm{t}} ^{(\lambda, \nu)}( q_{XY}, p_{XY} |W)$ for a fixed $p_{XY}$ is expressed explicitly. 
The condition is expressed by the following lemma~\cite{Tridenski2020arXiv}. 
\begin{lemma}[Lemma 3 in\cite{Tridenski2020arXiv}]
\label{lemma3:Tridenski2020arXiv}
Define 
$\mathcal{T} = \{ \bm{t}=(t_1, t_2, t_3, t_4): t_1\geq 0, t_2\geq 0, t_3\geq 0, t_4\geq 0\}$ and
\begin{align}
    \mathcal{T}_1 &=
    \{ (t_1, t_2, t_3, t_4) \in \mathcal{T}: t_1=t_2+1\}, \notag\\
    \mathcal{T}_2 &=
    \{ (t_1, t_2, t_3, t_4) \in \mathcal{T} : t_4=t_3+\lambda/(1-\lambda)\} . 
\end{align}
If $\bm{t} \in \mathcal{T}_1 \cup \mathcal{T}_2$, 
then for $\lambda \in (0,1)$, 
the optimal distribution $q_{XY}$ that attains the minimum of $J_{\bm{t}}^{(\lambda, \nu)}(q_{XY}, p_{XY}|W)$ for a fixed $p_{XY}$ is expressed explicitly. 
\end{lemma}

The parameterized surrogate objective function $ J_{\bm{t}}^{(\lambda, \nu)}(q_{XY}, p_{XY}|W) $ produces
a parameterized alternative expression of the exponent. 
To illustrate this, we define a set of representations of the exponent as
\begin{align}
{\mathcal{F}} & =\{ \min_{q_{XY}} \min_{p_{XY}} J_{ \bm{t} }^{(\lambda, \nu)} (q_{XY}, p_{XY} |W )\}_{\bm{t} 
\in \mathcal{T}} , \\
        \mathcal{F}_i &= \left\{ \min_{q_{XY}} \min_{p_{XY}} 
    J_{\bm{t}} ^{(\lambda, \nu) } ( q_{XY} , p_{XY} | W ) 
    \right\} _{\bm{t} \in \mathcal{T}_i }, \quad i=1,2. \label{def:F_i}
\end{align}
We also define 
\begin{align}
    \hat J_{ \bm{t} } ^{(\lambda, \nu)} (p_{XY} |W ) 
    = \min_{q_{XY}} J_{ \bm{t} } ^{(\lambda, \nu)}(q_{XY}, p_{XY} |W ),
    \label{def:hat_J_t}
\end{align}
\begin{align}
    \widehat{\mathcal{F}}_i &= \left\{  \min_{p_{XY}} 
    \hat J_{\bm{t}} ^{(\lambda, \nu) } ( p_{XY} | W ) 
    \right\} _{\bm{t} \in \mathcal{T}_i }, \quad i=1,2. \label{def:hat_F_i}
\end{align}

\begin{figure*}
\centering
\includegraphics[width=0.95\textwidth]{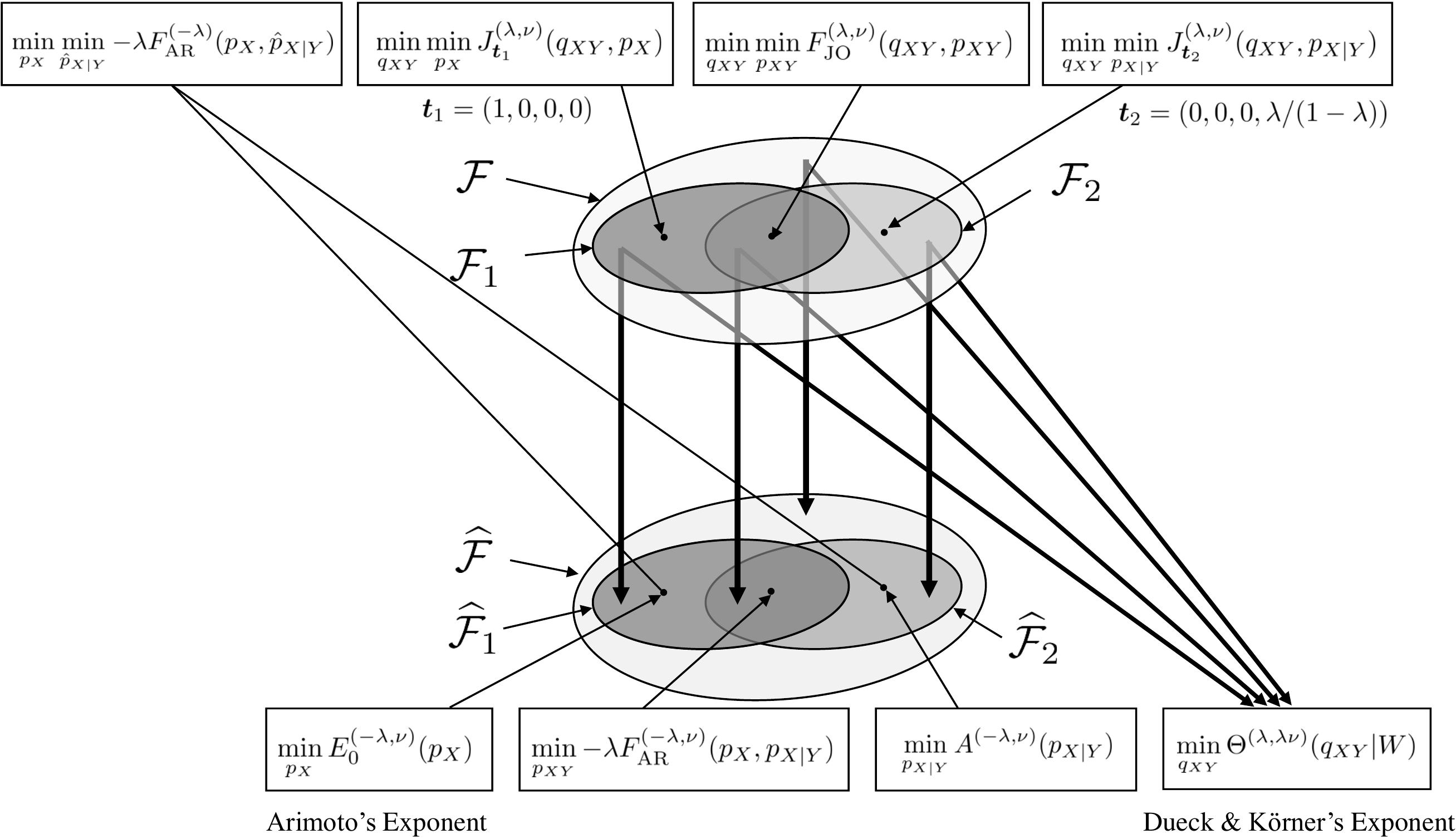}
\caption{
Tridenski et al.'s algorithm family corresponds to the set $\mathcal{F}$ of surrogate objective functions for Dueck and K\"orner's exponent. $\widehat{\mathcal{F}}$ is the set of alternative expressions of Dueck and K\"orner's exponent derived from the surrogate objective functions parameterized by $\bm{t}$, which includes the expression of Arimoto's error exponent, and two other expressions as special cases.
}
\label{fig:algorithm_family_channel_coding}
\end{figure*}

Fig.~\ref{fig:algorithm_family_channel_coding} illustrates the relation between
Dueck and K\"orner's exponent, 
$\mathcal{F}$, $\mathcal{F}_i$, $\widehat{\mathcal{F}}$,
$\widehat{\mathcal{F}}_i$, and some special cases. 
Note that $\mathcal{F}$ and $\widehat{\mathcal{F} }$ are the set of expressions of the minimization problems and 
the values of all elements in $\mathcal{F}$ and $\widehat{\mathcal{F}}$ 
coincide because 
$$
\min_{q_{XY}} \Theta^{(\lambda, \lambda \nu)}(q_{XY}|W)
= \min_{q_{XY}} \min_{p_{XY}} J_{ \bm{t} }^{(\lambda, \nu)} (q_{XY}, p_{XY} |W ) 
= \min_{p_{XY}} \hat J_{ \bm{t} }^{(\lambda, \nu)} (p_{XY} |W ) 
$$ 
holds for any $\bm{t} \in \mathcal{T}$.
Then, the large circle at the top of Fig.\ref{fig:algorithm_family_channel_coding} expresses
the set $\mathcal{F}$. 
The small circles $\mathcal{F}_1$ and $\mathcal{F}_2$ are the subsets of $\mathcal{F}$ for which $\bm{t}$ belongs
to $\mathcal{T}_1$ and $\mathcal{T}_2$, respectively. 
The large circle at the bottom of Fig.\ref{fig:algorithm_family_channel_coding} expresses
the set $\widehat{\mathcal{F}}$, and  $\widehat{\mathcal{F}}_1$
and $\widehat{\mathcal{F}}_2$ are the subsets of $\widehat{\mathcal{F}}$.
Tridenski and Zamir's algorithm~\cite{Tridenski2018arXiv}, described in Section~\ref{section:new_algorithms}, 
corresponds to the case 
$\bm{t} = \bm{t}_1 := 
(1, 0, 0, 0)\in \mathcal{T}_1$
and the authors' previous algorithm~\cite{Jitsumatsu_Oohama_IT_Trans2020, YutakaISIT2015}, described in Section~\ref{section:our_previous_algorithm}, corresponds to the case 
$\bm{t}= 
(1, 0, 0, \lambda/(1-\lambda))\in \mathcal{T}_1 \cap \mathcal{T}_2$.  
The expression $\min_{p_X} E_0^{(-\lambda, \nu)} (p_X|W)$ belongs to $\widehat{\mathcal{F}}_1$,
$\min_{ p_{X|Y} } A^{(-\lambda, \nu)} (p_{X|Y} |W)$ belongs to $\widehat{\mathcal{F}}_2$,
and $\min_{ p_{XY} } -\lambda F_{\rm AR}^{(-\lambda, \nu)} (p_X, p_{X|Y} |W)$ 
belongs to the intersection of $\mathcal{\widehat{F}}_1$ and $\mathcal{\widehat{F}}_2$. 
Each element in $\mathcal{F}$ has two arrows. The arrow heading straight down means that
$\hat J_{\bm{t}}^{(\lambda,\nu)} ( p_{XY}|W)$ is obtained by executing the minimization with respect to $ q_{XY} $,
whereas the arrow heading right down means that
$\Theta^{(\lambda, \lambda\nu)} ( q_{XY}|W)$ is obtained by executing the minimization with respect to $ p_{XY} $.

\begin{algorithm}
\caption{
Algorithm family of Tridenski et al.~\cite{Tridenski2020arXiv} 
}
\label{alg:TSZ_generalized}
\begin{algorithmic}
    \Require The conditional probability of the channel $W$, 
    the cost function $c$, 
    $\lambda \in (0,1)$ and $\nu\geq 0$.
    \State Choose $\bm{t} \in \mathcal{T}_1 \cup \mathcal{T}_2$ and
    choose initial joint probability distribution $q_{XY}^{[0]}$ such that
    the set $\{ q_{XY}: J_{\bm{t}}^{(\lambda, \nu)} (q_{XY}, q_{XY}^{[0]} | W ) < +\infty \}$ is non-empty. 

\For{ $i=0,1,2,\ldots$,}
\If{$\bm{t}\in \mathcal{T}_1$}, 
\begin{align}
    q_{XY}^{[i+1]}(x,y) 
    &= \frac{1}{L_1} 
    q_Y^{[i]}(y)^{ \frac{(1-\lambda) (t_2+t_3) }{ 1 + (1-\lambda) ( t_2+t_3) } } 
    \left\{ 
    q_X^{[i]}(x)  
    q_{X|Y}^{[i]}(x|y)^{ t_2 + t_4 } 
    \{ W(y|x) \mathrm{e}^{ -\lambda \nu c(x) }  \}^\frac{1}{1-\lambda} 
    \right\}^{ \frac{1}{1+t_2+t_4} }\notag\\
    &\cdot \left\{\sum_{x\in \mathcal{X} } \left\{ 
    q_X^{[i]}(x)  
    q_{X|Y}^{[i]}(x|y)^{t_2 + t_4} \{ W(y|x) \mathrm{e}^{ -\lambda \nu c(x) } \}^\frac{1}{1-\lambda} 
    \right\}^{ \frac{1}{1+t_2+t_4} }
    \right\}^{ \frac{ (1 - \lambda) ( t_4 - t_3 ) -\lambda }{ 1 + (1-\lambda) ( t_2+t_3) }  },  
    \label{eq.29}
\end{align}
\State{where $L_1$ is a normalization factor.}
\ElsIf{$\bm{t}\in \mathcal{T}_2$}, 
\begin{align}
    q_{XY}^{[i+1]}(x,y)
    &=\frac{1}{L_2}
    q_X^{[i]}(x)^{\frac{(1-\lambda) (t_1 + t_3) }{\lambda+(1-\lambda)(t_1+t_3)}} 
    \left\{
    q_{Y|X}^{[i]}(y|x) ^{(1-\lambda)(t_2+t_3) }
    q_{X|Y}^{[i]}(x|y) ^{\lambda}
    W(y|x) \mathrm{e}^{ -\lambda \nu c(x) } 
    \right\}^{ \frac{1}{ 1 + (1-\lambda)(t_2+t_3) } } \notag\\
    &  \cdot 
    \left\{ 
    \sum_{y\in \mathcal{Y} }
    \left\{
    q_{Y|X}^{[i]}(y|x) ^{(1-\lambda) (t_2+t_3) }
    q_{X|Y}^{[i]}(x|y) ^{ \lambda }
    W(y|x) \mathrm{e}^{ -\lambda \nu c(x) } 
    \right\}^{ \frac{1}{1+(1-\lambda)(t_2+t_3) } }
    \right\}^{ \frac{ (1 - \lambda) ( 1 - t_1 + t_2 ) }
        { \lambda+ (1-\lambda) ( t_1 + t_3) } }, \label{eq.30}
\end{align}
\State{where $L_2$ is a normalization factor.}
\EndIf 
\EndFor
\end{algorithmic}
\end{algorithm}

Algorithm~\ref{alg:TSZ_generalized} shows the Tridenski et al.'s algorithm family\footnote{They appear to have typographic errors in~\cite[Lemma 3]{Tridenski2020arXiv}. 
The parameters $t_2$, $t_3$, and $t_4$ were misspelled as $t_4$, $t_2$, and $t_3$, respectively.}.
Then, the convergence of the probability distribution was stated in~\cite[Theorem 2]{Tridenski2020arXiv} as
\begin{theorem}[\cite{Tridenski2020arXiv}]
\label{theorem.convergence.algorithm.family}
Let $\{ q_{XY}^{[i]} \}_{i=0}^{+\infty}$ be generated by Algorithm~\ref{alg:TSZ_generalized}.
Then, $ \hat J_{ \bm{t} } ^{(\lambda, \nu)}(q_{XY}^{[i]}|W)$ converges to 
\begin{align}
    \min_{ \genfrac{}{}{0pt}{} { q_{XY} \in\mathcal{P(X\times Y)}: }{ D_{ \bm{t} } (q_{XY}|| q_{XY}^{[0]})  < +\infty} } \hat J_{ \bm{t} }^{(\lambda, \nu)}(q_{XY}^{[i]}|W).
\end{align} 
\end{theorem}
See~\cite{Tridenski2020arXiv} for the proof. 

The above theorem states that $q_{XY}^{[i]}$ converges to the optimal 
distribution that minimizes 
$ \hat J_{ \bm{t} }^{(\lambda, \nu)}(q_{XY}^{[i]}|W)$. Does
this optimal distribution is the same as the one that minimizes 
$\Theta^{(\lambda, \lambda\nu)}(q_{XY}|W)$?
The following proposition provides the answer to this question.
\begin{proposition}
For any $\lambda\in (0,1)$, $\nu\geq 0$, and $\bm{t}\in \mathcal{T}_1 \cup \mathcal{T}_2$,
we have
\begin{align} 
& 
\hat J_{\bm{t} } ^{(\lambda, \nu)} (q_{XY}^{[i]} |W) 
= 
J_{ \bm{t} } ^{(\lambda, \nu)}( q_{XY}^{[i+1]}, q_{XY}^{[i]} |W )
\notag\\
&\geq
J_{ \bm{t} } ( q_{XY}^{[i+1]}, q_{XY}^{[i+1]} |W )
=\Theta^{(\lambda, \lambda \nu)}  ( q_{XY}^{[i+1]} | W) \notag\\
& 
\geq
\hat J_{\bm{t} } ^{(\lambda, \nu)} (q_{XY}^{[i+1]} |W) 
= 
J_{ \bm{t} } ^{(\lambda, \nu)}( q_{XY}^{[i+2]}, q_{XY}^{[i+1]} |W )
\notag\\
& 
\geq
\min_{p_{XY} } \hat J_{\bm{t}}^{(\lambda, \nu)} (p_{XY} | W )
=
\min_{p_{XY} } \min_{ q_{XY} } J_{\bm{t}}^{(\lambda, \nu)} (q_{XY}, p_{XY} | W ) 
\notag\\
    &=
    \min_{ q_{XY} } J_{\bm{t}}^{(\lambda, \nu)} (q_{XY}, q_{XY} | W )  
    = 
    \min_{ q_{XY} } \Theta^{(\lambda, \lambda \nu)} ( q_{XY} | W ) . 
\end{align}
\end{proposition}
\begin{IEEEproof}
The first inequality follows from the definition of $J_{\bm{t}}^{(\lambda, \nu)}(q_{XY}, p_{XY}|W)$
and the non-negativity of the divergence 
and the second inequality follows from (\ref{def:hat_J_t}). 
\end{IEEEproof}
From this proposition, $q_{XY}^{[i]}$ is guaranteed to converges to the
optimal distribution that minimizes $\Theta^{(\lambda, \lambda\nu)}(q_{XY}|W)$. 

Here we give three remarks on Algorithm~\ref{alg:TSZ_generalized}. 
The first one is that the probability update rules for $\mathcal{T}_1$ and $\mathcal{T}_2$ were parameterized with two parameters by letting $a=(1-\lambda)(t_2+t_3)$, $b=(1-\lambda)(t_2+t_4)$, and $c=(1-\lambda)(t_1+t_3)$ in\cite{Tridenski2020arXiv}. The reduction of the number of parameters is possible because we have $ D ( q_{XY} || p_{XY} ) = 
D ( q_{Y|X} || p_{Y|X} | q_X ) 
+ D( q_X || p_X ) 
= D(q_{X|Y}||p_{X|Y}|q_Y) 
+ D(q_Y||p_Y) $ 
and therefore $t_1, t_2, t_3$, and $t_4$ are redundant. 
Hence, the function $J_{\bm{t}}^{(\lambda, \nu)}(q_{XY}, p_{XY}|W)$ of $\bm{t}=(t_2+1, t_2, t_3, t_4)\in \mathcal{T}_1$ is the same 
as that of $\bm{t}=(1, 0, t_2+t_3, t_2+t_4)$ and that of 
$\bm{t} =(t_1, t_2, t_3, t_3 + \lambda/(1-\lambda) )\in \mathcal{T}_2$ is the same 
as that of $\bm{t}=(t_1+t_3, t_2+t_3, 0, \lambda/(1-\lambda))$.  
The second is that 
by selecting the initial distribution  $q_{XY}^{[0]}$ so that every element $(x,y) \in \mathcal{X\times Y}$
is non-negative, we can avoid the situation that $J_{\bm{t}}^{(\lambda, \nu)} (q_{XY}, q_{XY}^{[0]} | W ) = +\infty $ for some $q_{XY}$ and that $D_{ \bm{t} } (q_{XY}|| q_{XY}^{[0]}) =+\infty$ for some $q_{XY}$. 
A possible choice for $q_{XY}^{[0]}$ is the uniform distribution. 
The last remark is that 
%
%
The left hand sides of Eqs.(\ref{eq.29}) and (\ref{eq.30}) are joint distributions but they can be separated
$(q_{X|Y}^{[i+1]}, q_Y^{[i+1]})$ and $(q_{Y|X}^{[i+1]}, q_X^{[i+1]})$ respectively. 
The following lemma will be used in the following subsections.
\begin{lemma}
\label{lemma.4}
Assume $\lambda\in (0,1)$, $\lambda\geq 0$, and $p_{XY}$ are fixed. 
If $\bm{t}\in \mathcal{T}_1$, then $J_{\bm{t}}^{(\lambda, \nu)}(q_{XY}, p_{XY}|W) $ is minimized by
\begin{align}
    q_{X|Y}(x|y) 
    &= \frac1{ K_1(y) } 
     \left\{ 
     p_X(x) p_{X|Y}^{ t_2 + t_4 } (x|y)
    \{ 
      W(y|x) \mathrm{e}^{ -\lambda \nu c(x) }  
    \}^\frac{1}{1-\lambda} 
    \right\}^{ \frac{1}{1+t_2+t_4} }, \label{eq.233}\\
    q_Y(y) 
    &=
    \frac{ 
      \left\{ 
        p_Y^{ (1-\lambda) (t_2+t_3) } (y) 
        K_1^{ (1 - \lambda) ( 1 + t_2 + t_4 ) } (y) 
      \right\}^{ 
       \frac{ 1 }{ 1 + (1-\lambda) ( t_2 + t_3 ) }  
         }
    }
    {
      \sum_{ y' \in \mathcal{Y} } 
    \left\{
    p_Y^{ (1-\lambda) (t_2+t_3) } (y') 
    K_1^{ (1 - \lambda) ( 1 + t_2 + t_4 )  }(y')
    \right\}^{\frac{ 1}{ 1 + (1-\lambda) ( t_2+t_3) } }
    }, \label{eq.234}
\end{align}
where
\begin{align}
    K_1(y) &= \sum_{x\in \mathcal{X} } \left\{ p_X(x)  
    p_{X|Y}(x|y)^{t_2 + t_4} \{ W(y|x) \mathrm{e}^{ -\lambda \nu c(x) } \}^\frac{1}{1-\lambda} 
    \right\}^{ \frac{1}{1+t_2+t_4} }. 
\end{align}
The minimum value is
\begin{align}
    & 
    \hat J_{\bm{t}}^{(\lambda, \nu)} ( p_{XY} | W )
    \notag\\
    & =
    -\{ 1 + (1-\lambda) (t_2+t_3) \}
    \log 
    \sum_{y}
    p_Y^{ \frac{(1-\lambda) (t_2+t_3) }{ 1 + (1-\lambda) ( t_2+t_3) } }(y) \notag\\
    &\quad \cdot \left[
    \sum_{x} 
    \left\{
    p_X(x)  p_{X|Y}^{t_2 + t_4}(x|y) 
    \{ W(y|x) \mathrm{e}^{ -\lambda \nu c(x) } \} ^{\frac{1}{1-\lambda}} 
    \right\}^{ \frac{1}{ 1 + t_2 + t_4 }}
    \right]^{ \frac{ (1 - \lambda) ( 1 + t_2 + t_4 ) }{ 1 + ( 1-\lambda) ( t_2+t_3) } } . 
    \label{eq.232}
\end{align}

If $\bm{t}\in \mathcal{T}_2$, then 
$J_{\bm{t}}^{(\lambda, \nu)}(q_{XY}, p_{XY}|W) $ is minimized by
\begin{align}
    q_{Y|X}(y|x) 
    &=
    \frac{1}{ K_2(x) }
    \left\{
    p_{Y|X}^{(1-\lambda)(t_2+t_3) }(y|x) 
    p_{X|Y}^{\lambda} (x|y) 
    W(y|x) \mathrm{e}^{ -\lambda \nu c(x) } 
    \right\}^{ \frac{1}{ 1 + (1-\lambda)(t_2+t_3) } } , \label{eq.239}\\
    q_{X}(x)
    &= 
    \frac{ 
      \left\{
      p_X^{ (1-\lambda) (t_1 + t_3) } (x)
        K_2^{  1+ (1 - \lambda) ( t_2 + t_3 ) } (x) 
      \right\}^{ 
      \frac{1}{\lambda+(1-\lambda)(t_1+t_3) }
      }
    }
    { 
    \sum_{x' \in \mathcal{X} } 
      \left\{
      p_X^{ (1-\lambda) (t_1 + t_3) } (x')
        K_2^{  1+ (1 - \lambda) ( t_2 + t_3 ) } (x') 
      \right\}^{ 
      \frac{1}{\lambda+(1-\lambda)(t_1+t_3) }
      }
    } ,
    \label{eq.240} 
\end{align}
where
\begin{align}
        K_2(x) &= \sum_{y\in \mathcal{Y} }
    \left\{
    p_{Y|X}^{(1-\lambda) (t_2+t_3) }(y|x) 
    p_{X|Y}^{ \lambda }(x|y) 
    W(y|x) \mathrm{e}^{ -\lambda \nu c(x) } 
    \right\}^{ \frac{1}{1+(1-\lambda)(t_2+t_3) } }.
\end{align}
The minimum value is 
\begin{align}
    & 
    \hat J_{\bm{t}}^{(\lambda, \nu)} ( p_{XY} | W )
    \notag\\
    & =
    -\{ \lambda + (1-\lambda) (t_1+t_3) \}
    \log 
    \sum_{x}
    p_X(x)^{\frac{(1-\lambda)(t_1 + t_3) }{ \lambda + (1-\lambda)(t_1+t_3) }} \notag \\
    &\quad \cdot \left[
    \sum_{y}
    \left\{
    p_{Y|X}^{(1-\lambda) (t_2+t_3) }(y|x) 
    p_{X|Y}^{ \lambda }(x|y) 
    W(y|x) \mathrm{e}^{ -\lambda \nu c(x) }  
    \right\}^{ \frac{1}{1 + (1-\lambda)(t_2+t_3) } }
    \right]^{ \frac{ 1 + (1 - \lambda) ( t_2 + t_3 ) }
    { \lambda + (1-\lambda) ( t_1 + t_3) } } . 
\end{align}
\end{lemma}
The procedure of the derivation is the same as the proof of Theorem~\ref{theorem.convergence.algorithm.family}
and the proof is omitted here.

From this lemma, we have the following corollary,
which shows the relation between $J_{\bm{t}}^{(\lambda, \nu)}(q_{XY}, p_{X}|W)$ 
and Arimoto's exponent. 
This paper is the first to point out such an important relationship. 
\begin{corollary}
\label{corollary1}
For $\bm{t}=\bm{t}_1=(1,0,0,0)\in \mathcal{T}_1$, 
any fixed $ \lambda \in [ 0, 1 ) $, $\nu\geq 0$, 
and any fixed $p_X$, 
$ J_{\bm{t}_1}^{(\lambda, \nu)}(q_{XY}, p_X | W) $
is minimized by 
\begin{align}
q_{X|Y}(x|y) 
&= \frac{
p_X(x) \{ W(y|x) {\rm e}^{-\lambda \nu c(x)} \}^{\frac{1}{1-\lambda}}
}
{\sum_{x'}
p_X(x') \{ W(y|x') {\rm e}^{-\lambda \nu  c(x')} \}^{\frac{1}{1-\lambda}}
}, \label{eq.lemma6.1}\\
q_Y(y)
&=
\frac
{
\left\{
\sum_{x}
p_X(x) \{ W(y|x) {\rm e}^{-\lambda \nu  c(x)} \}^{\frac{1}{1-\lambda}}
\right\}^{1-\lambda}
}
{
\sum_{y'}
\left\{
\sum_{x}
p_X(x) \{ W(y'|x) {\rm e}^{-\lambda \nu c(x)} \}^{\frac{1}{1-\lambda}}
\right\}^{1-\lambda}
}. \label{eq.lemma6.2}
\end{align}
Denote the joint distribution computed from the above $q_{X|Y}$ and $q_Y$ by
$q_{XY}^*(p_X)$.
The minimum value of $J_{\bm{t}_1}^{(\lambda, \nu)}(q_{XY}, p_X | W)$ for a fixed $p_X$ 
is
\begin{align}
J_{\bm{t}_1}^{(\lambda, \nu)}( q_{XY}^*(p_X), p_X | W)
&=
-\log \sum_y \left[
\sum_{x} p_X(x) \{ W(y|x) {\rm e}^{-\lambda \nu c(x)} \}^{\frac{1}{1-\lambda} } 
\right]^{1-\lambda} 
\notag\\
&= E_0^{(-\lambda, \nu)}(p_X|W). 
\end{align}
This implies that 
\begin{align}
    &\min_{ q_{XY} } \min_{ p_X }
    J_{\bm{t}_1}^{(\lambda, \nu)}(q_{XY}, p_X | W)\notag \\
    &=
    \min_{ p_X }
    J_{\bm{t}_1}^{(\lambda, \nu)}(q_{XY}^*(p_X), p_X | W)\notag \\
&=\min_{ p_X } E_0^{(-\lambda, \nu)}(p_X|W).
\label{eq.in.lemma6}
\end{align}
\end{corollary}
This corollary will be used in the next subsection.
We also give the following corollary, which will be used Section~\ref{section:new_algorithm2}.
In order to describe the corollary, we use 
the following function that appears in Arimoto's algorithm~\cite{Arimoto1976}.
\begin{definition}
For a given transition probability $W(y|x)$ and 
$0< |\rho| \le 1$, $\nu\geq 0$, we define 
\begin{align}
& A^{(\rho, \nu)} ( \hat p_{X|Y} | W ) \notag \\
& = 
\rho \log  \sum_{x} \mathrm{e}^{-\nu c(x) }
\left[
\sum_{y} \hat p_{X|Y}^{-\rho}(x|y) W(y|x) 
\right]^{ -1/\rho } .
\label{A_rho}
\end{align}
\end{definition}

\begin{corollary}
\label{corollary2}
For $\bm{t}=\bm{t}_2 = (0,0,0,\lambda/(1-\lambda) )\in \mathcal{T}_2$, 
any fixed $\lambda\in(0,1]$, $\nu\geq 0$, and any fixed $p_{X|Y}$, 
$J_{\bm{t}_2}^{(\lambda, \nu)}(q_{XY}, \hat p_{X|Y}| W)$ is minimized by
\begin{align}
q_{Y|X}(y|x) 
&= \frac{
\hat p_{X|Y}^\lambda (x|y) W(y|x) 
}{
\sum_{y'}
\hat p_{X|Y}^\lambda (x|y') W(y'|x) 
}, \\
q_X(x)
	&=
\frac{
\mathrm{e}^{-\nu c(x) }
\left\{
\sum_{y}
\hat p_{X|Y}^\lambda (x|y) W(y|x) 
\right\}^{1/\lambda}
}
{
\sum_{x'}
\mathrm{e}^{-\nu c(x') }
\left\{
\sum_{y}
\hat p_{X|Y}^\lambda (x'|y) W(y|x') 
\right\}^{1/\lambda}
	}. 
\end{align}
Denote the joint distribution calculated from the above $q_{Y|X}$ and $q_X$ by $\tilde q_{XY}(p_{X|Y})$.
Then, the minimum value of $J_{\bm{t}_2}^{(\lambda, \nu)}(q_{XY}, \hat p_{X|Y}| W)$ for fixed $p_{X|Y}$ is 
\begin{align}
J_{\bm{t}_2}^{(\lambda, \nu )} (\tilde q_{XY}(p_{X|Y}), \hat p_{X|Y} | W )
&= -\lambda  \log \sum_x 
{\rm e}^{-\nu c(x) } 
\left[
\sum_{y} \hat p_{X|Y}^{\lambda}(x|y) W(y|x) 
\right]^{1/\lambda}\notag\\
&= A^{ ( -\lambda, \nu  ) }( \hat p_{X|Y} | W ) . 
\end{align}
This implies that 
\begin{align}
    \min_{q_{XY} } \min_{\hat p_{X|Y}} 
    J_{\bm{t}_2}^{(\lambda, \nu) } (q_{XY}, \hat p_{X|Y} |W)
    &= \min_{\hat p_{X|Y}}
    J_{\bm{t}_2}^{(\lambda, \nu) } ( \tilde q_{XY}(\hat p_{X|Y}), \hat p_{X|Y} |W)\notag\\
    &= \min_{\hat p_{X|Y}} 
    A^{ ( -\lambda, \nu ) }( \hat p_{X|Y} | W ). 
\label{eq.in.lemma8}
\end{align}
\end{corollary}

It is important to note that the algorithm family includes the algorithm of Tridenski and Zamir~\cite{Tridenski2018arXiv} and the algorithms previously presented by the authors~\cite{YutakaISIT2015,Jitsumatsu_Oohama_IT_Trans2020} as special cases, allowing multiple algorithms to be discussed from a unified point of view. 
On the other hand, it is more prospective to focus on individual algorithms in order to describe their relationship to the Arimoto algorithm. 
In the following subsections, the three special cases are described in detail.

\subsection{ Algorithm~\ref{new_algorithm_A}
}
\label{section:new_algorithms}
We call Tridenski and Zamir's algorithm~\cite{Tridenski2018arXiv} as Algorithm~\ref{new_algorithm_A}.
They proposed two types of algorithms. 
One is the algorithm for computing Tridenski and Zamir's exponent (\ref{G_TZ})
for fixed $R$ without using the slope parameter~\cite{Tridenski2018ISIT, Tridenski2018arXiv} and the other is for fixed slope\cite[Section VII]{Tridenski2018arXiv}. This subsection describes the latter.
Although this algorithm is a special case of the algorithm family (Algorithm~\ref{alg:TSZ_generalized}),
it has attractive properties and therefore we will discuss this algorithm separately.

For describing Tridenski and Zamir's algorithm, we need to 
show that $ G_{\rm TZ}(R, \Gamma | W) $ coincides with $ G_{\rm DK}(R, \Gamma | W) $.
Tridenski and Zamir introduced the following function\cite[Section VII]{Tridenski2018arXiv}:
\begin{align}
    F_{\rm TZ}^{(\lambda)} (q_{XY}, p_X | W )
    &= 
    D(q_{XY} || p_X \circ W) - \lambda D(q_{XY} || p_X \times q_Y) 
    \label{def:F_TZ}
\end{align}
The following lemma holds. 
\begin{lemma}
\label{lemma2}
For fixed $\lambda \in [0,1) $ and 
$q_{XY}$, $ F_{\rm TZ}^{(\lambda)}(q_{XY}, p_X | W) $ is minimized 
by $p_X=q_X$  and its minimum value is 
\begin{align}
F_{\rm TZ}^{(\lambda)}(q_{XY}, q_X | W)
= 
D(q_{Y|X} || W | q_X) - \lambda I(q_{X}, q_{Y|X} ).
\end{align}
\end{lemma}

\noindent 
\textit{Proof:} 
By the definition of the ordinary and the conditional divergences, we have 
\begin{align}
    F_{\rm TZ}^{(\lambda)} (q_{XY}, p_X | W )
    & 
    = \mathrm{E}_{ q_{XY} } 
    \left[ 
    \frac{ q_{XY}(X,Y) }{ p_X(X) W(Y|X) }
    \frac{ q_{X|Y}^{-\lambda}(X|Y) }{ p_X^{-\lambda}(X) }
    \right] \notag\\
    & 
    = \mathrm{E}_{ q_{XY} } 
    \left[ 
    \frac{ q_{Y|X}^{1-\lambda}(Y|X) q_Y^\lambda(Y)  }{ W(Y|X) }
    \frac{ q_{X}^{1-\lambda}(X) }{ p_X^{1-\lambda}(X) }
    \right] \notag \\
    &= D(q_{Y|X} || W | q_X) - \lambda I(q_{X}, q_{Y|X} ) + (1-\lambda)D(q_X||p_X) . 
\end{align}
Because $D(q_X||p_X)$ is non-negative and takes zero if and only if $p_X = q_X$, 
we have
$
\min_{p_X}
F_{\rm TZ}^{(\lambda)} (q_{XY}, p_X | W )
= D(q_{Y|X} || W | q_X) - \lambda I(q_{X}, q_{Y|X} ) 
$, which completes the proof. \hfill\IEEEQED

Then we have the following lemma. 
\begin{lemma}
\label{lemma:GDK_GTZ}
For any $R\geq 0$ and $\Gamma\ge 0$, we have
\begin{align}
    G_{\rm DK}(R, \Gamma |W) &= G_{\rm TZ}(R, \Gamma|W) . 
\end{align}
\end{lemma}
In~\cite{TridenskiISIT2017, Tridenski2018arXiv}, the proof of the
the match of $G_{\rm TZ}(R, \Gamma |W)$ and $G_{\rm AR}(R, \Gamma |W)$ for a DMC without input-cost was given.
Because the match of $G_{\rm AR}(R, \Gamma |W)$ and $G_{\rm DK}(R, \Gamma |W)$ is a known fact~\cite{OohamaIEICE2018}, this suggests
$G_{\rm TZ}(R, \Gamma |W) = G_{\rm DK}(R, \Gamma |W)$, too. 
However, it should be better if we can show this match directly.

\textit{Proof:} 
We have the following chain of equalities: 
\begin{align}
    &G_{\rm TZ}(R, \Gamma |W) \notag\\
    &=
    \min_{
    \substack{ q_{XY}: \\
    \mathrm{E}_{q_{X}} [c(X)]\leq \Gamma } 
    }
    \min_{p_X} 
    \left\{
    D(q_{XY} || p_X \circ W)
    + | R - D(q_{X|Y} || p_X | q_Y ) |^+
    \right\}\notag\\
    & 
    \stackrel{\rm (a)}
    =
    \min_{
    \substack{ q_{XY}: \\
    \mathrm{E}_{q_{X}} [c(X)]\leq \Gamma } 
    }
    \min_{p_X} 
    \left\{
    D(q_{XY} || p_X \circ W)
    + \max_{0\leq \lambda\leq 1} \lambda [ R - D(q_{XY} || p_X \times q_Y ) ]
    \right\} \notag\\
    & \stackrel{\rm (b)}
    =
    \min_{
    \substack{ q_{XY}: \\
    \mathrm{E}_{q_{X}} [c(X)]\leq \Gamma } 
    } 
    \max_{0\leq \lambda\leq 1} 
    \min_{p_X} 
    \left\{ \lambda R + 
    D(q_{XY} || p_X \circ W)
    - \lambda     
    D(q_{XY} || p_X \times q_Y ) 
    \right\}  \notag\\
    & \stackrel{\rm (c)}
    =
    \min_{
    \substack{ q_{XY}: \\
    \mathrm{E}_{q_{X}} [c(X)]\leq \Gamma } 
    }
    \max_{0\leq \lambda\leq 1} 
    \left\{ \lambda R + 
    D(q_{Y|X} || W | q_X) - \lambda I(q_{X}, q_{Y|X}) 
    \right\} \notag\\
    &
    = 
    \min_{
    \substack{ q_{XY}: \\
    \mathrm{E}_{q_{X}} [c(X)]\leq \Gamma } 
    }
    \{
    D(q_{Y|X} || W | q_X) + \max_{0\leq \lambda \leq 1}
    \lambda [ R- I(q_X, q_{Y|X}) ] \} \notag\\
    &
    \stackrel{\rm (d)}
    =
    \min_{
    \substack{ q_{XY}: \\
    \mathrm{E}_{q_{XY}} [c(X)]\leq \Gamma } 
    } \{ D(q_{Y|X} || W | q_X) + | R- I(q_X, q_{Y|X}) |^+\}
    \notag\\
    &= G_{\rm DK}(R, \Gamma |W)
\end{align}
Step (a) and (d) follows from the identity $[x]^+ = \max_{ 0\leq \lambda \leq 1} \lambda x$.
Step (b) follows from the mini-max theorem because the expression in the parenthesis is convex in $p_X$ and linear in $\lambda$ for a fixed $q_{XY}$.
Step (c) follows from Lemma~\ref{lemma2}. 
\hfill\IEEEQED

Now, we describe Tridenski and Zamir's algorithm for fixed slope parameter $\lambda$. 
This algorithm corresponds to the case of $\bm{t}=\bm{t}_1 \coloneqq (1,0,0,0) \in \mathcal{T}_1$. The objective function
of Tridenski and Zamir's algorithm is therefore 
\begin{align}
J_{\bm{t}_1}^{(\lambda, \nu)} (q_{XY}, p_X | W )
&=
 J_{(1,0,0,0)}^{ (\lambda, \nu) }( q_{XY}, p_{XY} | W ) \notag\\
& = 
\Theta^{(\lambda, \lambda \nu)} (q_{XY}|W) + (1-\lambda ) D(q_X||p_X) . 
\label{eq.def.J_1}
\end{align}
Trindenski and Zamir's algorithm is based on the double minimization
of $J_{\bm{t}_1}^{(\lambda, \nu)}(q_{XY}, p_X |W)$ with respect to $q_{XY}$ and $p_X$.
From Lemma~\ref{lemma2} and the definition of $\Theta^{(\lambda, \lambda \nu)}(q_{XY}|W)$,
for any $\lambda \in [0,1]$ and $\nu\geq0$, 
we have 
\begin{align}
    J_{\bm{t}_1}^{(\lambda, \nu)} (q_{XY}, p_X |W)
    &=
    F_{\rm TZ}^{(\lambda)}(q_{XY} , p_X | W) + \lambda \nu \mathrm{E}_{q_X} [ c(X) ].
    \label{eq.26}
\end{align}
Tridenski and Zamir's original algorithm in~\cite{Tridenski2018arXiv} is the algorithm for
the channel without cost constraint. 
Eq.(\ref{eq.26}) shows that the function $ J_{\bm{t}_1}^{(\lambda, \nu)} (q_{XY}, p_X |W) $
is the extension of $F_{\rm TZ}^{(\lambda)}(q_{XY} , p_X | W) $ to channels under input cost.

The function $J_{\bm{t}_1}^{(\lambda, \nu)}(q_{XY}, p_{X}|W)$ satisfies the following property.
\begin{property}
\label{property1}
For 
any fixed $-1\leq \lambda<1$, $\nu\geq 0$ and any fixed $p_{X}$, 
$J_{\bm{t}_1}^{(\lambda, \mu)}(q_{XY}, p_X|W)$ is convex in 
$q_{XY}$.
For any fixed $-1\leq \lambda<1$, $\nu\geq 0$ and any fixed $q_{XY}$,
$J_{\bm{t}_1}^{(\lambda, \nu )}(q_{XY}, p_X|W)$ is convex in 
$p_X$.
\end{property}
See Appendix~\ref{appendixA} for the proof.




Tridenskii  and Zamir's algorithm 
shown in Algorithm~\ref{new_algorithm_A} in obtained by setting $t_2=t_3=t_4=0$ in Lemma~\ref{lemma.4}.
The probability update rule follows from Corollary~\ref{corollary1}.
Here, Eq.(\ref{algorithm_3_update_c}) is a computation of the marginal distribution from $(q_Y^{[i]}, q_{X|Y}^{[i]})$.

\begin{algorithm}
\caption{
Tridenski and Zamir's algorithm~\cite{Tridenski2018arXiv,Tridenski2020arXiv}}
\label{new_algorithm_A}
\begin{algorithmic}
    \Require The conditional probability of the channel $W$, 
    the cost function $c$, 
    $\lambda \in (0,1)$ and $\nu\geq 0$.
    Choose any initial joint probability distribution $p_{X}^{[0]}$. 
\For{ $i=0,1,2,\ldots$,}
\State
\begin{align}
q_{X|Y}^{[i]}(x|y) 
&= \frac{
p_X^{[i]}(x) \{ W(y|x) \mathrm{e}^{- \lambda \nu c(x) } \}^{\frac{1}{1-\lambda}}
}
{\sum_{x'}
p_X^{[i]}(x') \{ W(y|x') \mathrm{e}^{- \lambda \nu c(x') } \}^{\frac{1}{1-\lambda}}
}, 
\label{Update3}
\\
q_Y^{[i]}(y)
&=
\frac
{
\left\{
\sum_{x}
p_X^{[i]}(x) \{ W(y|x) \mathrm{e}^{-\lambda \nu c(x)} \}^{\frac{1}{1-\lambda}}
\right\}^{1-\lambda}
}
{
\sum_{y'}
\left\{
\sum_{x}
p_X^{[i]}(x) \{ W(y'|x) \mathrm{e}^{-\lambda \nu c(x')} \}^{\frac{1}{1-\lambda}}
\right\}^{1-\lambda}
}, 
\label{algorithm_3_update_b}
\\
p_X^{[i+1]}(x)
&=
\sum_{y}
q_{X|Y}^{[i]}(x|y) q_Y^{[i]}(y) . 
\label{algorithm_3_update_c}
\end{align}
\EndFor
\end{algorithmic}
\end{algorithm}

Because Algorithm~\ref{new_algorithm_A} is a member of the algorithm family,
the convergence theorem of this algorithm is supported by Theorem~\ref{theorem.convergence.algorithm.family}.

Now, we can state the following attractive property of Tridenski and Zamir's algorthim:
\begin{proposition}
\label{proposition_new_alg_A}
For $i=1,2\ldots$, we have
\begin{align*}
 &  J_{\bm{t}_1}^{(\lambda, \nu)}(q_{XY}^{[0]}, p_X^{[0]}|W)
    \stackrel{\rm (a)} 
    \geq 
    J_{\bm{t}_1}^{(\lambda, \nu)}(q_{XY}^{[0]}, p_X^{[1]}|W)
    \stackrel{\rm (b)} 
    \geq 
    J_{\bm{t}_1}^{(\lambda, \nu)}(q_{XY}^{[1]}, p_X^{[1]}|W)
    \cdots\\
 &  \geq J_{\bm{t}_1}^{(\lambda, \nu)}(q_{XY}^{[i]}, p_X^{[i]}|W) = E_0^{(-\lambda, \nu)}(p_X^{[i]}|W) \\
 &  \stackrel{\rm (a)} 
    \geq J_{\bm{t}_1}^{(\lambda, \nu)}(q_{XY}^{[i]}, p_X^{[i+1]}|W) = \Theta^{(\lambda, \lambda \nu)}(q_{XY}^{[i]}|W)\\
 &  \stackrel{\rm (b)} 
    \geq J_{\bm{t}_1}^{(\lambda, \nu)}(q_{XY}^{[i+1]}, p_X^{[i+1]}|W) = E_0^{(-\lambda, \nu)}(p_X^{[i+1]}|W)\geq 
    \cdots\\
 & \geq 
 \min_{q_{XY} } \Theta^{(\lambda, \lambda \mu)}(q_{XY}|W)
 \stackrel{\rm (c)}
 =
 \min_{p_X} E_0^{(-\lambda, \mu)}(p_X|W).
\end{align*}
\end{proposition}

\textit{Proof:} Step (a) follows from Lemma~\ref{lemma5} and
step (b) follows from Corollary~\ref{corollary1}. 
Step (c) follows from (\ref{eq.in.lemma5}) in Lemma~\ref{lemma5} and
(\ref{eq.in.lemma6}) in Corollary~\ref{corollary1}.
This completes the proof. \hfill$\IEEEQED$

A feature of Algorithm~\ref{new_algorithm_A} indicated by Proposition~\ref{proposition_new_alg_A} is that the $E_0$-function for Arimoto's strong converse exponent and the $\Theta^{(\lambda)}(q_{XY}|W)$ for the Csisz\'ar and K\"orner's exponent appear alternately.
The convergence of Algorithm~\ref{new_algorithm_A} immediately gives
the proof of the fact that Dueck and K\"orner's and Arimoto's exponents
coincide, stated in the following proposition.
\begin{proposition}[\cite{Dueck_Korner1979}]
\label{proposition:GDKmatchGAR}
Suppose the channel transition probability $W\in \mathcal{P(Y|X)}$ is given.
Then, for any $R\geq 0$, $\Gamma\geq 0$, we have 
\begin{align}
  G_{\rm DK}(R, \Gamma|W) = G_{\rm AR}(R, \Gamma|W).  
\end{align}
\end{proposition}

\begin{IEEEproof}
We have the following chain of equalities:
\begin{align}
    G_{\rm AR}(R, \Gamma|W)
    &
    \stackrel{\rm (a)}
    = 
    \sup_{\lambda \in [0,1) } \sup_{\nu\geq 0}
    \left\{
    \min_{p_X} E_0^{(-\lambda, \nu)} ( p_X |W) + \lambda R - \lambda \nu \Gamma
    \right\} \notag \\
    &
    \stackrel{\rm (b)}
    =
    \sup_{\lambda \in [0,1) }  \sup_{\nu\geq 0}
    \left\{
    \min_{ q_{XY} } \Theta^{(\lambda, \lambda \nu)}(q_{XY} |W) + \lambda R - \lambda \nu \Gamma
    \right\} \notag \\
    &
    \stackrel{\rm (c)}
    =
    \max_{\lambda \in [0,1] }  \sup_{\nu\geq 0}
    \left\{
    \min_{ q_{XY} } \Theta^{(\lambda, \lambda \nu)}(q_{XY} |W) + \lambda R - \lambda \nu \Gamma 
    \right\} \notag \\
    &
    \stackrel{\rm (d)}
    = G_{\rm DK}(R, \Gamma |W).
\end{align}
Step (a) follows from definition~\ref{G_AR},
Step (b) follows from the equation (c) in Proposition~\ref{proposition_new_alg_A},
Step (c) holds because we have $\lim_{\lambda\to 1} \Theta^{ ( \lambda, \lambda \nu ) }(q_{XY}, p_X|W)
= \Theta^{(1,\nu)}(q_{XY}, p_X|W)$ for a fixed $q_{XY}$, 
and Step (d) follows from Definition~\ref{G_DK}.
\end{IEEEproof}

The match of Dueck and K\"orner's  and  Arimoto's strong converse exponents is well-known
but no proof was written in~\cite{Dueck_Korner1979}. An explicit proof for the channel 
with cost constraint is found in~\cite{OohamaIEICE2018}. 
In the above proof, the convergence of Tridenski and Zamir's algorithm plays a critical role. 
This proof is interesting because it differs significantly from previously known proofs.

In Section~\ref{section:arimoto_channel_coding}, we will compare Tridenski and Zamir's algorithm
with Arimoto's algorthm, where update rule (\ref{Update3}) is the same as in Arimoto algorithm.
Another update rule (\ref{algorithm_3_update_b}) is not explicitly used in the Arimoto algorithm, 
but we will give the view that (\ref{algorithm_3_update_b}) is a hidden update rule of the Arimoto algorithm.

\subsection{Algorithm \ref{new_algorithm_B}}
\label{section:new_algorithm2}
In this subsection, we consider another special case, $\bm{t}=\bm{t}_2\coloneqq (0, 0, 0, \lambda/(1-\lambda)) \in \mathcal{T}_2$. 
This case has not been discussed in depth, but is important in the sense that 
during the alternate updates of the probability distribution, 
the corresponding algorithm alternates between 
the form of the Dueck and K\"orner's exponents 
and the form of the alternative representation of the Arimoto's strong converse exponent.
By extending the domain of $\lambda$ from $[0,1]$ to $[-1,1]$, 
we will derive an important property for the two types of expressions of the error exponent, 
which is the topic of Section~\ref{section:Er_and_E_CK_match}.
Because $\bm{t}_2$ with negative $\lambda$ is not included in $\mathcal{T}$,
the property of $J_{\bm{t}_2}^{(\lambda, \nu)}(q_{XY}, \hat p_{X|Y}|W)$ in this case
is described in this chapter. 

We use the following objective function of $q_{XY}$ and $\hat {p}_{X|Y}$:
\begin{align}
J_{\bm{t}_2}^{(\lambda, \nu)}(q_{XY}, \hat p_{X|Y} |W) 
& = J_{(0,0,0,\frac{\lambda}{1-\lambda})} ^{(\lambda, \nu)}
(q_{XY}, \hat p_{XY}|W) \notag\\
& = \Theta^{(\lambda, \lambda \nu )} (q_{XY}|W) +
\lambda D(q_{X|Y} || \hat p_{X|Y} | q_Y) \notag\\
&=
{\rm E}_{q_{XY}} \left[
\log \frac{ q_{Y|X}(Y|X) q_X^{\lambda}(X)} 
{ \hat p_{X|Y}^{\lambda}(X|Y) W(Y|X) \mathrm{e}^{-\lambda \nu  c(X) } }
\right] \label{definition_J_2}
\end{align}
The reason why the conditional distribution is denoted as $\hat p_{X|Y}$ instead of $p_{X|Y}$ in (\ref{definition_J_2}) is to unify the notation of probability distributions in Fig.\ref{fig.1}.
The first thing to check is the convexity of this function.
Here, we check it for $\lambda\in [-1, 1]$. Positive $\lambda$ is for strong converse exponent
and negative $\lambda$ is for error exponent.
\begin{property}
\label{property2}
\begin{enumerate}
\item[a)]  
For any fixed $\lambda \in [0,1]$ (resp. $\lambda \in [-1,0]$), $\nu\geq 0$, and any fixed $q_{XY}$,
$J_{\bm{t}_2}^{(\lambda, \nu)}(q_{XY}, \hat p_{X|Y}|W)$ is convex (resp. concave) in $\hat p_{X|Y}$.
\item[b)]
For any fixed $\lambda \in [0,1]$ (resp. $\lambda \in [-1,0]$), $\nu\geq 0$, 
any fixed $\hat p_{X|Y}$, and any fixed $ q_{Y|X}$, 
$J_{\bm{t}_2}^{(\lambda, \nu)}((q_{X}, q_{Y|X}), \hat p_{X|Y}|W)$ is convex (resp. concave) in $q_{X}$.
\item[c)]
For any fixed $-1 \le \lambda \le 1$, $\nu\geq 0$,
any fixed $q_{X}$ and $p_{X|Y}$,
$J_{\bm{t}_2}^{(\lambda, \nu)}((q_{X}, q_{Y|X}), \hat p_{X|Y}|W)$ is 
convex in $q_{Y|X}$.
\end{enumerate}
\end{property}

See Appendix~\ref{appendixA} for the proof.

An algorithm for computing the strong converse exponent based on
Lemma~\ref{lemma5} and Lemma~\ref{lemma.4} with $\bm{t}=\bm{t}_2$
is shown in Algorithm~\ref{new_algorithm_B}. 
Eq.(\ref{algorithm_4_update_c}) is simply a
conditional probability of backward channel
computed from $ ( q_X^{[i]}, q_{Y|X}^{[i]} ) $.

\begin{algorithm}
\caption{Computation of $ \min_{q_{XY}} \min_{\hat p_{X|Y}} J_{\bm{t}_2}^{(\lambda, \nu)}(
q_{XY}, \hat p_{X|Y}|W)$}
\label{new_algorithm_B}
\begin{algorithmic}
    \Require The conditional probability of the channel $W$, 
    $\lambda \in (0,1)$, and $\nu\geq 0$.
    Choose any initial joint probability distribution $p_{X|Y}^{[0]}$.
    such that all $p_{X|Y}^{[0]}(x|y)$ are positive. 
\For{ $i=0,1,2,\ldots$,}
\State
\begin{align}
q_{Y|X}^{[i]}(y|x) 
&= \frac{
\hat p_{X|Y}^{[i]} (x|y)^\lambda W(y|x) 
}{
\sum_{y'}
\hat p_{X|Y}^{[i]}(x|y')^\lambda W(y'|x)  
} , 
\label{algorithm_4_update_a}
\\
q_X^{[i]}(x)
	&=
\frac{
\mathrm{e}^{-\nu c(x)}
\left\{
\sum_{y}
\hat p_{X|Y}^{[i]} (x|y)^\lambda W(y|x) 
\right\}^{1/\lambda}
}
{
\sum_{x'}
\mathrm{e}^{-\nu c(x')}
\left\{
\sum_{y}
\hat p_{X|Y}^{[i]} (x'|y)^\lambda W(y|x') 
\right\}^{1/\lambda}
	} , 
\label{algorithm_4_update_b}
	\\
\hat p_{X|Y}^{[i+1]}(x|y)
&=
\frac{
q_{Y|X}^{[i]}(y|x) q_X^{[i]}(x) 
}{ 
\sum_{x'}
q_{Y|X}^{[i]}(y|x') q_X^{[i]}(x') 
}. 
\label{algorithm_4_update_c}
\end{align}
\EndFor
\end{algorithmic}
\end{algorithm}

Similarly to the Proposition~\ref{proposition_new_alg_A}, 
the following proposition holds:
\begin{proposition}
\label{proposition_new_alg_B}
For $i=1,2\ldots$, we have
\begin{align*}
 &  J_{\bm{t}_2}^{(\lambda, \nu)}(q_{XY}^{[0]}, \hat p_{X|Y}^{[0]}|W)
    \stackrel{\rm (a)} 
    \geq 
    J_{\bm{t}_2}^{(\lambda, \nu)}(q_{XY}^{[0]}, \hat p_{X|Y}^{[1]}|W)
    \stackrel{\rm (b)} 
    \geq 
    J_{\bm{t}_2}^{(\lambda, \nu)}(q_{XY}^{[1]}, \hat p_{X|Y}^{[1]}|W)
    \cdots\\
 &  \geq 
    J_{\bm{t}_2}^{(\lambda, \nu)}(q_{XY}^{[i]}, \hat p_{X|Y}^{[i]}|W) 
    = A^{(-\lambda,\nu)}(\hat p_{X|Y}^{[i]}|W) \\
 &  \stackrel{\rm (a)} 
    \geq 
    J_{\bm{t}_2}^{(\lambda, \nu)}(q_{XY}^{[i]}, \hat p_{X|Y}^{[i+1]}|W) 
    = \Theta^{(\lambda,\lambda \nu)}(q_{XY}^{[i]}|W)\\
 &  \stackrel{\rm (b)} 
    \geq 
    J_{\bm{t}_2}^{(\lambda, \nu)}(q_{XY}^{[i+1]}, \hat p_{X|Y}^{[i+1]}|W) 
    = A^{(-\lambda,\nu)}(\hat p_{X|Y}^{[i+1]}|W)\geq 
    \cdots\\
 & \geq 
 \min_{q_{XY} } \Theta^{(\lambda,\lambda\nu)}(q_{XY}|W)
 \stackrel{\rm (c)}
 =
 \min_{p_{X|Y}} A^{(-\lambda,\nu)}(\hat p_{X|Y}|W).
\end{align*}
\end{proposition}

\textit{Proof:} Step (a) follows from Lemma~\ref{lemma5} and
step (b) follows from Corollary~\ref{corollary2}. 
Step (c) follows from (\ref{eq.in.lemma5}) in Lemma~\ref{lemma5} and
(\ref{eq.in.lemma8}) in Corollary~\ref{corollary2}.
This completes the proof. \hfill$\IEEEQED$

Proposition~\ref{proposition_new_alg_B} shows that 
the objective function takes the forms of the alternative expression 
of Arimoto's exponent and the parametric expression of Dueck and K\"orner's 
exponent alternately and that 
$J_{\bm{t}_2}^{(\lambda, \nu)}(q_{XY}^{[i]}, 
\hat p_{X|Y}^{[i+1]} |W)
= 
\Theta^{(\lambda, \lambda \nu)}(q_{XY}^{[i]} | W) 
$ monotonically decreases. 
Next theorem shows that 
$
\Theta^{(\lambda, \lambda\nu)}(q_{XY}^{[i]} | W) 
$ converges to 
$ \Theta^{(\lambda,\lambda \nu)}(W)$.
\begin{theorem}
\label{theorem_convergence_algorithm4}
For any $\lambda\in (0,1]$, $\nu\geq 0$ and $W\in \mathcal{P(Y|X)}$, 
the series of distributions $q_{XY}^{[i]}$ defined by (\ref{algorithm_4_update_a})-(\ref{algorithm_4_update_c}) converges 
to an optimal distribution $ q_{XY}^*$
that minimizes $\Theta^{(\lambda, \lambda\nu)}(q_{XY}|W)$.
The approximation error
$ \Theta^{(\lambda, \lambda\nu)}( q_{XY}^{[i]} | W ) - \Theta^{(\lambda,\lambda\nu)}(q_{XY}^*|W)$
is inversely proportional to the number of iterations.
\end{theorem}
See Appendix~\ref{appendixC} for the proof.

We give a remark that  
when $\lambda = 0$, Algorithm~\ref{new_algorithm_A} reduces to
$p_X^{[i+1]}(x) = p_X^{[i]}(x)$ and when $\lambda =1$
Algorithm~\ref{new_algorithm_B} reduces to
$\hat p_{X|Y}^{[i+1]}(x) = \hat p_{X|Y}^{[i]}(x)$. 
In these cases, the probability distributions are not updated.
Therefore, we cannot use Algorithm~\ref{new_algorithm_A} and 
Algorithm~\ref{new_algorithm_B} respectively when $\lambda = 0$ and 
$\lambda =1$. 
However, these algorithms work in the opposite cases.
In the limit of $\lambda\to 0$, (\ref{algorithm_4_update_b}) becomes  
$$
q_X^{[i]}(x) = 
\frac{ \mathrm{e}^{-\nu c(x) }
\prod_y \hat p_{X|Y}^{[i]} (x|y)^{W(y|x)}}
{\sum_{x'} \mathrm{e}^{- \nu c(x') }\prod_y \hat p_{X|Y}^{[i]} (x|y)^{W(y|x)}}
$$
and Algorithm~\ref{new_algorithm_B} reduces to
the Arimoto-Blahut algorithm for computing the channel capacity.

\subsection{Comparison with Arimoto algorithm for channel coding exponents}
\label{section:arimoto_channel_coding}

This subsection gives comparisons between Tridenski and Zamir's algorithm,
Algorithm~\ref{new_algorithm_B},  and Arimoto's algorithms for channel coding~\cite{Arimoto1976}. 
Arimoto's algorithm was derived from the optimization problem 
appeared in (\ref{E_r}) and (\ref{G_AR}).
Unlike Tridenski and Zamir's algorithm and Algorithm~\ref{new_algorithm_B}, Arimoto's algorithm can be applied to the computation of both the error and the strong converse exponents. 

For fixed $\rho$ and $\nu\geq 0$, the function $E_0^{(\rho, \nu)}(p_X|W)$ is concave in $p_X$ when $\rho$ is positive and is convex in $p_X$ when $\rho$ is negative.  
Arimoto chose $(1/\rho) E_0^{(\rho, \nu)}(p_X|W)$ as an objective function which 
is concave for positive and negative $\rho$.  
Arimoto~\cite{Arimoto1976} introduced the following function.
\begin{definition}
For a given transition probability of a channel $W(y|x)$, 
$-1< \rho \le 1$, and $\nu$, we define 
\begin{align}
& F_{\rm AR}^{(\rho, \nu)} ( p_X, \hat p_{X|Y} | W ) \notag \\
& = 
-\frac{1}{\rho}
\log  \sum_{x\in \mathcal{X}} \sum_{y \in \mathcal{Y}}
p_X^{1+\rho}(x) \hat p_{X|Y}^{-\rho}(x|y)
W(y|x) \mathrm{e}^{\rho \nu c(x)}. 
\label{F_rho}
\end{align}
\end{definition}

Arimoto's algorithm is an algorithm that finds 
\begin{align}
  \max_{p_X} \max_{p_{X|Y}} F_{\rm AR}^{(\rho,\nu)}(p_X, \hat p_{X|Y}|W) 
  \label{double_max_Arimoto}
\end{align}
by optimizing $p_X$ and $\hat p_{X|Y}$ alternately.
Arimoto proved the following theorem.
\begin{theorem}[Arimoto\cite{Arimoto1976}]
\label{theorem.Arimoto}
For a fixed $-1< \rho \le 1$, $\nu\geq 0$ and any fixed $p_X$, 
we have 
\begin{align}
  & \frac{1}{\rho} E_0^{(\rho, \nu)} (p_X|W)
  =
  \max_{\hat p_{X|Y} } F_{\rm AR}^{(\rho, \nu)} (p_X, \hat p_{X|Y}|W) 
\label{eq.8}
\end{align}
and the maximum value is attained by 
\begin{align}
\hat p_{X|Y}(x|y) 
&= \frac{
p_X(x) \{ W(y|x) \mathrm{e}^{\rho \nu c(x) }\}^{\frac{1}{1 + \rho}}
}
{\sum_{x'}
p_X(x') \{ W(y|x') \mathrm{e}^{\rho \nu c(x') } \} ^{\frac{1}{1 + \rho}}
}.
\label{Update1}
\end{align}
On the other hand, for a fixed 
$\rho\in [-1,0)\cup(0,1]$, $\nu\ge 0$ and any fixed $\hat p_{X|Y}$, we have 
\begin{align}
& \frac{1}{\rho} A^{(\rho, \nu)}(\hat p_{X|Y} | W) 
=
\max_{p_X}
F_{\rm AR}^{(\rho, \nu)} (p_X, \hat p_{X|Y}|W). 
\label{eq.10}
\end{align}
The maximum value is attained by 
\begin{align}
p_X(x)
	&=
\frac{
\left\{
\sum_{y}
\hat p_{X|Y}^{-\rho} (x|y) W(y|x) \mathrm{e}^{\rho \nu c(x) }
\right\}^{-1/\rho}
}
{
\sum_{x'}
\left\{
\sum_{y}
\hat p_{X|Y}^{-\rho} (x'|y) W(y|x') \mathrm{e}^{\rho \nu c(x') }
\right\}^{-1/\rho}
	}.
\label{Update2}
\end{align}
\end{theorem}

See~\cite{Arimoto1976} for the proof. A quick proof will be shown later. 
Based on Theorem~\ref{theorem.Arimoto}, Arimoto presented 
an iterative algorithm shown in Algorithm~\ref{alg:Arimoto}.

    \begin{algorithm}
    \caption{Arimoto's algorithm for error and strong converse exponents in channel coding~\cite{Arimoto1976}}
    \label{alg:Arimoto}
    \begin{algorithmic}
    \Require The conditional probability of the channel $W$, $\rho\in(-1,0)\cup(0,\infty)$ and $\nu\geq 0$. 
    Choose initial $p_X^{[0]}$ such that all components are nonzero.
    \For{ $i = 0, 1 , 2 , \ldots$ }
    \State 
    \begin{align}
\hat p_{X|Y}^{[i]}(x|y) 
&= \frac{
p_X^{[i]}(x) \{ W(y|x) {\rm e}^{ \rho \nu c(x) } \}^{\frac{1}{1 + \rho}}
}
{\sum_{x'}
p_X^{[i]}(x') \{ W(y|x') {\rm e}^{ \rho \nu c(x') }\}^{\frac{1}{1 + \rho}}
}, \label{update1a}
\\
p_X^{[i+1]}(x)
	&=
\frac{
\left\{
\sum_{y}
\hat p_{X|Y}^{[i]} (x|y)^{-\rho} W(y|x) {\rm e}^{ \rho \nu c(x) }
\right\}^{-1/\rho}
}
{
\sum_{x'}
\left\{
\sum_{y}
\hat p_{X|Y}^{[i]} (x'|y)^{-\rho}
W(y|x') {\rm e}^{ \rho \nu c(x') }
\right\}^{-1/\rho}
	} .\label{update1b}
\end{align}    
\EndFor
    \end{algorithmic}
    \end{algorithm}

When $\rho\to 0$, Algorithm~\ref{alg:Arimoto} reduces to the ordinary Arimoto-Blahut algorithm for computing the channel capacity. 
For the case of $\rho=0$, $(1/\rho) E_0^{(\rho, \nu)}(p_X|W)$ is  
interpreted as its limiting function, i.e., 
\begin{align}
\lim_{\rho\to 0} \frac{1}{\rho} E_0^{(\rho, \nu)}(p_X|W) 
= I(p_X, W) - \nu \mathrm{E}_{p_X}[ c(X) ].
\label{E_0rhoiszero}
\end{align}
When $\rho=0$, $F_{\rm AR}^{(\rho, \nu)}(p_X, \hat p_{X|Y}|W)$
is interpreted as 
\begin{align}
\lim_{\rho\to 0}
& F_{\rm AR}^{(\rho, \nu)}(p_X, \hat p_{X|Y}|W)\notag\\
&= \sum_{x \in \mathcal{X} } p_X(x) W(y|x)                                            
\log 
\frac{ \hat p_{X|Y}(x|y) \mathrm{e}^{ - \nu c(x) } }
{p_X(x)}. 
\label{Frho0}
\end{align}
The rhs of Eq.(\ref{Frho0}) is the function
that appears in Arimoto-Blahut algorithm for finding 
the channel capacity. 
Thus we can regard $F_{\rm AR}^{(\rho, \nu)}( p_X, \hat p_{X|Y} | W ) $ as 
a generalization of (\ref{Frho0}) to the case $\rho \neq 0$.

From Eqs. (\ref{eq.8}) and (\ref{eq.10}) in Theorem~\ref{theorem.Arimoto}, 
we have the following corollaries. 
\begin{corollary}
\label{corollary:alternative_form}
For any fixed $\rho \in (-1,0) \cup (0,1]$ and $\nu\ge 0$, we have 
\begin{align}
  & \max_{p_{X}} \frac{1}{\rho} E_0^{(\rho, \nu)} (p_X|W)
  =  
  \max_{p_{X} }
  \max_{\hat p_{X|Y} } F_{\rm AR}^{(\rho, \nu)} (p_X, \hat p_{X|Y}|W)\notag \\
  &= 
  \max_{\hat p_{X|Y} } \frac{1}{\rho} A^{(\rho, \nu)} ( \hat p_{X|Y} | W ).
  \label{eq.13}
\end{align}

\end{corollary}
\begin{corollary}
\label{corollary_2}
Define the following functions. 
\begin{align}
\tilde E_{\rm r}(R|W)
=&
\max_{ \rho \in [0,1] }
\inf_{\nu\ge 0}
\left\{
\max_{\hat p_{X|Y} \in\mathcal{P(X|Y)} } A^{(\rho, \nu)}( \hat p_{X|Y} |W) -\rho R 
+ \rho \nu \Gamma 
\right\}, 
\label{alternative_error_exponent} 
\\
\tilde G_{\rm AR}(R|W)
=&
\sup_{ \rho \in (-1,0] }
\sup_{ \nu \ge 0 }
\left\{
\min_{\hat p_{X|Y} \in\mathcal{P(X|Y)} } A^{(\rho, \nu)}(\hat p_{X|Y} |W) -\rho R 
+ \rho \nu \Gamma 
\right\} .
\end{align}
Then, we have 
\begin{align}
    \tilde E_{\rm r}(R, \Gamma |W)
    = E_{\rm r}(R, \Gamma |W), \quad 
    \tilde G_{\rm AR}(R, \Gamma |W)
    = G_{\rm AR}(R, \Gamma |W).
    \label{eq.24}
\end{align}
\end{corollary}
We call $\tilde E_{\rm r}(R, \Gamma |W)$ and $ \tilde G_{\rm AR}(R, \Gamma |W)$ alternative representations of the error exponent and the strong converse exponent. 

At the end of this subsection, we give a proof of Theorem~\ref{theorem.Arimoto} using the conditional R\'enyi divergence. 
Many attractive properties of R\'enyi entropy, R\'enyi divergence, and exponent function were described by Verd\'u~\cite{Verdu_alpha_mutual_information2015, Verdu_alpha_mutual_information2021} 
and Cai and Verd\'u~\cite{Cai_Verdu2019}.
One of the important properties discussed in~\cite{Verdu_alpha_mutual_information2015, Verdu_alpha_mutual_information2021, Cai_Verdu2019} is that the Gallagar's error exponent is expressed by using Arimoto's conditional R\'enyi entropy~\cite{Arimoto1977} as well as Sibson's mutual information of $\alpha$.
The R\'enyi divergence and the conditional R\'enyi divergence of order $\alpha$ are defined by 
\begin{align}
D_{\alpha}(p_X ||q_X) &=
\frac{1}{\alpha-1} \log \sum_{x\in \mathcal{X} } p_X^{\alpha}(x) q_X^{1-\alpha}(x),\\
D_{\alpha}(p_{X|Y} || q_{X|Y} | p_{Y} )
&=
\frac{1}{\alpha-1} \log \sum_{y\in \mathcal{Y} } p_Y(y) 
\sum_{x\in \mathcal{X} } p_{X|Y}^{\alpha}(x|y) q_{X|Y}^{1-\alpha}(x|y).
\end{align}
R\'enyi divergence and conditional R\'enyi divergence are non-negative 
and vanish if and only if $p_X=q_X$ and $p_{X|Y}=q_{X|Y}$, respectively.
To the best of our knowledge, the following relations have not been pointed out so far. 
\begin{lemma}
\label{lemma0}
For a given $W\in\mathcal{P(Y|X)}$, $p_X \in \mathcal{P(X)
}$, and $\hat p_{X|Y} \in \mathcal{P(X|Y)}$, set 
\begin{align}
    \hat p_{X|Y}^*(p_X)(x|y)
    &=
    \frac{ p_X(x) \{ W(y|x) 
    \mathrm{e}^{ -\mu c(x) }
    \}^{1/(1+\rho)} } 
    { \sum_{x'\in\mathcal{X} } p_X(x') \{ W(y|x') 
    \mathrm{e}^{ \rho \nu c(x') }
    \}^{1/(1+\rho)}
    }, \label{eq.lemma1.a}\\
    \hat p_Y^*(p_X)(y)
    &=
    \frac{
    \{ \sum_{x \in\mathcal{X} } p_X(x) 
    \{ W(y|x) \mathrm{e}^{ \rho \nu c(x) }
    \}^{1/(1+\rho)} \}^{1+\rho}
    }{
        \sum_{y'} 
        \{ \sum_{x \in\mathcal{X} } p_X(x) 
        \{ W(y'|x) \mathrm{e}^{ \rho \nu c(x) }
    \}^{1/(1+\rho)} \}^{1+\rho}
    }, \label{eq.lemma1.b}\\
    p_X^*(\hat p_{X|Y})(x) &=
    \frac{ \{ 
    \sum_{y} \hat p_{X|Y}^{-\rho}(x|y) \{ W(y|x)   
    \mathrm{e}^{ \rho \nu c(x) }
    \}    
    \}^{-1/\rho}
    }{\sum_{x'}
    \{ 
    \sum_{y} \hat p_{X|Y}^{-\rho}(x'|y) \{ 
    W(y|x') \mathrm{e}^{ \rho \nu c(x') }
    \}
    \}^{-1/\rho}
    }.\label{eq.lemma1.c}
\end{align}
Then we have 
\begin{align}
    F^{(\rho, \nu)}_{\rm AR}(p_X, \hat{p}_{X|Y}|W) 
    &=
    \frac1\rho E_0^{(\rho, \nu )}(p_X|W) - D_{1+\rho} ( \hat p_{X|Y}^*(p_X) || \hat p_{X|Y} | \hat p_{Y}^*(p_X) )
    \label{eq.lemma0.1}\\
    &=
    \frac1\rho A^{(\rho, \nu)}(\hat p_{X|Y}|W) - D_{1+\rho}( p_X || p_X^*(\hat p_{X|Y}) ). 
    \label{eq.lemma0.2}
\end{align}
\end{lemma} 
See Appendix~\ref{appendix0} for the proof. 

{\it Proof of Theorem~\ref{theorem.Arimoto}:}
Eqs.(\ref{eq.8}) and (\ref{eq.10}) are proved by the nonnegativity of R\'enyi divergence and the conditional R\'enyi divergence. 
The conditions for R\'enyi divergence and the conditional R\'enyi divergence
to take zero directly prove (\ref{Update1}) and (\ref{Update2}).
This complets the proof.
\hfill$\IEEEQED$



It should be noted that 
although Eq.(\ref{Update3}) in Algorithm~\ref{new_algorithm_A}
and Eq.(\ref{algorithm_4_update_b}) in Algorithm~\ref{new_algorithm_B}
were derived from Dueck and K\"orner's exponent function,
their functional structures are the same as Eqs. (\ref{update1a}) 
and (\ref{update1b}) in Arimoto's algorithm.
Note also that the structure of (\ref{eq.lemma1.b}) in Lemma~\ref{lemma0} is the same as (\ref{algorithm_3_update_b}). 
In Arimoto algorithm, $\hat p_Y$ is not used. 
However, this probability distribution appears when we express the difference between 
$F_{\rm AR}^{(\rho, \nu)}(p_X, \hat p_{X|Y}|W)$ and
$(1/\rho) E_0^{(\rho,\nu)}(p_X|W)$ by R\'enyi Divergence as in (\ref{eq.lemma0.1}).
We will see the same form as (\ref{eq.lemma1.b}) in the next section again.

\subsection{Comparison with our previous algorithm}
\label{section:our_previous_algorithm}
This subsection gives a comparison between Tridenski and Zamir's algorithm, 
Algorithms~\ref{new_algorithm_B}, and  
our previous algorithm in~\cite{YutakaISIT2015}. 
Before \cite{YutakaISIT2015}, no algorithm was known for 
computing the optimal distribution for Dueck and K\"orner's strong converse exponent. 
The algorithm for DMCs without input cost constraint was extended to the algorithm
for DMCs under input cost~\cite{Jitsumatsu_Oohama_IT_Trans2020}. 
%
%
%
%
We introduced the following function 
to obtain the minimum of 
$\Theta^{(\lambda, \lambda \nu)}( q_{XY} |W)$.
\begin{align}
& F_{\rm JO}^{(\lambda, \nu)} (q_{XY}, \hat q_{XY} | W) \notag\\
&=
\mathrm{E}_{q} \left[
\log \frac{\hat q_{Y|X}^{1 - \lambda}(Y|X)
\hat q_{Y}^{\lambda}(Y) 
}
{ W(Y|X) \mathrm{e}^{-\lambda \nu c(x)} }
\right]
+D( q_{XY} || \hat q_{XY})
\label{J}
\end{align}
Regarding this function,
the following two lemmas hold. 
\begin{lemma}[\cite{Jitsumatsu_Oohama_IT_Trans2020}]
\label{lemma.8}
For a fixed $\lambda \in [0,1]$, $\nu\ge 0$ and $q_{XY}$, 
$F_{\rm JO}^{(\lambda, \nu)} (q_{XY}, \hat q_{XY} | W)$
is minimized by $\hat q_{XY} = q_{XY}$ and its 
minimum value is 
\begin{align}
F_{\rm JO}^{(\lambda, \nu)} (q_{XY}, q_{XY} | W)
=
\Theta^{(\lambda, \lambda \nu)}(q_{XY}|W). 
\end{align}
This implies that 
\begin{align}
 \min_{q_{XY}} \min_{\hat q_{XY}} F_{\rm JO}^{(\lambda, \nu)} (q_{XY}, \hat q_{XY} | W)
=\min_{q_{XY}} F_{\rm JO}^{ (\lambda, \nu) } (q_{XY}, q_{XY} | W)
=\min_{q_{XY}} \Theta^{ (\lambda, \lambda \nu) }(q_{XY}|W). 
\end{align}
\end{lemma} 

\begin{lemma}[\cite{Jitsumatsu_Oohama_IT_Trans2020}]
\label{lemma.9}
For a fixed $\lambda \in [0,1]$, $\nu \ge 0$ and  
$\hat q_{XY}$, 
$ F_{\rm JO}^{(\lambda, \nu)} (q_{XY}, \hat q_{XY} | W) $
is minimized by 
\begin{align}
q_{XY}(x,y) &=
\frac{\hat q_X^{1-\lambda} (x) \hat q_{X|Y}^\lambda (x|y)
\{ W(y|x)
\mathrm{e}^{-\lambda \nu c(x) }
\} 
}
{ \sum_{x'} \sum_{y'} \hat q_X^{1-\lambda} (x') \hat q_{X|Y}^\lambda (x'|y') \{ W(y'|x') \mathrm{e}^{-\lambda \nu  c(x') } 
\}
} \notag\\
&\stackrel{\triangle}=
q_{XY}^{*}(\hat q_{XY})(x,y).
\end{align}
and its minimum value is 
\begin{align}
&
F_{\rm JO}^{(\lambda, \nu)} (q_{XY}^*(\hat q_{XY}) , \hat q_{XY} | W) \notag\\
&=
-\log \mathrm{E}_{\hat q_{XY} } 
\left[
\frac{ W(Y|X) \mathrm{e}^{- \lambda \nu c(x) } }
{ \hat q_{Y|X}^{1-\lambda}(Y|X) 
  \hat q_Y^{(\lambda)}(Y) }
\right] \notag \\
& =
- \log \sum_x \sum_y
\hat q_X^{1-\lambda} (x) \hat q_{X|Y}^\lambda (x|y)
W(y|x) \mathrm{e}^{- \lambda \nu c(x) } .
\label{min_J}
\end{align}
This implies that 
\begin{align}
&\min_{q_{XY}} \min_{\hat q_{XY}} F_{\rm JO}^{(\lambda, \nu)} (q_{XY}, \hat q_{XY} | W)
=\min_{\hat q_{XY}} F_{\rm JO}^{(\lambda, \nu)} (q_{XY}^*(\hat q_{XY}) , \hat q_{XY} | W)\notag \\
&=\min_{\hat q_{XY}} -\log \sum_x \sum_y
\hat q_X^{1-\lambda} (x) \hat q_{X|Y}^\lambda (x|y) W(y|x)
\mathrm{e}^{-\lambda \nu c(x) }. 
\end{align}
\end{lemma} 
See~\cite{Jitsumatsu_Oohama_IT_Trans2020} for the proof of Lemmas~\ref{lemma.8} and \ref{lemma.9}.

The algorithm for computing $\min_{q_{XY}} \Theta^{(\lambda, \lambda \nu)}(q_{XY}|W)$ 
is shown in Algorithm~\ref{alg:previous_work}.
In~\cite{Jitsumatsu_Oohama_IT_Trans2020}, the condition
for the termination of the iterative updating is mentioned
but is omitted here.
\begin{algorithm}
\caption{Computation of $ \min_{q_{XY}} \Theta^{(\lambda, \lambda \nu)}(q_{XY}|W) $~\cite{Jitsumatsu_Oohama_IT_Trans2020}}
\label{alg:previous_work}
\begin{algorithmic}
\Require 
The conditional probability of the channel $W$, 
$\lambda \in (0,1)$ and $\nu \ge 0$
Choose initial joint probability distribution
$q_{XY}^{[0]}$
such that $q_{XY}^{[0]}(x,y) = 0 $
if $W(y|x)=0$ and 
$q_{XY}^{[0]}(x,y) > 0 $
if $W(y|x)>0$.
\For{$i=0,1,2,\ldots$,}
\begin{align}
  & q_{XY}^{[i+1]}(x,y) \notag \\
  &=
\frac{q_X^{[i]} (x)^{1-\lambda} q_{X|Y}^{[i]}(x|y)^\lambda 
W(y|x) {\rm e}^{-\lambda \nu c(x) }}
{ \displaystyle \sum_{x'} \sum_{y'} 
q_X^{[i]} (x')^{1-\lambda} 
q_{X|Y}^{[i]}(x'|y')^\lambda  W(y'|x') {\rm e}^{-\lambda \nu c(x') } }
\end{align}
\EndFor 
\end{algorithmic}
\end{algorithm}

The important observation here is that the function in the rhs of (\ref{min_J}) is exactly the same as the function in the rhs of Eq. (\ref{F_rho}).
Then, it can be seen that the function 
$F_{\rm JO}^{(\lambda,\nu)}(q_{XY}, \hat q_{XY} |W)$ in (\ref{J}) satisfies 
\begin{align}
& F_{\rm JO}^{(\lambda, \nu)}(q_{XY}, \hat q_{XY}|W)\notag\\
& = \mathrm{E}_{q_{XY}} 
\left[
\log \frac{ q_{XY}(X,Y) }
{\hat q_X^{1-\lambda}(X) 
\hat q_{X|Y}^\lambda (X|Y) W(Y|X) {\mathrm{e}^{- \lambda \nu c(x) } }} 
\right]\notag
\\
&= 
\Theta^{(\lambda, \lambda \nu)}(q_{XY}|W) + (1-\lambda)
D( q_X \| \hat q_X )  + \lambda D( q_{X|Y} \| \hat q_{X|Y} | q_Y ) .
\label{J2}
\end{align}
From this equation, we observe that $F_{\rm JO}^{(\lambda, \nu)}(q_{XY}, \hat q_{XY}|W) $
is equal to $J_{ \bm{t}_1 + \bm{t}_2 }^{(\lambda, \nu)}(q_{XY}, \hat q_{XY}|W)$, 
where $\bm{t}_1+\bm{t}_2=(1, 0, 0, \frac{\lambda}{1-\lambda})$. 
Here, 
$\hat q_X$ and $\hat q_{X|Y}$ are 
the marginal and the conditional distribution 
derived from the joint distribution 
$\hat q_{XY}$. However, 
what if we can replace $\hat q_{X|Y}$
with another conditional distribution function $\hat p_{X|Y}$
that is not related to $\hat q_{X}$? 
By slightly modifying $F_{\rm JO}^{(\lambda, \nu)}(q_{XY}, \hat q_{XY}|W)$,
we have introduced the following function ($\hat q_X$ is also replaced with $p_X$)~\cite{Jitsumatsu_Oohama_IT_Trans2020}.
\begin{align}
&\tilde F_{\rm JO}^{ ( \lambda, \nu ) }(q_{XY}, p_X, \hat p_{X|Y}|W)\notag\\
&=
\Theta^{(\lambda, \lambda \nu)}(q_{XY}|W) + (1-\lambda)
D(q_X \| p_X) +
 \lambda 
D( q_{X|Y} \| \hat p_{X|Y} | q_Y)
\label{Ftilde}
\end{align}
Then, we can easily prove the following lemmas.
\begin{lemma}[\cite{Jitsumatsu_Oohama_IT_Trans2020}]
\label{lemma3}
For a fixed $\lambda \in [0,1]$, $\nu\ge 0$ and $q_{XY}$, we have 
\begin{align}
\min_{p_X} \min_{\hat p_{X|Y} } 
\tilde F_{\rm JO}^{(\lambda, \nu)} (q_{XY}, p_X, \hat p_{X|Y}|W)
=
\Theta^{(\lambda, \lambda \nu)}(q_{XY}|W). 
\end{align}
The minimum is attained if and only if 
$p_X = q_X$ and $\hat p_{X|Y} = q_{X|Y}$ holds.
\end{lemma}

\begin{lemma}[\cite{Jitsumatsu_Oohama_IT_Trans2020}]
\label{lemma4}
For a fixed $\lambda \in [0,1]$, $\nu\geq 0$, $p_X$, and $\hat p_{X|Y}$,
we have 
\begin{align}
\min_{q_{XY}}
\tilde F_{\rm JO}^{(\lambda, \nu)} (q_{XY}, p_X, \hat p_{X|Y}|W)
= -\lambda 
F_{\rm AR}^{(-\lambda, \nu)}(p_X, \hat p_{X|Y}|W).
\label{eq.84}
\end{align}
The minimum is attained if and only if
\begin{align}
q_{XY}(x,y) =
\frac{p_X^{1-\lambda} (x) \hat p_{X|Y}^\lambda (x|y) W(y|x) 
\mathrm{e}^{-\lambda \nu c(x) }
}
{ \displaystyle \sum_{x'} \sum_{y'} p_X^{1-\lambda} (x') \hat p_{X|Y}^\lambda (x'|y') 
W(y'|x') \mathrm{e}^{-\lambda \nu c(x') }} 
. \label{optimal_q_Lemma13}
\end{align} 
\end{lemma}

\begin{figure*}
\centering
\includegraphics[width=0.95\textwidth]{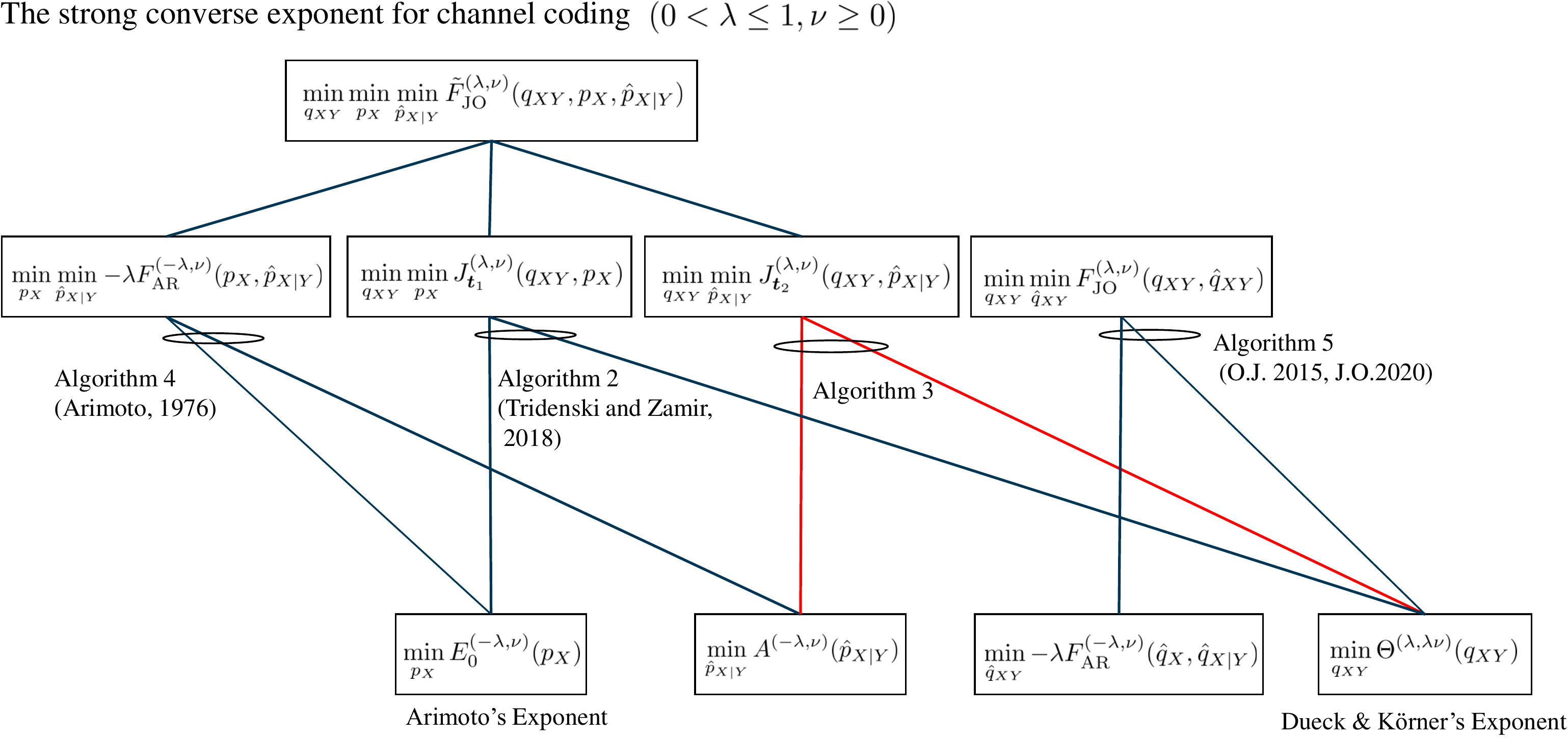}
\caption{Relation between the triple minimization form, four double minimization forms,
and four single minimization forms of the strong converse exponent for the channel coding of DMCs}
\label{fig.1L}
\end{figure*}

From Lemmas~\ref{lemma3} and~\ref{lemma4}, we have the following proposition:
\begin{proposition}
For a fixed $\lambda \in [0,1]$, $\nu\ge 0$, and $W\in \mathcal{P(Y|X)}$,
we have
\begin{align}
    \min_{q_{XY}} \Theta^{(\lambda, \lambda\nu)}( q_{XY} | W )
    =
    \min_{p_X} \min_{ \hat p_{X|Y} }
    -\lambda F_{A\rm R}^{(-\lambda, \nu)} ( p_X, \hat p_{X|Y} | W ).
\end{align}
\end{proposition}
This proposition clearly shows that the Dueck and K\"orner's exponent
is equal to the double minimization form of Arimoto's alglrithm. 
It is an important property of the function $\tilde F_{\rm JO}^{(\lambda, \nu)} (q_{XY}, p_X, \hat p_{X|Y}|W)$ that the double minimization form is derived from this function. 
Moreover, we immediately obtain the following two lemmas:
\begin{lemma}
For fixed $\lambda \in [0,1]$, $\nu \ge 0$, $p_X$, and $q_{XY}$, we have 
    \begin{align}
        \min_{ \hat p_{X|Y} } 
        \tilde F_{\rm JO}^{(\lambda, \nu )} (q_{XY}, p_X, \hat p_{X|Y}|W)
        &=
        \Theta^{(\lambda, \lambda \nu )}(q_{XY}|W)+ (1-\lambda) D(q_X || p_X) \notag \\
        &= J_{\bm{t}_1}^{(\lambda, \nu)} (q_{XY}, p_X |W). 
        \label{def:J1}
    \end{align}
\end{lemma}
\begin{lemma}
For a fixed $\lambda \in [0,1]$, $\hat p_{X|Y} $, and $q_{XY}$, we have 
    \begin{align}
        \min_{ p_{X} } 
        \tilde F_{\rm JO}^{(\lambda, \nu )} (q_{XY}, p_X, \hat p_{X|Y}|W)
        &=
        \Theta^{( \lambda, \lambda \nu )}(q_{XY}|W) + \lambda D( q_{X|Y} || \hat p_{X|Y} | q_Y ) \notag\\
        &=J_{\bm{t}_2}^{(\lambda, \nu)} (q_{XY}, \hat p_{X|Y} |W).
        \label{def:J2}
    \end{align}
\end{lemma}
Fig.\ref{fig.1L} shows the relationship between 
the triple minimization $\min_{q_{XY}} \min_{p_X} \min_{\hat p_{X|Y} } 
\tilde F_{\rm JO}^{(\lambda, \nu )}(q_{XY}, p_X, \hat p_{X|Y}|W)$, 
the double minimization form of Arimoto's algorithm,
$J_{\bm{t}_1}^{(\lambda, \nu )} (q_{XY}, p_X |W)$, and 
$J_{\bm{t}_2}^{(\lambda, \nu )} (q_{XY}, \hat p_{X|Y} |W)$.
This figure shows that the three double minimization forms are obtained by evaluating the minimum value with respect to one probability distribution for the triple minimization form and removing one minimization operation.


Because the objective function $F_{\rm JO}^{(\lambda,\nu)}(q_{XY}, \hat q_{XY}|W)$
is a special case of the parameterized surrogate objective function 
$F_{\bm{t}}^{(\lambda, \nu)}(q_{XY}, \hat q_{XY})$, 
we can construct a parameterized function such that 
$\tilde F_{\rm JO}^{(\lambda,\nu)}(q_{XY}, p_X, \hat p_{X|Y}|W)$ is a special case
of the function. 
We define such a parameterized function as 
\begin{align}
    \tilde F_{ {\rm JO}, \bm{t}}^{(\lambda, \nu)}
    (q_{XY}, p_{XY}, \hat p_{XY} |W)
    & = \Theta^{(\lambda, \lambda,\nu)} (q_{XY}|W)
    + (1-\lambda) \{ t_1 D(q_{X}|| p_{X} ) + t_2 D(q_{Y|X}|| p_{Y|X} |q_X )\notag\\
    & \quad +  t_3 D(q_{Y}|| \hat p_{Y} ) + t_4 D(q_{X|Y}|| \hat p_{X|Y} | q_Y) \}.
    \label{generalized_objective_function_three_joint}
\end{align}
This function reduces to
$\tilde F_{\rm JO}^{(\lambda,\nu)}(q_{XY}, p_X, \hat p_{X|Y}|W)$
when $\bm{t} = \bm{t}_1 + \bm{t}_2 = (1, 0, 0, \lambda/(1-\lambda))$. 
By introducing this function, the following interesting things can be seen.
The function $F_{\rm AR}^{(\lambda, \nu)}(p_X, \hat p_{X|Y}|W)$ in the double minimization form, on which Arimoto's algorithm is based, is replaced by a parameterized function $F_{{\rm AR}, \bm{t} }^{(\lambda, \nu)}(p_{XY}, \hat p_{XY}|W)$,
from which a parameterized algorithm is derived. 
In Arimoto's algorithm $p_X$ and $\hat p_{Y|X}$ are alternately updated, while 
$(p_X, p_{Y|X})$ and $(\hat p_Y, \hat p_{X|Y})$ are alternately updated in the parameterized algorithm.
Details of the parameterized algorithm is shown in Appendix~\ref{appendix:parameterized_Arimoot}.
A special case of $\bm{t} = \bm{t}_1 + \bm{t}_2$ is interesting.
We expect the parameterized algorithm reduces to Algorithm~\ref{alg:Arimoto}.
It actually does so, but it also has extra updating rules for $\hat p_Y$ and $p_{Y|X}$.
It is explicitly written as 
\begin{align}
&\hat p_{X|Y}^{[i]}(x|y) 
= \frac{
p_X^{[i]}(x) \{ W(y|x) {\rm e}^{ -\lambda \nu c(x) } \}^{\frac{1}{1 - \lambda}}
}
{\sum_{x'}
p_X^{[i]}(x') \{ W(y|x') {\rm e}^{ -\lambda \nu c(x') }\}^{\frac{1}{1 - \lambda}}
},\label{eq.103}
\\
&\hat p_Y^{[i]}(y) = \frac{
\left[
\sum_x p_X^{[i]}(x) \{ W(y|x) {\rm e}^{-\lambda \nu c(x)} \}^{1/(1-\lambda)} 
\right]^{1-\lambda}
}{
\sum_{y'} \left[
\sum_x p_X^{[i]}(x) \{ W(y'|x) {\rm e}^{-\lambda \nu c(x)} \}^{1/(1-\lambda)} 
\right]^{1-\lambda}
}, \label{eq.118}\\
& 
p_X^{[i+1]}(x)
	=
\frac{
\left\{
\sum_{y}
\hat p_{X|Y}^{[i]} (x|y)^{\lambda} W(y|x) {\rm e}^{ -\lambda \nu c(x) }
\right\}^{1/\lambda}
}
{
\sum_{x'}
\left\{
\sum_{y}
\hat p_{X|Y}^{[i]} (x'|y)^{\lambda}
W(y|x') {\rm e}^{ -\lambda \nu c(x') }
\right\}^{1/\lambda}
	}, \label{eq.105}\\
&p_{Y|X}^{[i+1]}(y|x) = 
\frac{
\hat p_{X|Y}^{[i]}(x|y)^{\lambda} W(y|x) 
}{
\sum_{y'}
\hat p_{X|Y}^{[i]}(x|y')^{\lambda} W(y'|x) 
}. \label{eq.119}
\end{align}

Then, 
the relationship between Algorithms~\ref{new_algorithm_A}, \ref{new_algorithm_B},
and~\ref{alg:Arimoto} becomes clearer.
Eqs.(\ref{eq.103}) and (\ref{eq.105}) are equal to (\ref{update1a}) and (\ref{update1b}) with $\rho=-\lambda$.
Eqs.(\ref{eq.118}) and (\ref{eq.119}) do not appear in Arimoto's algorithm.
However, they appear in (\ref{algorithm_3_update_b}) and (\ref{algorithm_4_update_a}) in
Algorithms~\ref{new_algorithm_A} and \ref{new_algorithm_B}, respectively. 
The $\hat p_Y^{[i]}(y)$ and $p_{Y|X}^{[i+1]}(y|x)$ in (\ref{eq.118}) and (\ref{eq.119})
are considered to be natural distribution updating rules, which are derived 
from the parameterized objective function
$F_{ {\rm AR}, \bm{t}} ^{(-\lambda, \nu)}(p_{XY}, \hat p_{XY}|W)$.


\section{The match of Gallager's and Csisz\'ar and K\"orner's error exponents}
\label{section:Er_and_E_CK_match}

Arimoto's algorithm is based on the double maximization expression in
Eq.(\ref{double_max_Arimoto}) for positive and negative $\rho$ 
and therefore can be applied for computing both the Gallager's error exponent
and the Arimoto's strong converse exponent.
Algorithms~\ref{new_algorithm_A},~\ref{new_algorithm_B}, and~\ref{alg:previous_work},
were derived based on the Dueck and K\"orner's expression of
the strong converse exponent. 
Unfortunately, we cannot apply Algorithms~\ref{new_algorithm_A},~\ref{new_algorithm_B}, and~\ref{alg:previous_work} for computing the Csisz\'ar and K\"orner's 
error exponent because of the following reasons. 
Comparing Eq.(\ref{E_CK}) with (\ref{G_DK}), 
Eq. (\ref{G_DK}) is a minimum value of a functional with respect to a joint distributions, whereas Eq.(\ref{E_CK}) is a saddle point of a functional.
At the saddle point, the functional is maximized with respect to the input probability distribution $q_X$ and it is minimized with respect to a conditional distribution $q_{Y|X}$. 
In order to derive Arimoto-Blahut type algorithm, 
we must find an expression that can be defined as a joint maximization problem or a joint minimization problem.
%
%

This section shows that we can prove that Csisz\'ar and K\"orner's error exponent matches with the alternative expression of error exponent defined in (\ref{alternative_error_exponent}). 
Because the alternative expression matches with Gallager's error exponent, the proof immediately 
implies that Csisz\'ar and K\"orner's exponent matches with Gallager's exponent.
For the proof, we use the surrogate objective function $J_{\bm{t}_2}^{(\lambda, \nu)}(q_{XY}, \hat p_{X|Y}|W)$ that was used to derive  Algorithm~\ref{new_algorithm_B}. Thus, the result of this section is an application of Algorithm~\ref{new_algorithm_B}. It should be noted that the objective functions that are used in Algorithm~\ref{new_algorithm_A} and \ref{alg:previous_work} cannot be applied to prove the match of the two error expressions of error exponent. 

The following proposition shows that Csisz\'ar and K\"orner's exponent matches with Gallager's exponent.
\begin{proposition}
\label{proposition4}
For any $R\geq 0$ and $\Gamma \ge 0$, we have 
\begin{align}
E_{\rm CK}(R, \Gamma |W) 
= E_{\rm r}(R, \Gamma |W). 
\end{align}
\end{proposition}

In this section, we first give an overview of the standard method of proving Proposition~\ref{proposition4}. Then, a new proof is presented.

\subsection{A proof shown in Exercise 10.24 in~\cite{Csiszar-KornerBook}}
The following steps are suggested in~\cite[Exercise 10.24]{Csiszar-KornerBook} to
prove\footnote{
To be precise, it shows the proof that the two forms of the sphere-packing exponent function match.} 
Proposition~\ref{proposition4} which states
that Csisz\'ar and K\"orner's exponent and Gallager's exponent match.
In order to explicitly discuss the difference between this steps and 
our proof, we describe the steps in detail. 
Cost constraints essentially have no effect on the proof step. Thus, 
as in~\cite[Exercise 10.24]{Csiszar-KornerBook}, 
we consider a DMC without cost constraint in this subsection. 
See also Lapidoth and Miliou~\cite{Lapidoth2006}, 
showing that these two exponents are Lagrange dual each other. 

\begin{enumerate}
\item Prove the equation
\begin{align}
& E_{\rm CK}(R|W) \notag \\
&=
\max_{q_X} \max_{\rho\in [0,1]}
\{
\min_{q_{Y|X}}
\Theta^{(-\rho)}(q_{XY}|W) - \rho R\}.
\label{eq.54}
\end{align}

\item 
Define
\begin{align}
    J_{\rm CK}^{(\rho)}  ( q_{XY}, \hat p_Y |W)
    = \Theta^{(-\rho)} (q_{XY}|W) + \rho D(q_Y|| \hat p_Y ) 
\end{align}
and prove the equation 
\begin{align}
  \min_{q_{Y|X}} \Theta^{(-\rho)}(q_{XY}|W) 
 = \min_{\hat p_Y} \tilde J_{\rm CK}^{(\rho)} (q_X, \hat p_Y|W), 
 \label{eq.46}
\end{align}
where
\begin{align}
    & \tilde J_{\rm CK}^{(\rho)} (q_X, \hat p_Y |W) \notag\\
	&= \min_{q_{Y|X}} J_{\rm CK}^{(\rho)} (q_{XY}, \hat p_{Y} |W) \notag\\
&=
-(1+\rho) \sum_{x} q_X(x) \log \sum_{y} W^{\frac{1}{1+\rho}} (y|x) \hat p_Y^{\frac{\rho}{1+\rho}}(y). 
\end{align}
\item In the following, we give an upper
and a lower bounds on 
$
\max_{q_X}
\min_{\hat p_{Y}}
\tilde J_{\rm CK}^{(\rho)} (q_X, \hat p_Y|W)$.
If the upper and the lower bounds are equal, the bound is found the desired saddle point.
First, by the concavity of $\log$ function and Jensen's inequality, 
we have 
\begin{align}
&\tilde 
J_{\rm CK}^{(\rho)} (q_X, \hat p_Y|W)\notag \\
&\geq 
-(1+\rho) \log \sum_{x} q_X(x) \sum_{y} W^{\frac{1}{1+\rho}} (y|x) \hat p_Y^{\frac{\rho}{1+\rho}}(y).
\label{eq.48}
\end{align}
Then, show that the right hand side of (\ref{eq.48}) is minimized by 
\begin{align}
\hat p_Y(y) = \Lambda^{-1} 
\left[
\sum_x q_X(x) W^{\frac1{1+\rho}}(y|x) 
\right]^{1+\rho}, 
\label{eq.49}
\end{align}
where $\Lambda$ is a normalization factor. 
Denote this $\hat p_Y$ by $\hat p_Y^*(q_X)$. 

\item From Step 3, we have 
\begin{align}
& \min_{\hat p_Y} \tilde J_{\rm CK}^{(\rho)} (q_X, \hat p_Y | W) \notag\\ 
& \stackrel{\rm (a)} 
  \geq 
  \min_{\hat p_Y} 
-(1+\rho) \log \sum_{x} q_X(x) \sum_{y} W^{\frac{1}{1+\rho}} (y|x) \hat p_Y^{\frac{\rho}{1+\rho}}(y)
\notag\\  
& \stackrel{\rm (b)}
= - \log \sum_y \left[
\sum_x q_X(x) W^{\frac{1}{1+\rho}} (y|x)
\right]^{1+\rho} .
\label{eq.50}
\end{align}
Step (a) follows from Eq.(\ref{eq.48}). 
In Step (b), $\hat p_Y = \hat p_Y^*(q_X)$ is substituted. 
Using Lagrange's undetermined multiplier method, 
show that the $q_X$ that maximizes (\ref{eq.50}) satisfies 
\begin{align}
&
\sum_{y} W^{\frac{1}{1+\rho}}(y|x) 
\left[
\sum_{x'} q_X(x') 
W^{\frac{1}{1+\rho}}(y|x')
\right]^\rho \notag \\
&
\stackrel{\rm (a)}
\geq 
\sum_{y} \left[
\sum_{x'} q_X(x') W^{\frac{1}{1+\rho}}(y|x')
\right]^{1+\rho}= \Lambda \label{eq.51} . 
\end{align}
Inequality (a) holds with equality for every $x$ such that $q_X(x)>0$.

\item 
Fix $\hat p_{Y}$ to some distribution
to give an upper bound of 
$
\max_{q_X}
\min_{\hat p_{Y}}
\tilde J_{\rm CK}^{(\rho)} (q_X, \hat p_Y |W)$.
For this purpose, assume 
$\hat p_{Y} = \hat p_{Y}^*$. Then,
\begin{align}
&
\max_{q_X}
\min_{\hat p_{Y}}
\tilde J_{\rm CK}^{(\rho)} (q_X, \hat p_Y|W)\notag \\
& \leq 
\max_{q_X}
\tilde J_{\rm CK}^{(\rho)} (q_X, \hat p_Y^*(q_X)|W) \notag \\
&=
\max_{q_X}
-(1+\rho) \sum_{x} q_X(x) \log  
\Lambda^{-\frac{\rho}{1+\rho}} \notag\\
& \quad \cdot \sum_{y} W^{\frac{1}{1+\rho}} (y|x) 
\left[
\sum_{x'} q_X(x') W^{\frac1{1+\rho}}(y|x') 
\right]^{\rho}\notag\\
&\stackrel{(a)}
\leq
\max_{q_X}
-(1+\rho) \sum_{x} q_X(x) \log  
\Lambda^{-\frac{\rho}{1+\rho} + 1} 
\notag\\
&= 
\max_{q_X}
-\log 
\sum_{y} \left[
\sum_{x'} q_X(x') W^{\frac{1}{1+\rho}}(y|x')
\right]^{1+\rho} . 
\end{align}
Step (a) follows from Eq.(\ref{eq.51}). 
Hence, 
$\max_{q_X} E_0^{(\rho)}(p_X|W)$
is found to be an upper as well as lower bounds
of 
$
\max_{q_X}
\min_{\hat p_{Y}}
\tilde J_{\rm CK}^{(\rho)} (q_X, \hat p_Y |W)$.  
Therefore
\begin{align}
& E_{\rm CK}(R|W) \notag \\
&=
\max_{q_X} \max_{\rho\in [0,1]}
\{
\min_{q_{Y|X}}
\Theta^{(-\rho)}(q_{XY}|W) - \rho R\}\notag\\
&=
\max_{\rho\in [0,1]}
\{
\max_{q_X}
\min_{\hat p_{Y}}
\tilde J_{\rm CK}^{(\rho)} (q_X, \hat p_Y | W) - \rho R\}\notag\\
&=
\max_{\rho\in [0,1]}
\{
\max_{q_X}
E_0^{(\rho)}(q_X|W) 
- \rho R\}
\notag\\
&=E_{\rm r}(R|W).
\end{align}

\end{enumerate}
As shown above, 
Step 3 uses Jensen's inequality and steps 4 and 5 evaluate the upper and lower bounds.
In the next subsection, we show another procedure to show the match of the two error exponents. 
In this new procedure, the evaluation of each step is explicit and thus the new proof is more elementary than the standard proof. 
The interesting point is that 
we use the function $J_{\bm{t}_2}^{(\lambda, \nu)} (q_{XY}, \hat p_{X|Y} | W )$ that was used to derive Algorithm~\ref{new_algorithm_B}.

\subsection{New proof for the match of random coding error exponents}
Here, we give a new proof for the match of Gallager's and Csisz\'ar and K\"orner's error exponents.
In the previous section, we removed the cost constraint for simplicity and to match the description in~\cite[Exercise 10.24]{Csiszar-KornerBook}. 
In this section, we consider DMCs with cost constraints. 
The proof steps are the same with or without cost constraints.

Csisz\'ar and K\"orner's exponent is expressed using the function $\Theta^{(\lambda, \lambda \nu )}(q_{XY}|W)$ but 
with negative $\lambda$ (we put $\rho = -\lambda$).
When $0\leq \rho \leq 1$, $\Theta^{(-\rho, -\rho \nu)}(q_{XY}|W)$ satisfies the following property.
\begin{property}
\label{property:convex_concave}
Suppose $\rho \in [0,1]$ and $\nu\geq 0$ are fixed. 
Then $\Theta^{(-\rho, -\rho\nu)}(q_{XY}|W)$ is 
concave in $q_{X}$ for a fixed $q_{Y|X}$ 
and convex in $q_{Y|X}$ for a fixed $q_{X}$. 
\end{property}

\textit{Proof:}
By definition, we have $\Theta^{(-\rho, -\rho\nu)}(q_{XY}|W)
= I(q_X, q_{Y|X}) + \rho D(q_{Y|X}||W| q_X) -\rho\nu \mathrm{E}_{q_X}[c(X)]$.
The first term is concave in $q_X$ for a fixed $q_{Y|X}$
and is convex in $q_{Y|X}$ for a fixed $q_X$. 
The second term is convex in $q_{Y|X}$ for a fixed $q_X$
and is a linear function of $q_X$ for a fixed $q_{Y|X}$.
The last term is linear in $q_{XY}$. 
Therefore, the property holds. \hfill$\IEEEQED$

We use $J_{\bm{t}_2}^{(-\rho, \nu)}(q_{XY}, \hat p_{X|Y}|W)$ defined in (\ref{definition_J_2})
for $\lambda = -\rho$ with $0<\rho<1$. 
Define 
\begin{align}
    &\tilde J_{\bm{t}_2}^{(-\rho,\nu)} ( q_X, \hat p_{X|Y} | W )\notag \\
    &= \sum_{x} q_X(x) \log 
    \frac{q_X^{-\rho}(x)}
    {\sum_{y} W(y|x) \mathrm{e}^{-\lambda\nu c(x)} \hat p_{X|Y}^{-\rho}(x|y) } . 
\end{align}

Property~\ref{property:convex_concave} is used to prove the following two lemmas.
\begin{lemma}
\label{lemma10}
For any fixed $0<\rho\leq 1$ and any fixed $q_{XY}$, 
$J_{\bm{t}_2}^{( -\rho, \nu )} (q_{XY}, \hat p_{X|Y} |W )
$ is maximized by $\hat p_{X|Y} = q_{X|Y}$
and the maximum value is 
\begin{align}
J_{\bm{t}_2}^{( -\rho, \nu )} (q_{XY}, q_{X|Y} |W )
= \Theta^{(-\rho, -\rho \nu )}(q_{XY}|W). 
\end{align}
This implies that 
\begin{align}
\max_{q_X} \min_{q_{Y|X}} 
\max_{\hat p_{X|Y}} 
  J_{\bm{t}_2}^{( -\rho, \nu )} (q_{XY}, \hat p_{X|Y} |W )
&= 
\max_{q_X} \min_{q_{Y|X}} 
  J_{\bm{t}_2}^{( -\rho,\nu )} (q_{XY}, q_{X|Y} |W ) \notag \\
&=
\max_{q_X} \min_{q_{Y|X}} 
\Theta^{(-\rho, -\rho\nu )}(q_{XY}|W)  . 
\end{align}
\end{lemma}
\textit{Proof:} It is obvious from the definition of 
$J_{\bm{t}_2}^{(\lambda, \nu)}(q_{XY}, \hat p_{X|Y} |W) $. 

\begin{lemma}
\label{lemma11}
For any fixed $0< \rho \leq 1$, $\nu \ge 0$, $\hat p_{X|Y}\in \mathcal{P(X|Y)}$ and $q_X\in \mathcal{P(X)}$, 
$J_{\bm{t}_2}^{(-\rho, \nu )} (q_{XY}, \hat p_{X|Y} |W )$ 
is minimized by
\begin{align}
q_{Y|X}(y|x) 
&= \frac{
W(y|x) \mathrm{e}^{-\lambda\nu c(x)} \hat p_{X|Y}^{-\rho } (x|y)
}{
\sum_{y'}
W(y'|x) \mathrm{e}^{-\lambda\nu c(x')} \hat p_{X|Y}^{-\rho } (x|y')
} 
\stackrel{\triangle}{=}
q_{Y|X}^*(\hat p_{X|Y} ) (y|x) 
\end{align}
and the minimum value is 
\begin{align}
J_{\bm{t}_2}^{(-\rho,\nu )} ((q_X, q_{Y|X}^*(\hat p_{X|Y} ) ), \hat p_{X|Y} |W )
= \tilde J_{\bm{t}_2}^{(-\rho, \nu)}(q_X, \hat p_{X|Y}|W). 
\end{align}
This implies that 
\begin{align}
\max_{q_X} \max_{\hat p_{X|Y}} \min_{q_{Y|X}} 
J_{\bm{t}_2}^{(-\rho, \nu )} ( q_{XY}, \hat p_{X|Y} |W )
&=
\max_{q_X} \max_{\hat p_{X|Y}} 
J_{\bm{t}_2}^{(-\rho, \nu )} ( (q_X, q_{Y|X}^*(\hat p_{X|Y} ) ), p_{X|Y} |W ) \notag\\
&= 
\max_{q_X} \max_{\hat p_{X|Y}} 
\tilde J_{\bm{t}_2}^{(-\rho, \nu )}(q_X, \hat p_{X|Y}|W). 
\end{align}

\end{lemma}

\begin{lemma}
\label{lemma12}
For any fixed $0< \rho \leq 1$  and any fixed $\hat p_{X|Y}$,
$ \tilde J_{\bm{t}_2}^{(-\rho, \nu )}(q_X, \hat p_{X|Y}|W) $ is maximized by 
\begin{align}
q_X(x)
	&=
\frac{
\left\{
\sum_{y}
W(y|x) \mathrm{e}^{-\lambda\nu c(x)} \hat p_{X|Y}^{-\rho} (x|y)
\right\}^{-1/\rho}
}
{
\sum_{x'}
\left\{
\sum_{y}
W(y|x') \mathrm{e}^{-\lambda\nu c(x')} \hat p_{X|Y}^{-\rho} (x'|y)
\right\}^{-1/\rho}
	} 
	\stackrel{\triangle}{=}
q_X^*(\hat p_{X|Y})(x) 
\end{align}
and the maximum value is 
\begin{align}
    \tilde J_{\bm{t}_2}^{(-\rho, \nu)}(q_X^*(\hat p_{X|Y}), \hat p_{X|Y}|W)
    = A^{(\rho, \nu)} (\hat p_{XY} |W).
\end{align}
This implies that 
\begin{align}
    &\max_{q_X} \max_{\hat \hat p_{X|Y} } \tilde J_{\bm{t}_2}^{(-\rho, \nu)}(q_X), \hat p_{X|Y}|W)\notag\\
    &= \max_{ \hat p_{X|Y} } \tilde J_{\bm{t}_2}^{(-\rho, \nu)}(q_X^*(\hat p_{X|Y}), \hat p_{X|Y}|W)\notag\\
    &= \max_{ \hat p_{X|Y} } A^{(\rho, \nu)} (\hat p_{XY} |W). 
\end{align}

\end{lemma}
See Appendix \ref{appendixB} for the proof of Lemma~\ref{lemma11} and \ref{lemma12}.

\noindent 
\textit{Proof of Proposition \ref{proposition4}:}
For any fixed $\rho \in (0,1]$, $\nu\ge 0$, we have 
\begin{align}
& \max_{q_X} \min_{q_{Y|X}} \Theta^{(-\rho, -\rho \nu)}(q_{XY}|W) \notag \\
& \stackrel{\rm (a)}
= 
\max_{q_X} \min_{q_{Y|X}} \max_{\hat p_{X|Y} } J_{\bm{t}_2}^{(-\rho, \nu)} (q_{XY}, \hat p_{X|Y} |W) \notag \\
& \stackrel{\rm (b)}
= 
\max_{q_X} \max_{\hat p_{X|Y} } \min_{q_{Y|X}} J_{\bm{t}_2}^{(-\rho, \nu)} (q_{XY}, \hat p_{X|Y} |W) \notag \\
& \stackrel{\rm (c)}
= 
\max_{q_X} \max_{\hat p_{X|Y} } \tilde J_{\bm{t}_2}^{(-\rho, \nu)} (q_{X}, \hat p_{X|Y} |W) \notag \\
& \stackrel{\rm (d)}
= 
\max_{\hat p_{X|Y} } A^{(\rho, \nu)} (\hat p_{X|Y} |W) . 
\label{eq.70}
\end{align}
Step (a) follows from Lemma~\ref{lemma10}, 
In step (b), the interchange of the order of $\min$ and $\max$ is justified because of Property~\ref{property2}. 
Step (c) follows from Lemma~\ref{lemma11} and 
step (d) follows from Lemma~\ref{lemma12}. 
Therefore we have
\begin{align}
    E_{\rm CK}(R, \Gamma |W) 
    & \stackrel{\rm (e)}
    = \sup_{\rho \in (0,1] } 
      \inf_{\nu \ge 0}
    \left\{ \max_{q_X} \min_{q_{Y|X}} \Theta^{(-\rho, -\rho \nu)} (q_{XY}|W) - \rho R 
    +\rho \nu \Gamma \right\} \notag\\
    & \stackrel{\rm (f)}
    = \sup_{\rho \in (0,1] } \inf_{\nu \ge 0}
     \left\{ \max_{ \hat p_{X|Y} } A^{(\rho, \nu )} ( \hat p_{X|Y} |W)  -\rho R + \rho \nu \Gamma \right\}\notag\\
    & = \tilde E_{\rm r}(R, \Gamma|W) 
    \stackrel{\rm (g)}= E_{\rm r}(R, \Gamma|W) . 
\end{align}
Step (e) follows from (\ref{eq.54}), 
step (f) follows from (\ref{eq.70}), and 
step (g) follows from Corollary~\ref{corollary_2}.
This completes the proof.  \hfill$\IEEEQED$

\begin{figure*}
\centering
\includegraphics[width=0.75\columnwidth]{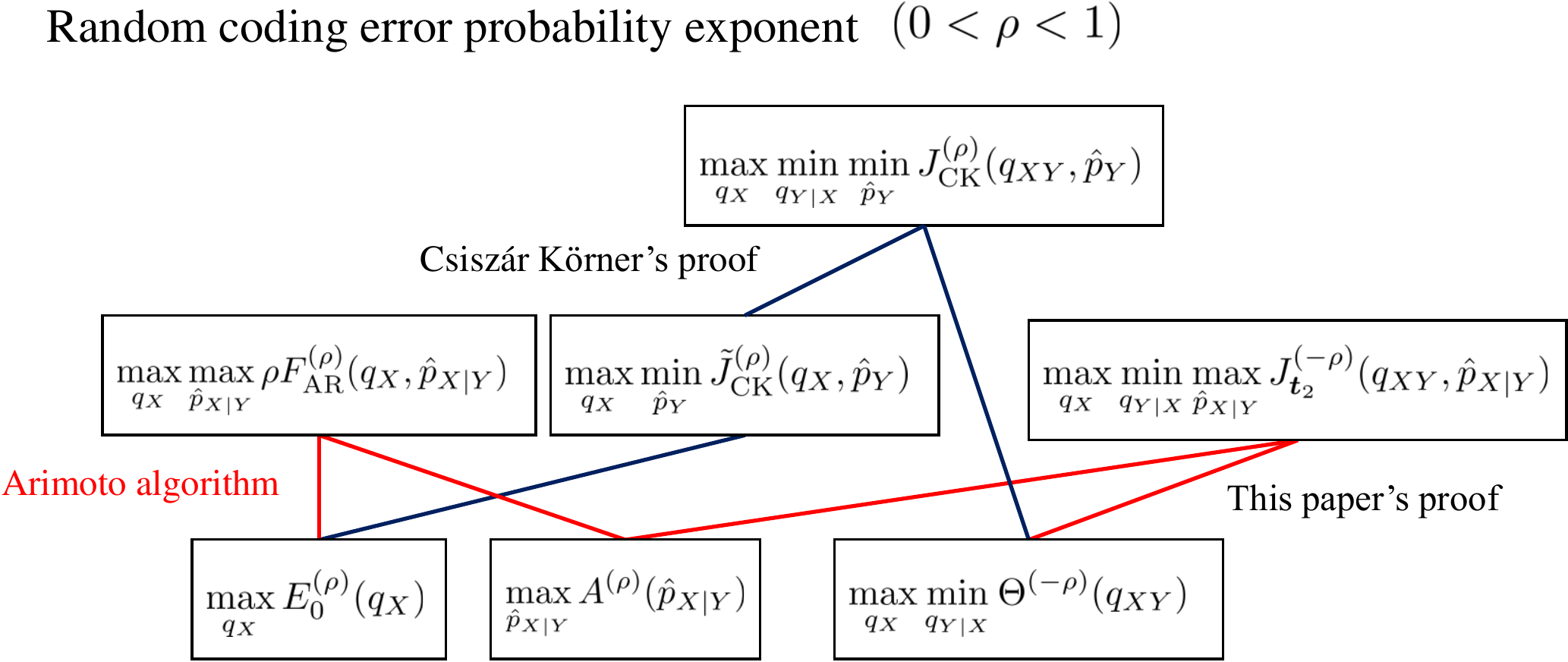}
\caption{Relation between several expressions of
the random coding error probability exponent}
\label{fig.2}
\end{figure*}

A comparison of the proof in~\cite[Exercise 10.24]{Csiszar-KornerBook} and our proof is shown in Fig.~\ref{fig.2}.
The guideline shown in~\cite[Exercise 10.24]{Csiszar-KornerBook} is a standard method to prove the match of 
Csisz\'ar and K\"orner's exponent and Gallager's exponent.
It is interesting that our procedure uses Corollary \ref{corollary_2}, i.e., the match of $\tilde E_{\rm r}$ and $E_{\rm r}$, which was derived by Arimoto's algorithm.   
It should be noted that due to the structural difference with $J_{\bm{t}_2}^{(\lambda, \nu)}(q_{XY}, \hat p_{X|Y}|W)$,
the function $J_{\bm{t}_1}^{(\lambda, \nu)}(q_{XY}, p_X|W)$ cannot be used to prove the match between the two exponents. 
The difference is that for $J_{\bm{t}_2}^{(\lambda, \nu)}(q_{XY}, \hat p_{X|Y}|W)$,
the joint distribution $q_{XY}$ can be separated $q_X$ and $q_{Y|X}$ as shown above, while
it is separated as $q_Y$ and $q_{X|Y}$ for 
$J_{\bm{t}_1}^{(\lambda, \nu)}(q_{XY}, p_X|W)$

If we think the double maximization in the third line of (\ref{eq.70}) will lead to an algorithm, that's a misunderstanding.
In fact, if the optimum solution when one of the variables is fixed cannot be written explicitly, it will not lead to a computation algorithm for alternating optimization like the Arimoto algorithm.
The reason why it is difficult to derive the computation algorithm for the Csisz\'ar and K\"orner's error exponent is that the the exponent is defined as the minimax problem, unlike the Dueck and K\"orner's strong converse exponent.
Even if the form of the double maximization problem can be derived, it is still necessary to be able to explicitly write the solution when one of the variables is fixed.
This condition makes it difficult to derive the computation algorithm for Csisz\'ar and K\"orner's error exponent.

\section{Algorithms for the lossy source-coding strong converse exponent}

\begin{figure*}
\centering
\includegraphics[width=0.95\textwidth]{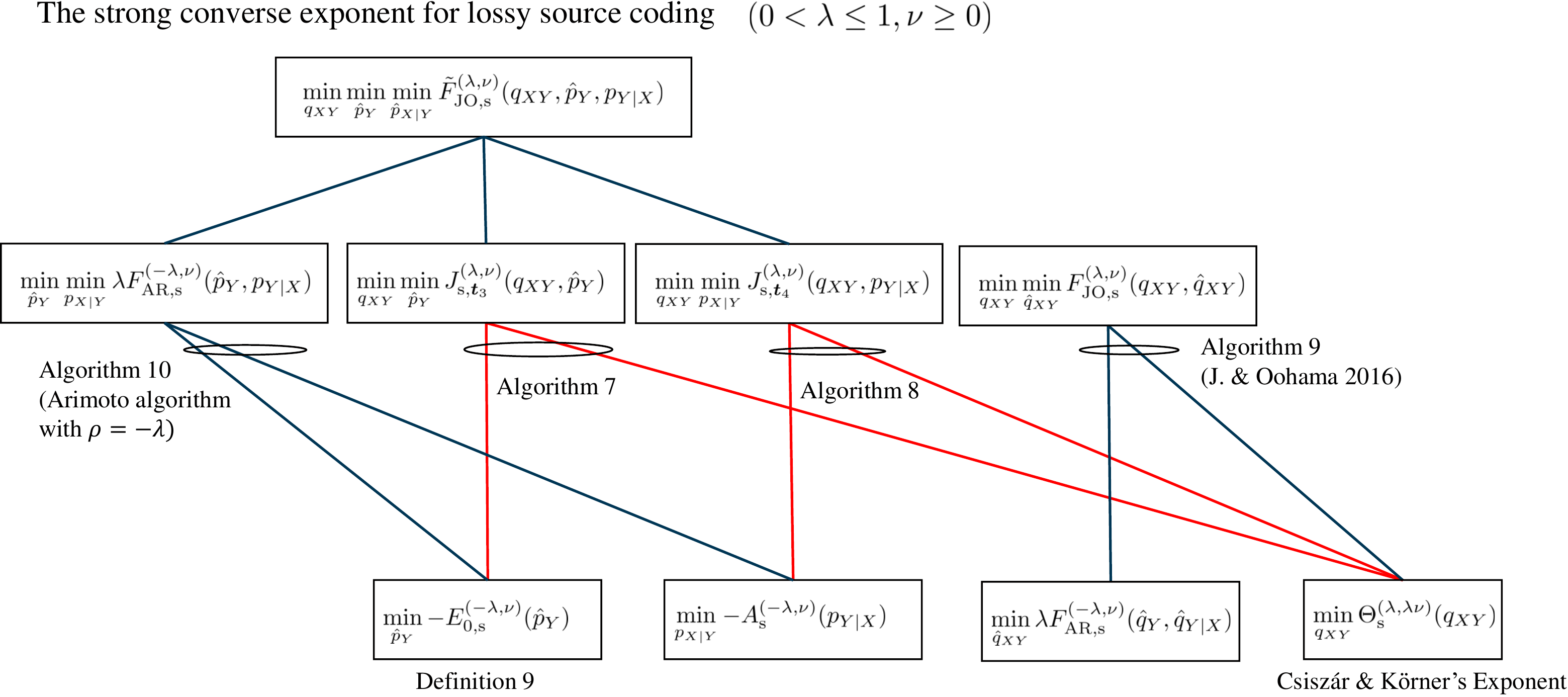}
\caption{Relation between seven expressions
for the strong converse exponent for lossy source coding of DMSs}
\label{fig:st_conv_source_codingL}
\end{figure*}

So far, we have discussed computation algorithms 
for the correct and error decoding probability exponents in channel coding.
This section describes algorithms for computing
exponents in lossy source coding. 
The first algorithm for computing 
Csisz\'ar and K\"orner's strong converse exponent
was due to the authors~\cite{YutakaISIT2016b}. 
In the following subsection, we propose a family of algorithms for computing the strong converse exponent in lossy source coding that is similar to the algorithm family of Tridenski et al. 
The previously proposed algorithm is included in the new algorithm family. 
Then, two special members of the algorithm family that have important properties are described in detail. 
Fig.~\ref{fig:st_conv_source_codingL} shows the relation between the algorithms. 
The structure of the diagram of Fig.~\ref{fig:st_conv_source_codingL} is quite similar to Fig.~\ref{fig.1L}. 
In~Fig.~\ref{fig:st_conv_source_codingL}, nine expressions defined by minimization problems are shown. They are the quantities used to express the strong converse exponent with parameter $\lambda \in [0,1]$ and $\nu\geq 0$. 
The function $\tilde F_{\rm JO, s}^{(\lambda, \nu)}(q_{XY}, \hat p_Y, p_{Y|X})$ will be defined in Section~\ref{section:comparison_lossy_source_coding}.

In Arimoto's paper~\cite{Arimoto1976}, an algorithm for computing Blahut's error exponent was given. The validity of computing the strong converse exponent with Arimoto algorithm has not been clarified. 
The issue is discussed in this section and Section~\ref{section_sourcecoding_Arimotoalgorithm}
and its validity is clarified. 


The rate-distortion function $R(\Delta|q_X)$ is generally not convex in $q_X$ and therefore it is difficult to develop an algorithm directly from the expression of (\ref{G_DK}).
This motivated us to define the following function: 
\begin{definition}
For any fixed 
$R\geq 0$, $\Delta\geq 0$, we have 
\begin{align}
    \tilde G_{\rm CK} ( R, \Delta | P_X)  
    = \min_{ \genfrac{}{}{0pt}{2}{ q_{XY} \in \mathcal{P(X,Y)}:}{{\rm E}[d(X,Y)] \leq \Delta } } 
    \{ D(q_X||P_X)  + | I( q_X, q_{Y|X} ) - R |^+ \} .
\end{align}
\end{definition}
Then, we have the following lemma:
\begin{lemma}
\label{lemma:tilde_G_CK}
For any fixed $R\geq 0$, $\Delta\geq 0$, 
and any fixed $P_X\in \mathcal{P(X)}$,
we have 
\begin{align}
  G_{\rm CK}(R, \Delta|P_X) = 
  \tilde G_{\rm CK}(R, \Delta|P_X). 
\end{align}
\end{lemma}
See Appendix~\ref{appendixD} for the proof of Lemma~\ref{lemma:tilde_G_CK}.

Lemma~\ref{lemma:tilde_G_CK} assures that $G_{\rm CK}(R,\Delta|P_X)$ can be
computed by evaluating $\tilde G_{\rm CK}(R,\Delta|P_X)$, the form of which
is quite similar to $G_{\rm CK}(R,\Delta|P_X)$. 
In the representation of $\tilde G_{\rm CK}(R,\Delta|P_X)$, we could eliminate the rate-distortion function. 
Based on this representation, we can derive an algorithm. 

The function $\tilde G_{\rm CK}(R, \Delta|P_X)$ satisfies the following 
{property, which is required to derive} its parametric expression:
\begin{property}\label{pr:G_CK}
\
\begin{itemize}
\item[a)] 
$\tilde G_{\rm CK}(R, \Delta|P_X)$ is a monotone non-increasing
function of $R \geq 0$
for a fixed $\Delta \geq 0$ 
and is a monotone non-increasing function of $\Delta \geq 0$
for a fixed $R\geq 0$.
\item[b)] $\tilde G_{\rm CK}(R, \Delta|P_X)$ is a convex function of $(R,\Delta)$.
\item[c)] 
$\tilde G_{\rm CK}(R, \Delta|P_X)$ takes positive value
for $0 \leq R < R(\Delta|P_X)$.
For $R\geq R(\Delta|P_X)$, 
$\tilde G_{\rm CK}(R, \Delta|P_X)=0$. 
\item[d)]
For $R' \geq R \geq 0$, we have
$
\tilde G_{\rm CK}(R, \Delta | P_X ) - \tilde G_{\rm CK}( R', \Delta | P_X )
\leq R' - R.
$
\end{itemize}
\end{property}
See Appendix~\ref{appendixE} for the proof.

Then, we define the following function that will
be used as an objective function of the algorithm. 
\begin{definition}
For any fixed $\lambda\in [0,1]$, we define 
\begin{align}
    G_{\rm CK}^{(\lambda)} ( \Delta | P_X)  
    = \min_{ \genfrac{}{}{0pt}{2}{ q_{XY} \in \mathcal{P(X,Y)}:}{{\rm E}[d(X,Y)] \leq \Delta } } 
    \{ D(q_X||P_X)  + \lambda I( q_X, q_{Y|X} ) \} .
\end{align}
\end{definition}

\begin{definition}
For any fixed 
$q_{XY}\in \mathcal{P(X\times Y)}$, 
$\lambda\in [0,1]$, and $\mu\geq 0$, we define 
\begin{align}
    \Theta^{(\lambda, \mu)}_{\rm s}(q_{XY}|P_X)
    &=
    D(q_X||P_X) + \lambda I(q_X, q_{Y|X})
    + \mu {\rm E}_{q_{XY}} [ d(X,Y) ] \notag\\
    &= 
    {\rm E}_{q_{XY} }\left[ \log 
    \frac{ q_X^{1-\lambda}(X) q_{X|Y}^{ \lambda }(X|Y)}{ P_X(X) {\rm e}^{-\mu d(X,Y) } } \right], 
        \label{Theta_s} \\
    \Theta^{(\lambda, \mu)}_{\rm s}(P_X)    
    &= \min_{q_{XY}}  \Theta^{(\lambda, \mu)}_{\rm s}(q_{XY}|P_X). 
\end{align}

\end{definition}

The subscript `$\mathrm{s}$' means that this function is the source coding counterpart of the corresponding function in channel coding. 
One can observe
that Eq.(\ref{Theta_s}) for 
$\Theta_{\rm s}^{(\lambda, \mu)}(q_{XY}|P_X)$ is similar to Eq.(\ref{Theta})
for $\Theta^{(\lambda)}(q_{XY}|W)$ for the channel coding. 

\begin{lemma} 
\label{lemma17}
For any fixed $\lambda\in[0,1]$, $\Delta\geq 0$, 
and $P_X\in \mathcal{P(X)}$,
we have 
\begin{align}
G_{\rm CK}^{(\lambda)} (\Delta| P_X )
&=
  \sup_{\mu \geq 0}
  \left\{
  \Theta_{\rm s}^{(\lambda, \mu)} ( P_X ) 
    - \mu \Delta 
  \right\}. 
  \label{eq.lemma17.1}
\end{align}
For any fixed $R\geq 0$, $\Delta\geq 0$, 
and $P_X\in \mathcal{P(X)}$,
we have 
\begin{align}
\tilde G_{\rm CK}(R, \Delta | P_X)
&= \max_{0\leq \lambda \leq 1}
  \left\{
  G_{\rm CK}^{(\lambda)} (\Delta| P_X ) 
    -\lambda R   
  \right\}
    \label{eq.lemma17.2}\\
&= \max_{0\leq \lambda \leq 1}
  \sup_{\mu \geq 0}
  \left\{
  \Theta_{\rm s}^{(\lambda, \mu)} ( P_X ) 
    - \mu \Delta -\lambda R 
  \right\} . 
  \label{eq.lemma17.3}
\end{align}
\end{lemma}
See Appendix~\ref{appendixF} for the proof of Lemma~\ref{lemma17}.

\begin{figure*}
    \centering
    \includegraphics[scale=0.6]{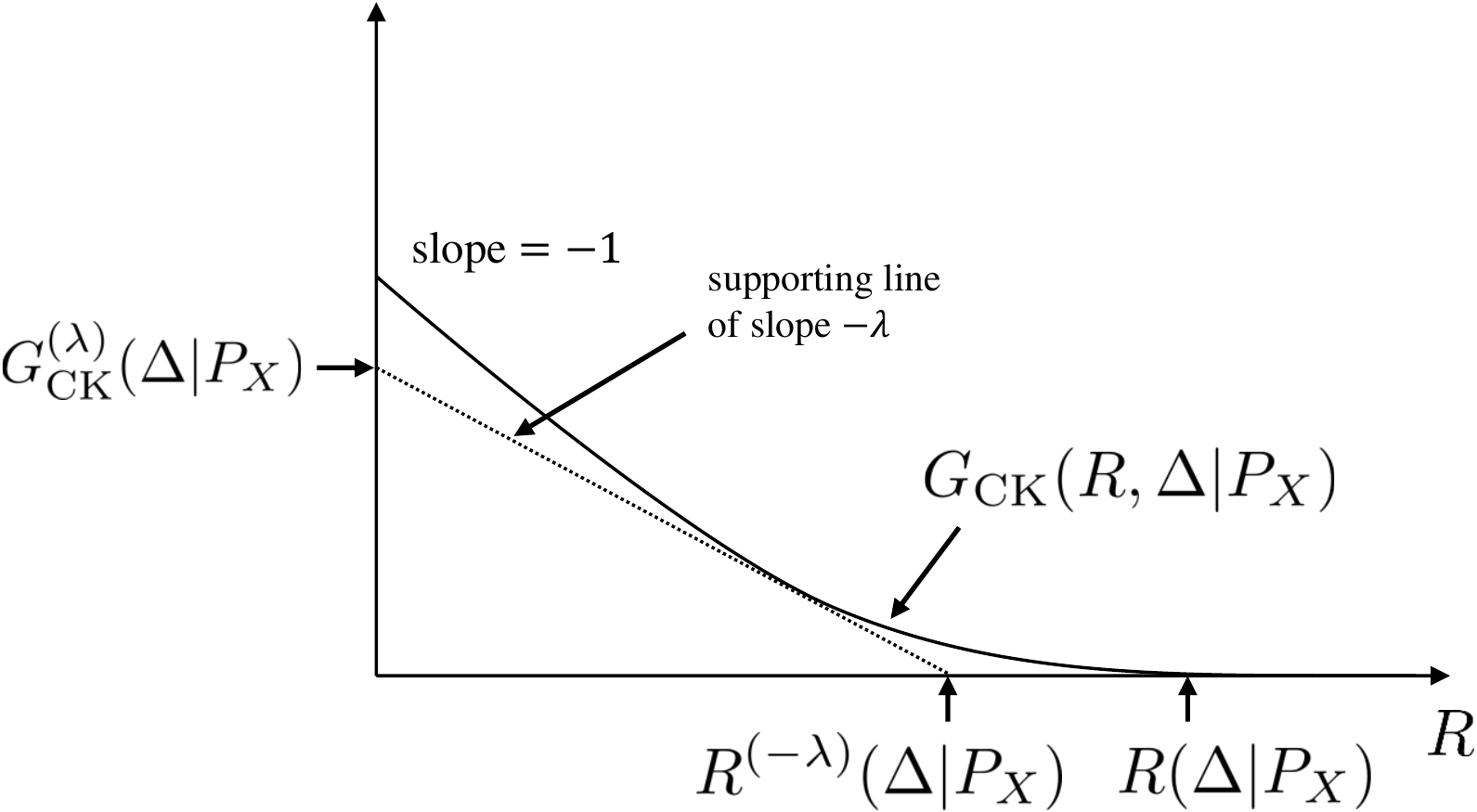}
    \caption{Csisz\'ar and K\"orner's strong converse exponent function and the generalized cutoff rate in lossy source coding.}
    \label{fig:strong_converse_exponent_source}
\end{figure*}

Eq.(\ref{eq.lemma17.1}) shows that $G_{\rm CK}^{(\lambda)}(\Delta |P_X)$
is computed by the Legendre transform of $-\Theta_{\rm s}^{(\lambda, \mu)}(P_X)$
with respect to $-\mu\leq 0$.
Then, Eq.(\ref{eq.lemma17.2}) shows that 
$\tilde G_{\rm CK}(R,\Delta |P_X)$ is
computed by the Legendre transform of $-G_{\rm CK}^{(\lambda)}(\Delta |P_X)$
with respect to $-\lambda \in [-1,0]$.
This immediately implies that $\tilde G_{\rm CK}(R,\Delta |P_X)$ is computed by the two-dimensional Legendre transform of $\Theta_{\rm s}^{(\lambda, \mu)}(P_X)$ whose domain is $(\lambda, \mu)\in [0,1]\times[0,\infty)$, as shown in Eq.(\ref{eq.lemma17.3}).
An image of the shape of function $G_{\rm CK}(R, \Delta |P_X)$ is shown in Fig.~\ref{fig:strong_converse_exponent_source}. The supporting line
to this function of slope $-\lambda $ is expressed by
$ -\lambda R + G_{\rm CK}^{(\lambda)}(\Delta|P_X)$. 
We can easily see that
$ G_{\rm CK}^{(0)}(\Delta|P_X) = \min_{q_{XY}: \mathrm{E}[d(X,Y)]\leq \Delta} D(q_X||P_X) = 0$
by choosing $q_X=P_X$ and $q_{Y|X} (y|x) = 1$ for every $x$ and $y$ with $d(x,y)=0$.
For $\lambda = 1$, we have
\begin{align}
    \Theta^{(1, \nu)}_{\rm s} (P_X) = -\log \max_y \sum_{x} P_X(x) {\rm e}^{-\nu d(x,y)}. 
\end{align}

The generalized cutoff rate, denoted by $R^{(-\lambda)}(\Delta|P_X)$ 
is the $R$-axis intercept of this supporting line. Therefore
$$
R^{(-\lambda)}(\Delta|P_X) = \frac1{\lambda} G_{\rm CK}^{(\lambda)}(\Delta|P_X).
$$

We now perform a change of variables: let $\mu = \lambda \nu$, and instead of the parameter pair $(\lambda, \mu)$, let $(\lambda, \nu)$ be the variable parameter pair.
By doing so, in the limit of $\lambda \to 0$, one of the new algorithms is reduced to the Arimoto-Blahut algorithm that computes the rate distortion function. 
From (\ref{eq.lemma17.1}), we have
\begin{align}
    R^{(-\lambda)}(\Delta|P_X)
    &= \frac{1}{\lambda} \sup_{\mu\geq0}\{ \Theta_{\rm s}^{(\lambda, \mu)} (P_X) - \mu \Delta\} \notag\\
    &=
    \sup_{\mu\geq0}\{ \frac{1}{\lambda} \Theta_{\rm s}^{(\lambda, \mu)} (P_X) - \frac{\mu}{\lambda}  \Delta\} \notag\\
    &=
    \sup_{ \nu \geq0}\{ \frac{1}{\lambda} \Theta_{\rm s}^{(\lambda, \lambda  \nu )} (P_X) -  \nu   \Delta\} .
    \label{eq.46a}
\end{align}

In the following subsections, we describe algorithms shown in Fig.~\ref{fig:st_conv_source_codingL}. 
The following discussions for source coding are parallel with the discussions for channel coding.
In Section~\ref{sec:generalized_algotirhm_source_coding}, we define a generalization algorithm for lossy  source coding inspired by the algorithm family of Tridenski et al.

\subsection{The algorithm family for lossy source coding}
\label{sec:generalized_algotirhm_source_coding}
We define the objective function for the family of algorithms for source coding as 
\begin{align}
    J_{{\rm s}, \bm{t}}^{(\lambda, \nu)} (q_{XY}, p_{XY} |P_X) 
    &= \Theta_{\rm s}^{(\lambda, \lambda \nu)}(q_{XY} |P_X)
    + (1-\lambda) D_{ \bm{ t } }(q_{XY}, p_{XY}),
    \label{def:J_t_s}
\end{align} 
where $\bm{t}=(t_1, t_2, t_3, t_4)$ and $t_i\geq 0$. 
The second term of the above equation is exactly the same as that of (\ref{def:F_TSZ}).
We have the following lemma:
\begin{lemma}
\label{lemma.min_p_Jst}
For fixed $\lambda\in[0,1]$, $\nu\geq 0$, $\bm{t}\in \mathcal{T}$ 
and $q_{XY}\in \mathcal{P(X\times Y)}$,
$J_{{\rm s}, \bm{t}}^{(\lambda, \nu)} (q_{XY}, p_{XY} | W)$ 
is minimized by $p_{XY} = q_{XY}$ and the minimum value is
\begin{align}
    J_{{\rm s}, \bm{t}}^{(\lambda, \nu)} ( q_{XY}, q_{XY} | P_X )
    =
    \Theta_{{\rm s}}^{(\lambda, \nu)} ( q_{XY} | P_X ).
\end{align}
This implies that
\begin{align}
&    \min_{q_{XY}} \min_{p_{XY}} 
    J_{{\rm s}, \bm{t}}^{(\lambda, \nu)} ( q_{XY}, p_{XY} | P_X )
    =
    \min_{q_{XY}} 
    J_{{\rm s}, \bm{t}}^{(\lambda, \nu)} ( q_{XY}, q_{XY} | P_X ) \notag\\
&    =
    \min_{q_{XY}} \Theta_{{\rm s}}^{(\lambda, \nu)} ( q_{XY} | P_X )
    =\Theta_{{\rm s}}^{(\lambda, \nu)} ( P_X )
    \label{eq.146}
\end{align}
holds for any $\bm{t}\in \mathcal{T}$.
\end{lemma}

\begin{IEEEproof}
We have $D_{\bm{t}}(q_{XY}, p_{XY})\geq 0$ because
of the non-negativity of the divergences and conditional divergences.
Equality holds if $q_{XY} = p_{XY}$. This completes the proof.
\end{IEEEproof}

\begin{lemma} 
\label{lemma.min_q_Jst}
For fixed $\lambda\in[0,1]$, $\nu\geq 0$, $\bm{t}\in \mathcal{T}$ 
and $p_{XY}\in \mathcal{P(X\times Y)}$,
$J_{{\rm s}, \bm{t}}^{(\lambda, \nu)} (q_{XY}, p_{XY} | P_X)$ 
is minimized by $q_{XY} = q_{XY}^{*}(p_{XY}) \coloneqq 
\arg\min_{ q_{XY} }
J_{{\rm s}, \bm{t}}^{(\lambda, \nu)} (q_{XY}, p_{XY} | P_X)$ 
and the minimum value is 
\begin{align}
    J_{{\rm s}, \bm{t}}^{(\lambda, \nu)} ( q_{XY}^{*}(p_{XY}), p_{XY} | P_X )
    =
    \hat J_{{\rm s}, \bm{t}}^{(\lambda, \nu)} ( p_{XY} | P_X ).
\end{align}
This implies that
\begin{align}
 \min_{q_{XY}} \min_{p_{XY}}
    J_{{\rm s}, \bm{t}}^{(\lambda, \nu)} ( q_{XY}, p_{XY} | P_X )
    =
    \min_{p_{XY}}  
    J_{{\rm s}, \bm{t}}^{(\lambda, \nu)} ( q_{XY}^{*}(p_{XY}), p_{XY} | P_X )
    =
    \min_{p_{XY}}
    \hat J_{{\rm s}, \bm{t}}^{(\lambda, \nu)} ( p_{XY} | P_X ).
    \label{eq.148}
\end{align}
\end{lemma}

In Lemma~\ref{lemma.min_q_Jst}, optimal $q^*_{XY}$ is expressed implicitly. 
Explicit form for the optimal $q^*_{XY}$ is needed to use $J_{{\rm s}, \bm{t}}^{(\lambda, \nu)} (q_{XY}, p_{XY} | P_X)$ as a surrogate function for $\Theta_{\rm s}^{(\lambda, \nu)}(q_{XY}|P_X)$.
It is easy to verify that function (\ref{def:J_t_s}) satisfies a) and b) among the three conditions 
listed in Section~\ref{section:generalized_algorithm_TSZ} for any $\bm{t} \in \mathcal{T}$. 
Regarding the condition c), we have the following lemma.
\begin{lemma}
\label{lemma:two_conditions_lossy}
Define
\begin{align}
\mathcal{T}_3 &= \{ (t_1, t_2, t_3, t_4) \in \mathcal{T}: t_3 = t_4 + \lambda /(1-\lambda) \},\\    
\mathcal{T}_4 &= \{ (t_1, t_2, t_3, t_4) \in \mathcal{T}: t_2 = t_1 + 1 \}.
\end{align}
If $\bm{t} \in \mathcal{T}_3 \cup \mathcal{T}_4$, then for $\lambda \in [0,1)$, the optimal distribution $q_{XY}$ that attains 
the minimum of $J_{{\rm s}, \bm{t}}^{(\lambda, \nu)}(q_{XY}, p_{XY}|W)$ for a fixed $p_{XY}$ is expressed explicitly. 
\end{lemma}

The proof of Lemma \ref{lemma:two_conditions_lossy} 
is provided in Appendix~\ref{section:update_rule_of_generalized_algorithm_source_coding}.
The algorithm family derived from $ J_{{\rm s}, \bm{t}}^{(\lambda, \nu)}(q_{XY}, p_{XY}|W) $ is given 
in Algorithm~\ref{alg:family_lossy_source_coding}.

\begin{algorithm}
\caption{
A family of algorithms for computing the lossy source coding strong converse exponent
}
\label{alg:family_lossy_source_coding}
\begin{algorithmic}
    \Require The probability distribution of the source $P_X$, 
    the distortion measure $d(x,y)$, 
    $\lambda \in (0,1)$ and $\nu\geq 0$.
    Choose initial joint probability distribution $p_{XY}^{[0]}$ such that
    the set $\{ q_{XY}: J_{{\rm s}, \bm{t}}^{(\lambda, \nu)} (q_{XY}, p_{XY}^{[0]} | P_X ) < +\infty \}$ is non-empty. Choose $\bm{t} \in \mathcal{T}_3 \cup \mathcal{T}_4$.  
\For{ $i=0,1,2,\ldots$,}
\If{$\bm{t}\in \mathcal{T}_3$},     
\begin{flalign}
    &K_3^{[i]}(x) =
    \sum_y 
        \left\{
    {\rm e}^{-\lambda \nu d(x,y) } p_{Y|X}^{[i]} (y|x)^{(1-\lambda) (t_2+t_4)} p_Y^{[i]}(y)^{\lambda}
    \right\} ^{\frac{1}{(1-\lambda)(\lambda + t_2 + t_4) }}, \\
    & q_{Y|X}^{[i]}(y|x) = \frac{1}{K_3^{[i]}(x)}
    \left\{
    {\rm e}^{-\lambda \nu d(x,y) } p_{Y|X}^{[i]} (y|x)^{(1-\lambda) (t_2+t_4)} p_Y^{[i]}(y)^{\lambda}
    \right\} ^{\frac{1}{(1-\lambda)(\lambda + t_2 + t_4) }} ,\\
    & q_{X}^{[i]} (x) = 
    \frac{ 
    \left\{ 
    P_X(x)
    p_X^{[i]}(x)^{  (1-\lambda) (t_1+t_4) }  
    K_3^{[i]}(x)^{ \lambda + (1-\lambda) (t_2+t_4) }
    \right\}^{ \frac{1}{ 1+(1-\lambda)(t_1+t_4) } }
    }
    {
    \sum_{x'} 
        \left\{ 
    P_X(x')
    p_X^{[i]}(x')^{  (1-\lambda) (t_1+t_4) }  
    K_3^{[i]}(x')^{ \lambda + (1-\lambda) (t_2+t_4) }
    \right\}^{ \frac{1}{ 1+(1-\lambda)(t_1+t_4) } }
    },
\end{flalign}
\ElsIf{$\bm{t}\in \mathcal{T}_4$}, 
\begin{align}
    & K_4^{[i]}(y) = \sum_{x}\left\{
    P_X(x) {\rm e}^{-\lambda \nu d(x,y)} p_{Y|X}^{[i]}(y|x)^{1-\lambda} 
    p_{X|Y}^{[i]}(x|y)^{(1-\lambda)(t_1+t_4)}
    \right\}^{\frac{1}{1+(1-\lambda)(t_1+t_4)}}, \\
    &q_{X|Y}^{[i]}(x|y) 
    = \frac{1}{K_4^{[i]}(y)} 
    \left\{
    P_X(x) {\rm e}^{-\lambda \nu d(x,y)} p_{Y|X}^{[i]}(y|x)^{1-\lambda} 
    p_{X|Y}^{[i]}(x|y)^{(1-\lambda)(t_1+t_4)}
    \right\}^{\frac{1}{1+(1-\lambda)(t_1+t_4)}}, \\
    & q_Y^{[i]}(y) 
    = 
    \frac{
    \left\{
    p_Y^{[i]}(y) ^{t_1+t_3}
    K_4^{[i]}(y)^{ \frac{1}{1-\lambda} + t_1+t_4 } 
    \right\}^{ \frac{1}{1+t_1+t_3} }
    }
    { 
    \sum_{y'}
    \left\{
    p_Y^{[i]}(y') ^{t_1+t_3}
    K_4^{[i]}(y')^{ \frac{1}{1-\lambda} + t_1+t_4 } 
    \right\}^{ \frac{1}{1+t_1+t_3} }
    }. 
\end{align}
\EndIf
\begin{flalign}
    & 
    p_{XY}^{[i+1]}(x,y) = q_{XY}^{[i]}(x,y) 
    \label{update2_Algorithm_Family_Source}
\end{flalign}
\EndFor
\end{algorithmic}
\end{algorithm}

\begin{figure*}
    \centering
    \includegraphics[width=0.9\textwidth]{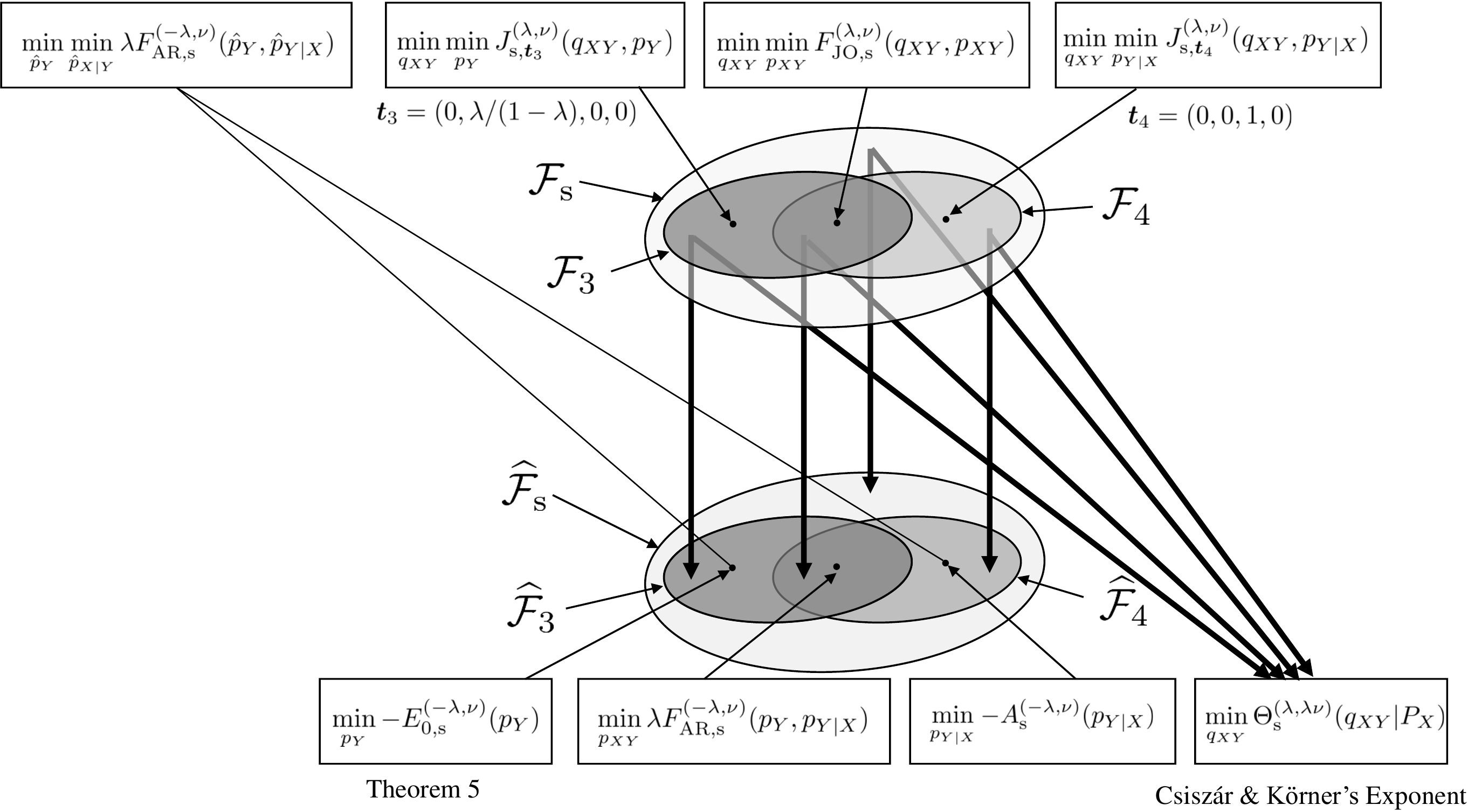}
    \caption{ The set $\mathcal{F}_{\rm s}$ of double minimization forms of the strong converse exponent for lossy source coding
    and the set $\mathcal{F}_{\rm s}$ of single minimization forms. 
    An algorithm parameterized by $\bm{t}$ is obtained if $\bm{t}\in \mathcal{T}_3 \cup \mathcal{T}_4$. 
    }
    \label{fig:generalized_algorithm_source_coding}
\end{figure*}

Fig.~\ref{fig:generalized_algorithm_source_coding} illustrates the relation 
between the expression of Csisz\'ar and K\"orner's exponent,
the double minimization form of the surrogate objective function with parameter $\bm{t}$,
and the expressions derived from the parameterized objective function by fixing $p_{XY}$
and optimizing the objective function with respect to $q_{XY}$.
Define
\begin{align}
    \hat J_{{\rm s}, \bm{t}} ^{(\lambda, \nu)} (p_{XY}|P_X)
    = \min_{ q_{XY} } J_{{\rm s}, \bm{t}} ^{(\lambda, \nu)} ( q_{XY}, p_{XY} |P_X) .
    \label{def:hat_J_t_s}
\end{align}
Define four sets of minimization problems as follows:
\begin{align}
    \mathcal{F}_{i} &=
    \left\{ \min_{ q_{XY} } \min_{ p_{XY} } 
    J_{{\rm s}, \bm{t}} ^{(\lambda, \nu)} ( q_{XY}, p_{XY} |P_X) 
    \right\}_{\bm{t} \in \mathcal{T}_i} \label{F_i_source}\\
    \widehat{\mathcal{F}}_{i} &=
    \left\{ \min_{ q_{XY} } 
    \hat{J}_{{\rm s}, \bm{t}} ^{(\lambda, \nu)} ( q_{XY}, p_{XY} |P_X) 
    \right\}_{\bm{t} \in \mathcal{T}_i} \label{hat_F_i_source}
\end{align}
for $i=3,4$. Replace the $\mathcal{T}_i$ in (\ref{F_i_source}) and (\ref{hat_F_i_source})
with $\mathcal{T}$ to define the sets $\mathcal{F}_{\rm s}$ and $\widehat{\mathcal{F}}_{\rm s}$.  
Then, the large circle at the top of Fig.~\ref{fig:generalized_algorithm_source_coding} expresses the
set $ \mathcal{F}_{\rm s} =  \{ \min_{q_{XY} } \min_{p_{XY} } J_{{\rm s}, \bm{t}} ^{(\lambda, \nu)} (q_{XY}, p_{XY}|P_X)\}_{\bm{t} \in \mathcal{T}} $. 
The two small circles within $\mathcal{F}_{\rm s}$ express $\mathcal{F}_3$ and $\mathcal{F}_4$.
The expression $\min_{p_Y} - E_{0,\rm s}^{(-\lambda,\nu)}(p_Y)$ that appeared in (\ref{def:eq.GCK_check}) 
is included in $\mathcal{F}_3$
and the expression $\min_{\hat p_{Y|X}} - A_{\rm s}^{(-\lambda,\nu)}(p_{Y|X})$ that will appear in Section~\ref{section:algorithm6} is included in $\mathcal{F}_4$. 
The expression $\min_{q_{YX}} \lambda F_{\rm AR,s}^{(-\lambda, \nu)}(q_Y, q_{Y|X})$ which
will appear in Section~\ref{section:comparison_lossy_source_coding} is included in the intersection of 
$\mathcal{F}_3$ and $\mathcal{F}_4$.

\begin{proposition}
\label{proposition_Algorithm_Family_Source}
For $i=1,2,3, \ldots$, we have
\begin{align*}
    & J_{{\rm s}, \bm{t}}^{(\lambda,\nu)} (q_{XY}^{[0]}, p_{XY}^{[0]} |P_X)
    \stackrel{\rm (a) }
    \geq
    J_{{\rm s}, \bm{t}}^{(\lambda,\nu)} (q_{XY}^{[0]}, p_{XY}^{[1]} |P_X)
    \stackrel{\rm (b) }
    \geq
    J_{{\rm s}, \bm{t}}^{(\lambda,\nu)} (q_{XY}^{[1]}, p_{XY}^{[1]} |P_X)\ge \cdots 
    \notag\\
    & 
    \geq
    J_{{\rm s}, \bm{t}}^{(\lambda,\nu)} (q_{XY}^{[i]}, p_{XY}^{[i]} |P_X)
    = \Theta_{\rm s}^{(\lambda, \lambda \nu) } ( q_{XY}^{[i]} | P_X ) \notag\\
    &
    \stackrel{\rm (a) }
    \geq
    J_{{\rm s}, \bm{t}}^{(\lambda,\nu)} (q_{XY}^{[i]}, p_{XY}^{[i+1]} |P_X)
    = \hat J_{{\rm s}, \bm{t}}^{(\lambda, \nu) } ( p_{XY}^{[i+1]} | P_X )\notag\\
    &\stackrel{\rm (b) }
    \geq
    J_{{\rm s}, \bm{t}}^{(\lambda,\nu)} (q_{XY}^{[i+1]}, q_{XY}^{[i+1]} |P_X)
    \geq \cdots \notag\\
    &
    \geq
    \min_{q_{XY}} \Theta_{\rm s}^{(\lambda, \lambda \nu)} ( q_{XY} | P_X )
    \stackrel{\rm (c)}
    =
    \min_{p_{XY}} \hat J_{{\rm s}, \bm{t}}^{(\lambda, \nu)}(p_{XY} | P_X ).
\end{align*}
\end{proposition}

\begin{IEEEproof}
Step (a) follows from Lemma~\ref{lemma.min_p_Jst} and 
step (b) follows from Lemma~\ref{lemma.min_q_Jst}.
Step (c) follows from (\ref{eq.146}) in Lemma~\ref{lemma.min_p_Jst} and (\ref{eq.148})
in Lemma~\ref{lemma.min_q_Jst}. This completes the proof.
\end{IEEEproof}

The convergence of the algorithm is stated as follows:

\begin{theorem}
\label{theorem.convergence.family.algorithm.source}
For any $\lambda \in (0,1)$, $\nu \ge 0$, $\bm{t} \in \mathcal{T}_3 \cup \mathcal{T}_4$, and $P_X \in \mathcal{P(X)}$, 
the series of distributions $p_{XY}^{[i]}$ defined by Lemma~\ref{alg:family_lossy_source_coding}  converges to an optimal distribution
$p_{XY}^*$ that minimizes $\hat J_{{\rm s}, \bm{t}}(p_{XY} | P)$. 

\end{theorem}
See Appendix~\ref{appendix.convergence.family.algorithm.source} for the proof.

In the following two subsections, two special cases of the algorithm family are discussed. 
The two cases are $\bm{t}=\bm{t}_3\coloneqq (0,0,\lambda/(1-\lambda),0)$
and $\bm{t}=\bm{t}_4\coloneqq (0,1,0,0)$. 
The algorithm of the first case is called Algorithm~\ref{algorithm_GCK1}.
Using Algorithm~\ref{algorithm_GCK1}, we can show that Csisz\'ar and K\"orner's 
strong converse exponent coincides with a function that is quite similar to Blahut's error exponent
but the slope parameter $\rho$ takes negative value.
The objective function of these case are 
\begin{align}
    J_{{\rm s}, \bm{t}_3 }^{(\lambda, \nu)}(q_{XY}, \hat p_{Y} |P_X) 
    &= J_{{\rm s}, (0,0,\frac{\lambda}{1-\lambda}, 0)}^{(\lambda,\nu)} (q_{XY}, p_{XY} | P_X) \notag\\
    & =  \Theta_{\rm s}^{(\lambda, \lambda \nu)}(q_{XY} |P_X)
    + \lambda D( q_Y || \hat p_Y) \notag\\
    &= \mathrm{E}_{q_{XY}} \left[
    \log \frac{q_X(X)  q_{Y|X}^\lambda(Y|X) }{P_X(X) \mathrm{e}^{- \lambda \nu d(X,Y)} \hat p_Y^\lambda(Y)}
    \right], 
    \label{J_1s}
    \\
    J_{{\rm s}, \bm{t}_4 }^{(\lambda, \nu)}(q_{XY}, p_{Y|X} |P_X)
    &= J_{{\rm s}, (0,1,0, 0)}^{(\lambda,\nu)} (q_{XY}, p_{XY} | P_X) \notag\\
    & =  \Theta_{\rm s}^{(\lambda, \lambda \nu)}(q_{XY} |P_X)
    + (1-\lambda) D( q_{Y|X} ||p_{Y|X} | q_X )\notag \\
    &= \mathrm{E}_{q_{XY}} \left[
    \log \frac{q_Y^{1-\lambda}(Y)  q_{X|Y}(X|Y) }{P_X(X) \mathrm{e}^{- \lambda \nu d(X,Y)}  p_{Y|X}^{1-\lambda}(Y|X)}
    \right]. \label{J_2s}
\end{align}
In Section~\ref{section:algorithm5}, we give Algorithm~\ref{algorithm_GCK1} 
that is derived from $J_{{\rm s}, \bm{t}_3 }^{(\lambda,\nu)}(q_{XY}, \hat p_{Y} |P_X)$ and 
in Section~\ref{section:algorithm6}, we give Algorithm~\ref{algorithm_GCK2} 
that is derived from $J_{{\rm s}, \bm{t}_4 }^{(\lambda, \nu)}(q_{XY}, p_{Y|X} |P_X)$. 

\subsection{Algorithm~\ref{algorithm_GCK1}} \label{section:algorithm5}

In this subsection we describe Algorithm~\ref{algorithm_GCK1}. This algorithm is a member of the algorithm family defined in the previous section and has an important application. 
In Section~\ref{section:def_exponent_source_coding}, we defined a new representation 
$G_{\rm JO}(R, \Delta |P_X)$ of the strong converse exponent, that is basically the same as Blahut exponent but the parameter $\rho$ takes negative values. 
The coincidence of this new exponent function with Csisz\'ar and K\"orner's exponent is proved by the convergence of Algorithm~\ref{algorithm_GCK1}. 
This is one of the main contribution of this paper and is 
analogous to the proof of the match of Arimoto's exponent and Dueck and K\"orner's exponent by the convergence of Algorithm~\ref{new_algorithm_A}. 

The function $J_{{\rm s}, \bm{t}_3 }^{(\lambda, \nu)}(q_{XY}, \hat p_{Y} |P_X)$ satisfies the following two lemmas.
\begin{lemma}
\label{lemma19}
For any fixed $\lambda\in[0,1]$, $\nu\geq 0$, 
and any fixed $q_{XY}$, 
$ J_{{\rm s}, \bm{t}_3 }^{(\lambda, \nu)} (q_{XY}, \hat p_{Y} |P_X)$
is minimized by $\hat p_Y = q_Y$ and the minimum value
is 
\begin{align}
J_{{\rm s}, \bm{t}_3 }^{(\lambda, \nu)} (q_{XY}, q_{Y} |P_X)
= \Theta_{\rm s}^{(\lambda, \lambda \nu)}(q_{XY} |P_X).
\end{align}
This implies that
\begin{align}
\min_{q_{XY} } \min_{\hat p_Y}
&J_{{\rm s}, \bm{t}_3 }^{(\lambda, \nu)} (q_{XY}, \hat p_{Y} |P_X)
=\min_{q_{XY} } J_{{\rm s}, \bm{t}_3 }^{(\lambda, \nu)} (q_{XY}, q_{Y} |P_X) \notag\\
&=\min_{q_{XY} } \Theta_{\rm s}^{(\lambda, \lambda \nu)}(q_{XY} |P_X)
= \Theta_{\rm s}^{(\lambda, \lambda \nu)}(P_X).
\end{align}
\end{lemma}
\begin{lemma}
\label{lemma20}
For any fixed $\lambda\in[0,1]$, $\nu\geq 0$, 
and any fixed $ \hat p_{Y}$, 
$J_{{\rm s}, \bm{t}_3 }^{(\lambda, \nu)}(q_{XY}, \hat p_{Y} |P_X)$ is minimized by 
\begin{align}
q_{Y|X}(y|x) 
& = \frac{ \hat p_Y(y) {\rm e}^{- \nu d(x,y) } } 
{ \sum_{y'} \hat p_Y(y') {\rm e}^{- \nu d(x,y') } } 
\label{eq.lemma20.2}\\
q_X (x) 
& = 
\frac{
P(x) \{ \sum_{y} \hat p_Y(y) {\rm e}^{-\nu d(x,y)} \}^\lambda
}{
\sum_{x'}
P(x') \{ \sum_{y} \hat p_Y(y) {\rm e}^{ - \nu d(x',y)} \}^\lambda
}.
\label{eq.lemma20.3}
\end{align}
Denote the joint distribution calculated from the above $q_{Y|X}$ and $q_X$ by
$q_{XY}^* ( \hat p_Y )$. 
Then, the minimum value of $J_{{\rm s}, \bm{t}_3 }^{(\lambda, \nu)}(q_{XY}, \hat p_{Y} |P_X)
 $ for fixed $\hat p_Y$ is 
\begin{align}
J_{{\rm s}, \bm{t}_3 }^{(\lambda, \nu)}(q_{XY}^*(\hat p_Y), \hat p_{Y} |P_X)
    &= -\log\sum_x P_X(x) \left\{
    \sum_y
    \hat p_Y(y) {\rm e}^{- \nu d(x, y) }
    \right\}^{\lambda}\notag \\
 &=-E_{0, \rm s}^{(-\lambda, \nu )} ( \hat p_Y | P_X ). 
    \label{eq.107}
\end{align}

This implies that 
\begin{align}
\min_{ \hat p_{Y} } 
\min_{q_{XY} }  
J_{{\rm s}, \bm{t}_3 }^{(\lambda, \nu)}(q_{XY}, \hat p_{Y} |P_X)
= \min_{ \hat p_{Y} } 
J_{{\rm s}, \bm{t}_3 }^{(\lambda, \nu)}(q_{XY}^*(\hat p_Y), \hat p_{Y} |P_X)
= 
\min_{ \hat p_{Y} } -E_{0, \rm s}^{(-\lambda, \nu )} ( \hat p_Y | P_X ).
\label{eq.139}
\end{align}
\end{lemma}
See Appendix~\ref{appendixB} for the proof.

An algorithm derived from Lemmas \ref{lemma19} and \ref{lemma20} is shown in Algorithm \ref{algorithm_GCK1}.

\begin{algorithm}
\caption{Algorithm for computing $\min_{q_{XY}} \min_{\hat p_Y} J_{{\rm s}, \bm{t}_3}^{(\lambda,\nu)}(q_{XY}, \hat p_Y| P_X)$ }
\label{algorithm_GCK1}
\begin{algorithmic}
    \Require 
    The probability distribution $P_X$ on $\mathcal{P(X)}$, the distortion measure $d(x,y)$, 
    $\lambda \in [0,1]$ and $\nu\geq 0$.
\State Choose any initial distribution $\hat p_Y^{[0]} $ such that $\hat p_Y^{[0]}(y)>0$ for all $y\in \mathcal{Y}$.
\For{$i=0,1,2,\ldots$} 
\begin{align}
    q_{Y|X}^{[i]}(y|x) 
& = \frac{ \hat p_Y^{[i]}(y) {\rm e}^{- \nu d(x,y) } } 
{ \sum_{y'} \hat p_Y^{[i]}(y') {\rm e}^{- \nu  d(x,y') } } ,
\label{algorithm_8_update_a}
\\
q_X^{[i]} (x) 
& = 
\frac{
P(x) \{ \sum_{y} \hat p_Y^{[i]}(y) {\rm e}^{-\nu  d(x,y)} \}^\lambda
}{
\sum_{x'}
P(x') \{ \sum_{y} \hat p_Y^{[i]}(y) {\rm e}^{ -\nu d(x',y)} \}^\lambda
},
\label{algorithm_8_update_b}
\\
\hat p_Y^{[i+1]}(y) 
&= \sum_{x} q_X^{[i]}(x) q_{Y|X}^{[i]}(y|x).
\label{algorithm_8_update_c}
\end{align}
\EndFor 
\end{algorithmic}
\end{algorithm}

We have the following proposition:
\begin{proposition}
\label{proposition_new_alg_source_B}
For $i=1,2,3,\ldots$, we have
\begin{align*}
 &  J_{{\rm s}, \bm{t}_3 }^{(\lambda, \nu)}(q_{XY}^{[0]}, \hat p_{Y}^{[0]}|P_X)
    \stackrel{\rm (a)} 
    \geq 
    J_{{\rm s}, \bm{t}_3 }^{(\lambda, \nu)}(q_{XY}^{[0]}, \hat p_{Y}^{[1]}|P_X)
    \stackrel{\rm (b)} 
    \geq 
    J_{{\rm s}, \bm{t}_3 }^{(\lambda, \nu)}(q_{XY}^{[1]}, \hat p_{Y}^{[1]}|P_X)
    \cdots\\
 &  \geq J_{{\rm s}, \bm{t}_3 }^{(\lambda, \nu)}(q_{XY}^{[i]}, \hat p_{Y}^{[i]}|P_X) 
 = -E_{0,\rm s}^{(-\lambda, \nu )}(\hat p_{Y}^{[i]}|P_X) \\
 &  \stackrel{\rm (a)} 
    \geq J_{{\rm s}, \bm{t}_3 }^{(\lambda, \nu)}(q_{XY}^{[i]}, \hat p_{Y}^{[i+1]}|P_X) 
    = 
    \Theta_{\rm s}^{(\lambda, \lambda \nu)}(q_{XY}^{[i]}|P_X)\\
 &  \stackrel{\rm (b)} 
    \geq 
    J_{{\rm s}, \bm{t}_3 }^{(\lambda, \nu)}(q_{XY}^{[i+1]}, \hat p_{Y}^{[i+1]}|P_X) 
    = -E_{0,\rm s}^{(-\lambda, \nu)}(\hat p_{Y}^{[i+1]}|P_X)\geq 
    \cdots\\
 & \geq 
 \min_{\hat p_Y} -E_{0,\rm s}^{(-\lambda, \nu)}(\hat p_{Y}|P_X)
 \stackrel{\rm (c)}
 =
 \min_{q_{XY} } \Theta_{\rm s}^{(\lambda, \lambda \nu)}(q_{XY}|P_X) = 
 \Theta_{\rm s}^{(\lambda, \lambda \nu)}(P_X) .
\end{align*}
\end{proposition}

\textit{Proof:} 
Step (a) follows from Lemma~\ref{lemma21} and
step (b) follows from Lemma~\ref{lemma22}. 
Step (c) follows from (\ref{eq.in.lemma21}) in Lemma~\ref{lemma21} and
(\ref{eq.in.lemma22}) in Lemma~\ref{lemma22}.
This completes the proof. \hfill$\IEEEQED$

This proposition shows that in Algorithm~\ref{algorithm_GCK1},
the objective function for Blahut's exponent with $\rho=-\lambda$ and
the objective function for Csisz\'ar and K\"orner's exponent appear alternately.

When $\lambda = 0$, Algorithm~\ref{algorithm_GCK1}
reduces to Blahut's algorithm for computing the rate distortion function~\cite{Blahut1972}.

Because the Algorithm~\ref{algorithm_GCK1} is a member of the family of the algorithms (Algorithm~\ref{alg:family_lossy_source_coding}),
the convergence of $q_{XY}^{[i]}$ defined by (\ref{algorithm_8_update_a})-(\ref{algorithm_8_update_c}) 
to the optimal distribution that minimizes $\Theta_{\rm s}^{(\lambda, \lambda \nu )} (q_{XY}|P_X)$
is guaranteed by Theorem~\ref{theorem.convergence.family.algorithm.source}.

We have the following proposition that can be used as a termination criterion by verifying that the $p_Y^{[i]}$ is sufficiently 
close to the optimal distribution. 
\begin{proposition}
A necessary and sufficient condition for $\hat p_Y^*$ to minimize 
$-E_{0, \rm s}^{(-\lambda, \nu)}(\hat p_Y | P_X)$ is that
\begin{align}
&    -\log \sum_x P_X(x) {\rm e}^{-\nu d(x,y)}
    \left\{
    \sum_{y'} \hat p_Y^*(y') {\rm e}^{-\nu d(x,y')}
    \right\}^{\lambda-1}\notag\\
&   \quad \begin{cases}
    = \Theta_{\rm s}^{(\lambda, \lambda \nu)}(P_X) \text{ for $y$ such that } \hat p_Y^*(y)>0,\\
    \geq \Theta_{\rm s}^{(\lambda, \lambda \nu)}(P_X) \text{ for $y$ such that } \hat p_Y^*(y)=0.\\
    \end{cases}
\end{align}
\end{proposition}
\textit{Proof:} This is exactly the KKT condition for $\hat p_Y^*$. 
Its derivation is straightforward and thus is omitted. 

As shown in Proposition~\ref{proposition_new_alg_source_B}, $\Theta_{\rm s}^{(\lambda, \lambda \nu )}(P_X)$ equals to the minimum value of $-E_{0,\rm s}^{(-\lambda, \nu )}(\hat p_Y|P_X)$ over $\hat p_Y$. 
Using the latter expression, we can show that Csisz\'ar and K\"orner's strong converse exponent is equal to (\ref{def:eq.GCK_check}) which is similar to Blahut's error exponent, except that it has a negative $\rho$. 
We have the following Theorem.
\begin{theorem}
\label{theorem5}
For any $R\geq 0$, and $\Delta\geq 0$, we have
\begin{align}
    G_{\rm CK}(R, \Delta|P_X)
=  G_{\rm JO} (R, \Delta|P_X).  
\end{align}
\end{theorem}
\textit{Proof:} 
We have the following equations. 
\begin{align}
& G_{\rm CK}(R, \Delta | P_X ) \notag\\
&\stackrel{\rm (a)}
=\tilde G_{\rm CK}(R, \Delta | P_X ) \notag \\
&
\stackrel{\rm (b)}
=
\max_{\lambda\in[0,1]} \sup_{\mu\geq 0} \{ 
\Theta_{\rm s}^{(\lambda, \mu)}( P_X ) 
-\lambda R -\mu \Delta\} \notag \\
&
\stackrel{\rm (c)}
= 
\sup_{ \lambda\in (0,1]} \sup_{\nu \geq 0} \{ 
\min_{ \hat p_Y } (-
E_{0, \rm s}^{(-\lambda, \nu )} ( \hat p_Y | P_X ) )
-\lambda R -\lambda \nu \Delta\}  \\ 
&
\stackrel{\rm (d)}
= 
\sup_{\rho\in [-1,0) } \sup_{\nu \geq 0} \min_{ \hat p_Y } 
\{ 
-E_{0, \rm s}^{(\rho, \nu )} ( \hat p_Y | P_X )
+ \rho  R +\rho \nu \Delta 
\} \notag\\
&=G_{\rm JO}(R, \Delta | P_X )
\label{eq.110}
\end{align}
Step (a) follows from Lemma~\ref{lemma:tilde_G_CK},
Step (b) follows from Lemma~\ref{lemma17},
Step (c) follows from the equation (c) in Proposition~\ref{proposition_new_alg_source_B} 
setting  $\mu = - \rho \nu $
and we put $\lambda=-\rho$ in Step (d). 
\hfill$\IEEEQED$

Theorem~\ref{theorem5} says that 
Csisz\'ar and K\"orner's strong converse exponent 
matches with 
$G_{\rm JO} (R, \Delta|P_X)$, which 
is written in a form similar to Blahut's error exponent. 

At the last of this subsection, we give a remark that 
the generalized cutoff rate is expressed by the following form.
\begin{property}
For a fixed $\lambda \in (0,1]$ and a fixed $\Delta\geq 0$, 
let $\nu^* = \nu^*(\lambda,\Delta)$ be an optimal $\nu$ that attains the rhs of (\ref{eq.46a})
and let $\hat p_{Y}^* = \hat p_{Y}^*(\lambda, \Delta)$ be an optimal distribution that attains 
$\min_{\hat p_Y} - E_{0,s}^{ (\lambda, \nu^*(\lambda, \Delta) )}(\hat p_Y|P_X)$. Then we have 
\begin{align}
    R^{(-\lambda)}(\Delta|P_X) 
    = \frac{-1}{\lambda} \log\sum_x P_X(x)
    \left\{ \sum_{y} \hat p_Y^*(y) 
    {\rm e}^{- \nu^* [d(x,y)-\Delta] }
    \right\}^\lambda .
  \label{eq.65}
  \end{align}
\end{property}

\textit{Proof:}
We have the following chain of equalities:
\begin{align}
    R^{(-\lambda)}(\Delta|P_X) 
    & 
    \stackrel{\rm (a)}
    = \frac1{\lambda } G_{\rm CK}^{(\lambda)} (\Delta |P_X) \notag\\
    & 
    \stackrel{\rm (b)}
    = \frac1{\lambda} \left\{
    \Theta_{\rm s}^{(\lambda, \lambda \nu^* )} (P_X) - \lambda \nu^* \Delta
    \right\} \notag\\
    & 
    \stackrel{\rm (c)}
    = \frac1{\lambda} \left\{
    - E_{0, \rm s}^{(-\lambda, \nu^* )} ( \hat p_{Y}^* |P_X) - \lambda \nu^* \Delta
    \right\} \notag\\
    & 
    \stackrel{\rm (d)}
    = \frac{-1}{\lambda} \log\sum_x P_X(x)
    \left\{ \sum_{y} \hat p_Y^*(y) 
    {\rm e}^{- \nu^* [d(x,y)-\Delta] }
    \right\}^\lambda 
\end{align}
Step (a) follows from Lemma~\ref{lemma17},
Step (b) follows from (\ref{eq.46a}) and the assumption of $\nu^*$,
Step (c) follows from the assumption of $\hat p_Y^*$, and
Step (d) follows from the definition of $E_{0,\rm s}^{(\rho, \nu)}(\hat p_Y|P_X)$.
This completes the proof. \hfill$\IEEEQED$

\subsection{Algorithm~\ref{algorithm_GCK2}}
\label{section:algorithm6}
In this subsection, we discuss another special case of the algorithm family, which
uses $J_{{\rm s}, \bm{t}_4 }^{(\lambda, \nu)}(q_{XY}, p_{Y|X} |P_X)$ as a surrogate objective function.
The following function that appears in Arimoto's algorithm will plays an important role in this special case. 
\begin{definition}
For a given source distribution $P_X$, transition probability
of a test channel $\hat{p}_{Y|X}$, and parameters
$\rho$ and $\nu\geq 0$, define 
\begin{align}
    A_{\rm s}^{(\rho, \nu)}(p_{Y|X} |P_X)
    =
    (1+\rho)\log\sum_{y}
\left[ 
 \sum_{x}
 P_X(x) {\rm e}^{\rho \nu d(x,y)}
 p_{Y|X}^{1+\rho}(y|x) 
\right]^{1/(1+\rho)}. 
\label{definition_As}
\end{align}
\end{definition}

\begin{lemma}
\label{lemma21}
For any fixed $\lambda \in [0,1]$ and $\nu \geq 0$,
$ J_{{\rm s}, \bm{t}_4 }^{(\lambda, \nu)} (q_{XY}, p_{Y|X} | P_X) $
is minimized if and only if $ p_{Y|X} = q_{Y|X}$ and 
the minimum value is
\begin{align}
    J_{{\rm s}, \bm{t}_4 }^{(\lambda, \nu)} (q_{XY}, q_{Y|X} | P_X)
    =
    \Theta_{\rm s}^{(\lambda, \lambda \nu )} (q_{XY}|P_X). 
\end{align}
This implies 
\begin{align}
    &\min_{q_{XY}} \min_{p_{Y|X}}
    J_{{\rm s}, \bm{t}_4 }^{(\lambda, \nu)} (q_{XY}, p_{Y|X} | P_X)
    =
    \min_{q_{XY}} 
    J_{{\rm s}, \bm{t}_4 }^{(\lambda, \nu)} (q_{XY}, q_{Y|X} | P_X) \notag\\
    &=
    \min_{q_{XY}} 
    \Theta_{\rm s}^{(\lambda, \lambda \nu)} (q_{XY}|P_X) =\Theta_{\rm s}^{(\lambda, \lambda \nu )}(P_X).
    \label{eq.in.lemma21}
\end{align}
\end{lemma}

\textit{Proof:} It is obvious from the nonnegativity of the relative entropy
$ D(q_{Y|X}|| p_{Y|X} | q_X)$. 

\begin{lemma}
\label{lemma22}
For any fixed $\lambda \in [0,1)$ and $\nu \geq 0$, 
$ J_{{\rm s}, \bm{t}_3 }^{(\lambda, \nu)} (q_{XY}, p_{Y|X} | P_X) $
is minimized by 
\begin{align}
    q_{X|Y}(x|y) &= \frac{ P_X(x) {\rm e}^{-\lambda \nu d(x,y)} p_{Y|X}^{1-\lambda}(y|x) } 
    { \sum_{x'} P_X(x') {\rm e}^{-\lambda \nu d(x',y)} p_{Y|X}^{1-\lambda}(y|x') } ,
    \label{eq.lemma22.2}
    \\
    q_Y(y) &= \frac{ \left[ \sum_{x} P_X(x) {\rm e}^{-\lambda\nu d(x,y)}  p_{Y|X}^{1-\lambda}(y|x) \right]^{1/(1-\lambda) } }{
    \sum_{y'}
    \left[ \sum_{x} P_X(x) {\rm e}^{-\lambda\nu d(x,y')} p_{Y|X}^{1-\lambda}(y'|x) \right]^{1/(1-\lambda) }
    }. \label{eq.lemma22.3}
\end{align}
Denote the joint distribution computed from the above $q_{X|Y}$ and $q_Y$ by $q_{XY}^* (p_X)$.
The minimum value of $ J_{{\rm s}, \bm{t}_3 }^{(\lambda, \nu)} (q_{XY}, p_{Y|X} | P_X)$ is 
\begin{align}
J_{{\rm s}, \bm{t}_4 }^{(\lambda, \nu)} ( q_{XY}^* (p_X), p_{Y|X} | P_X)
&=
-(1-\lambda) \log
\sum_{y \in \mathcal{Y}}
\left[ 
\sum_{x \in \mathcal{X} } 
P_X(x) {\rm e}^{-\lambda\nu d(x,y) }
p_{Y|X}^{1-\lambda}(y|x) 
\right] ^{1/(1-\lambda) }\notag\\
&=-A_{\rm s}^{(-\lambda, \nu )} ( p_{Y|X} | P_X). 
\label{eq.116}
\end{align}
This implies that
\begin{align}
    \min_{q_{XY}} \min_{ p_{Y|X} } J_{{\rm s}, \bm{t}_4 }^{(\lambda, \nu)} ( q_{XY}, p_{Y|X} | P_X)
= \min_{ p_{Y|X} } J_{{\rm s}, \bm{t}_4 }^{(\lambda, \nu)} ( q_{XY}^* (p_X), p_{Y|X} | P_X)
= \min_{ p_{Y|X} } -A_{\rm s}^{(-\lambda, \nu )} ( p_{Y|X} | P_X).
\label{eq.in.lemma22}
\end{align}
\end{lemma}

See Appendix~\ref{appendixB} for the proof. 

Note that the rhs of Eq.(\ref{eq.116}) has the same form of the rhs of Eq.~(\ref{definition_As}). 
We have the following equation. 
\begin{align}
    & \min_{q_{XY}} \Theta^{(\lambda, \lambda \nu)}( q_{XY}|P_X) \notag\\ 
    & =
    \min_{q_{XY}} \min_{p_{Y|X} } J_{{\rm s}, \bm{t}_4 }^{(\lambda, \nu)}(q_{XY}, p_{Y|X} |P_X)
    \notag\\
&=    \min_{p_{Y|X} } \{ 
- A_{\rm s}^{(-\lambda, \nu )}(p_{Y|X} |P_X) \}
\label{eq.118b}
\end{align}

For a fixed $\nu\geq 0$, we can show that the following proposition holds. 
\begin{proposition}
\label{limit_of_A_s}
We have
\begin{align}
    \lim_{\rho\to 0} \frac1{\rho} 
    A_{\rm s}^{(\rho, \nu)}( p_{Y|X} | P_X)
    =
    I(P_X, p_{Y|X}) + \nu 
    \mathrm{E}_{ (P_X, p_{Y|X} )} [ d(X,Y) ]. 
\end{align}
\end{proposition}

See Appendix~\ref{appendixH} for the proof. 
Thus, the function $(1/\rho) A_{\rm s}^{(\rho, \nu)}(p_{Y|X}|P_X)$, which we call an alternative form, has a natural connection to information theoretic quantities. 

An algorithm derived from Lemmas~\ref{lemma21} and~\ref{lemma22} is shown in Algorithm~\ref{algorithm_GCK2}.
Algorithm~\ref{algorithm_GCK2} is a member of the algorithm family 
and therefore its convergence to the optimal distribution is guaranteed by
Theorem\ref{theorem.convergence.family.algorithm.source}.

\begin{algorithm}[ht]
\begin{algorithmic}
\caption{Algorithm for computing $\min_{q_{XY}} \min_{q_{Y|X} } 
J_{{\rm s}, \bm{t}_4 }^{(\lambda, \nu)}(q_{XY}, p_{Y|X} |P_X)$ }
\label{algorithm_GCK2}
    \Require 
    The probability distribution $P_X$ on $\mathcal{P(X)}$, the distortion measure $d(x,y)$, 
    $\lambda \in (0,1)$ and $\nu\geq 0$.
    \State 
    Choose any initial distribution $ p_{Y|X}^{[0]} $ such 
    that $ p_{Y|X}^{[0]}(y|x)>0$ for all $x\in\mathcal{X}, y\in \mathcal{Y}$.
    \For{$i=0,1,2,$}  
\begin{align} 
    q_{X|Y}^{[i]}(x|y) &= \frac{ P_X(x) {\rm e}^{- \lambda \nu d(x,y)} p_{Y|X}^{[i]}(y|x)^{1-\lambda} } 
    { \sum_{x'} P_X(x') {\rm e}^{-\lambda\nu d(x',y)} p_{Y|X}^{[i]}(y|x')^{1-\lambda} }, \label{algorithm_7_update_a}\\ 
    q_Y^{[i]}(y) &= \frac{ \left[ \sum_{x} P_X(x) {\rm e}^{-\lambda\nu d(x,y)}  p_{Y|X}^{[i]}(y|x)^{1-\lambda} \right]^{1/(1-\lambda) } }{
    \sum_{y'}
    \left[ \sum_{x} P_X(x) {\rm e}^{-\lambda\nu d(x,y')} p_{Y|X}^{[i]}(y'|x)^{1-\lambda} \right]^{1/(1-\lambda) }
    }, \label{algorithm_7_update_b}\\
p_{Y|X}^{[i+1]}(y|x) 
&= \frac{ q_Y^{[i]}(y) q_{X|Y}^{[i]}(x|y) }
{ \sum_{y'} q_Y^{[i]}(y') q_{X|Y}^{[i]}(x|y') }.  \label{algorithm_7_update_c}
\end{align}
\EndFor 
\end{algorithmic}
\end{algorithm}

The following proposition illustrates an important property of Algorithm~\ref{proposition_new_alg_source_A},
that $-A_{\rm s}^{(-\lambda, \nu)}(q_{Y|X}|P_X)$ and the objective function of Csisz\'ar and K\"orner's exponent appear alternately.
\begin{proposition}
\label{proposition_new_alg_source_A}
For $i=1,2,3,\ldots$, we have
\begin{align*}
 &  J_{{\rm s}, \bm{t}_4 }^{(\lambda, \nu)}(q_{XY}^{[0]}, p_{Y|X}^{[0]}|P_X)
    \stackrel{\rm (a)} 
    \geq 
    J_{{\rm s}, \bm{t}_4 }^{(\lambda, \nu)}(q_{XY}^{[0]}, p_{Y|X}^{[1]}|P_X)
    \stackrel{\rm (b)} 
    \geq 
    J_{{\rm s}, \bm{t}_4 }^{(\lambda, \nu)}(q_{XY}^{[1]}, p_{Y|X}^{[1]}|P_X)
    \cdots\\
 &  \geq J_{{\rm s}, \bm{t}_4 }^{(\lambda, \nu)}(q_{XY}^{[i]}, p_{Y|X}^{[i]}|P_X) 
 = -A_{\rm s}^{(-\lambda, \nu)}(p_{Y|X}^{[i]}|P_X) \\
 &  \stackrel{\rm (a)} 
    \geq J_{{\rm s}, \bm{t}_4 }^{(\lambda, \nu)}(q_{XY}^{[i]}, p_{Y|X}^{[i+1]}|P_X) = \Theta_{\rm s}^{(\lambda, \lambda \nu)}(q_{XY}^{[i]}|P_X)\\
 &  \stackrel{\rm (b)} 
    \geq J_{{\rm s}, \bm{t}_4 }^{(\lambda, \nu)}(q_{XY}^{[i+1]}, p_{Y|X}^{[i+1]}|P_X) = -A_{\rm s}^{(-\lambda, \nu)}(p_{Y|X}^{[i+1]}|P_X)\geq 
    \cdots\\
 & \geq 
 \min_{q_{XY} } \Theta^{(\lambda, \lambda \nu)}(q_{XY}|P_X)
 \stackrel{\rm (c)}
 =
 \min_{p_X} -A_{\rm s}^{(-\lambda, \nu )}(p_{Y|X}|P_X).
\end{align*}
\end{proposition}

\textit{Proof:} Step (a) follows from Lemma~\ref{lemma21} and
step (b) follows from Lemma~\ref{lemma22}. 
Step (c) follows from (\ref{eq.in.lemma21}) in Lemma~\ref{lemma21} and
(\ref{eq.in.lemma22}) in Lemma~\ref{lemma22}.
This completes the proof. \hfill$\IEEEQED$

\subsection{Comparison with our previous algorithm}
\label{section:comparison_lossy_source_coding}
The first algorithm for computing the strong converse exponent for the lossy source coding
was given by the authors in~\cite{YutakaISIT2016b}. 
This algorithm is also a member of the algorithm family (Algorithm~\ref{alg:family_lossy_source_coding}).
In this subsection, we describe this special case.
Proofs for the theorems and lemmas omitted in \cite{YutakaISIT2016b} are provided in Appendix of this paper. 

In~\cite{YutakaISIT2016b}, we defined the following function. 
\begin{align}
    F_{\rm JO, s}^{(\lambda, \nu)}(q_{XY}, p_{XY} |P_X)
    &= 
    \Theta_{\rm s}^{(\lambda, \lambda\nu)} ( q_{XY} | P_X ) 
    +D( q_{XY}|| p_{XY} ) \notag\\
    &=
    {\rm E}_{X_{XY}} \left[
    \log 
    \frac{q_{XY}(X,Y)}
    {P_X(X) {\rm e}^{-\lambda \nu d(X,Y)} p_Y^{\lambda}(Y) p_{Y|X}^{1-\lambda} (Y|X) }
    \right]    
\end{align}
The following equation holds for this function.
\begin{align}
    F_{\rm JO, s}^{(\lambda, \nu)}(q_{XY}, p_{XY} |P_X)
    =  \Theta_{\rm s}^{(\lambda, \lambda \nu)}(q_{XY} |P_X)
    + (1-\lambda) D( q_{Y|X} || p_{Y|X} | q_X )
    + \lambda D( q_Y || p_Y)
    \label{Eq:J_s}
\end{align}
Therefore, $F_{\rm JO, s}^{(\lambda, \nu)}(q_{XY}, p_{XY} |P_X)$ corresponds to 
the case $\bm{t} = \bm{t}_3+\bm{t}_4 = (0, 1, \lambda/(1-\lambda), 0)$ of 
$J_{{\rm s}, \bm{t}}^{(\lambda, \nu)} (q_{XY}, p_{XY}|P_X)$. 

We have the following Lemmas. 
\begin{lemma}
\label{lemma_GCK_minimization1}
For any fixed $\lambda\in [0,1]$, $\nu\geq 0$, 
and any fixed $P_X\in \mathcal{P(X)}$, $q_{XY} \in \mathcal{P(X\times Y)}$, 
$F_{\rm JO, s}^{(\lambda, \nu)}(q_{XY}, p_{XY} |P_X)$ is minimized
if and only if $\hat q_{XY} = q_{XY}$ and its minimum value is
\begin{align}
    F_{\rm JO, s}^{(\lambda, \nu)}( q_{XY}, q_{XY} |P_X)
    = \Theta_{\rm s}^{(\lambda, \lambda \nu)} (q_{XY} | P_X ).
\end{align}
This implies that 
\begin{align*}
    \min_{ q_{XY} } \min_{ p_{XY}} 
    F_{\rm JO, s}^{(\lambda, \nu)}( q_{XY}, p_{XY} |P_X)
    &=
    \min_{ q_{XY} } 
    F_{\rm JO, s}^{(\lambda, \nu)}( q_{XY}, q_{XY} |P_X)\\
    &=
    \min_{ q_{XY} } 
    \Theta_{\rm s}^{(\lambda, \lambda \nu)} (q_{XY} | P_X ).
\end{align*}
\end{lemma}

\begin{lemma}
\label{lemma_GCK_minimization2}
For any fixed $\lambda\in [0,1]$, $\mu\geq 0$, $P_X\in \mathcal{P(X)}$, 
and any fixed $\hat q_{XY} \in \mathcal{P(X\times Y)}$,
$ F_{\rm JO, s}^{(\lambda, \nu)}(q_{XY}, p_{XY} |P_X)$ is minimized 
if and only if 
\begin{align*}
q_{XY}(x,y) &= 
\frac{
    P_X(x) {\rm e}^{-\lambda \nu d(x,y) }  p_Y^{\lambda}(y) p_{Y|X}^{1-\lambda}(y|x) 
}
{
    \sum_{x', y'}
    P_X(x') {\rm e}^{-\lambda \nu d(x',y') } p_Y^{\lambda}(y') p_{Y|X}^{1-\lambda}(y'|x') 
}\\
&\stackrel{\triangle}=
\hat q_{XY}(p_{XY}) (x,y)
\end{align*}
and its minimum value is
\begin{align}
    &  
    F_{\rm JO, s}^{(\lambda, \nu)}(\hat q_{XY}(p_{XY}), p_{XY} |P_X) \notag\\
    &= -\log \sum_{x,y} 
    P_X(x) {\rm e}^{-\lambda \nu d(x,y)} 
    p_{Y}^\lambda (y)
    p_{Y|X}^{1-\lambda}(y|x)
    . 
\end{align}
This implies that
\begin{align*}
    &\min_{q_{XY} } \min_{p_{XY} }
    F_{\rm JO, s}^{(\lambda, \nu)}( q_{XY}, p_{XY} |P_X)\\
    &= \min_{ p_{XY} } F_{\rm JO, s}^{(\lambda, \nu)}(\hat q_{XY}(p_{XY}), p_{XY} |P_X)\\
    &= \min_{ p_{XY} }
    \left(
    -\log \sum_{x,y} 
    P_X(x) {\rm e}^{-\lambda \nu d(x,y)}
    p_{Y}^\lambda (y)
    p_{Y|X}^{1-\lambda}(y|x)
    \right).
\end{align*}

\end{lemma}
See Appendix~\ref{appendixE} for the proof.

The algorithm based on Lemmas~\ref{lemma_GCK_minimization1} 
and~\ref{lemma_GCK_minimization2} is shown in Algorithm~\ref{algorithm_GCK}.

\begin{algorithm}
\caption{Algorithm for computing $\min_{q_{XY}} \Theta_{\rm s}^{(\lambda, \lambda \nu)} ( q_{XY} | P_X )$ \cite{YutakaISIT2016b} }
\label{algorithm_GCK}
\begin{algorithmic}
\Require The probability distribution $P_X$ of the information source,
 $\lambda\in (0,1)$, and $\nu\geq 0$. 
 The distortion measure $d(x,y)$ is also given. 
 \State 
 Choose any initial joint distribution $q_{XY}^{[0]}$ such that 
every element is strictly positive.
\For{$i=0,1,2, \ldots$} 
\begin{align}
q_{XY}^{[i+1]}(x,y) = 
\frac{
P_X(x) {\rm e}^{-\lambda \nu d(x,y) } 
q_Y^{[i]}(y)^{\lambda} q_{Y|X}^{[i]}(y|x)^{1-\lambda} 
}
{\displaystyle 
\sum_{x'\in \mathcal{X}} \sum_{y' \in \mathcal{Y} }
P_X(x') {\rm e}^{-\lambda \nu d(x',y') } 
q_Y^{[i]}(y')^{\lambda} 
q_{Y|X}^{[i]}(y'|x')^{1-\lambda} 
}.
\end{align}
\EndFor 
\end{algorithmic}
\end{algorithm}
Algorithm~\ref{algorithm_GCK} for $G_{\rm CK}(R,\Delta|P_X)$ is analogous to Algorithm~\ref{alg:previous_work} for $G_{\rm DK}(R,\Gamma|W)$.
As is discussed below, this observation motivated the development of Algorithms~\ref{algorithm_GCK1} and \ref{algorithm_GCK2}, which are similar to Algorithms~\ref{new_algorithm_A} and \ref{new_algorithm_B}.

Similar to Eq.(\ref{Ftilde}) in Section \ref{section:our_previous_algorithm}, 
we replace $p_Y$ in the right hand side of (\ref{Eq:J_s}) with $\hat p_Y$ to define 
\begin{align}
    \tilde F_{\rm JO, s}^{(\lambda, \nu)}(q_{XY}, \hat p_Y, p_{Y|X} |P_X)
    &=  \Theta_{\rm s}^{(\lambda, \lambda \nu)}(q_{XY} |P_X)
    + \lambda D( q_Y || \hat p_Y)
    + (1-\lambda) D( q_{Y|X} || p_{Y|X} | q_X ) \notag \\
    &=
    {\rm E}_{X_{XY}} \left[
    \log 
    \frac{q_{XY}(X,Y)}
    {P_X(X) {\rm e}^{-\lambda \nu d(X,Y)} \hat p_Y^{\lambda}(Y) p_{Y|X}^{1-\lambda} (Y|X) }
    \right]    .
    \label{eq.104}
\end{align}
Here, we use the following function introduced by 
Arimoto~\cite{Arimoto1976} to derive his algorithm.
\begin{definition}
For a given source probability distribution $P_X$,
an output distribution of the test channel $p_Y$, 
and parameters $\rho$ and $t$, define\footnote{
More precisely, Arimoto defined (\ref{definition_F_ARs})
multiplied $-1$ as the objective function and
its maximization was discussed. 
} 
\begin{align}
    F_{\rm AR, s}^{(\rho, \nu)} (\hat p_Y, p_{Y|X} |P_X) 
    =
    \frac{1}{\rho} \log 
    \sum_{x} \sum_{y} P_X(x) 
    {\rm e}^{\rho \nu d(x,y)} 
    \hat p_Y^{-\rho}(y) 
    p_{Y|X}^{1+\rho}(y|x). 
    \label{definition_F_ARs}
\end{align}
\end{definition}
We observe the similarity in (\ref{F_rho}) and (\ref{definition_F_ARs}).

Then, we have the following lemma.
\begin{lemma} 
For a fixed $\lambda\in[0,1]$ and $q_{XY}$, we have 
\begin{align}
    \min_{\hat p_Y} \min_{p_{X|Y}} 
    \tilde F_{\rm s}^{(\lambda, \nu)}
    (q_{XY}, \hat p_Y, p_{X|Y} |P_X) 
    = 
    \Theta^{(\lambda, \lambda \nu)}_{\rm s}
    (q_{XY}|P_X). 
\end{align}
\end{lemma} 
\textit{Proof:} It is obvious from the definition of $    
\tilde F_{\rm s}^{(\lambda, \nu)}
(q_{XY}, \hat p_Y, p_{X|Y} |P_X) 
$ and non-negativity of the relative entropy. 

\begin{lemma}
\label{lemma:F_s_tilde}
For a fixed $\lambda\in [0,1]$, $\nu\geq 0$, $\hat p_Y$, and $p_{Y|X}$, we have 
\begin{align}
    \min_{q_{XY} }  
    \tilde F_{\rm JO, s}^{(\lambda, \nu)}
    (q_{XY}, \hat p_Y, p_{X|Y} |P_X) 
    &= -\log \sum_{x,y} P_X(x) {\rm e}^{-\lambda \nu d(x,y)} \hat p_Y^{\lambda}(y) p_{Y|X}^{1-\lambda}(y|x)
    \notag\\
    &= \lambda 
    F_{\rm AR, s}^{(-\lambda, \nu)}
    (\hat p_Y, p_{Y|X}|P_X). \label{eq.127}
\end{align}
\end{lemma} 

See Appendix~\ref{appendixG} for the proof.

The above-mentioned relationship is illustrated in Fig.~\ref{fig:st_conv_source_codingL}.
The triple minimization $\min_{q_{XY}} \min_{\hat p_Y} \min_{p_{X|Y}} 
\tilde F_{\rm JO, s}(q_{XY}, \hat p_Y, p_{X|Y})$ is shown in the top level and 
is connected to three double minimization, as described in the following proposition.

\begin{proposition}
For any fixed $\lambda \in [0,1]$, $\nu\geq 0$, and $P_X \in \mathcal{P(X)}$,
we have
\begin{align}
    &\min_{q_{XY}} \min_{\hat p_Y} \min_{p_{X|Y}}
    \tilde F_{\rm JO, s}^{(\lambda, \nu)}
    (q_{XY}, \hat p_Y, p_{X|Y} |P_X) \notag\\ 
    &= 
    \min_{q_{XY}} \min_{\hat p_Y} 
    J_{{\rm s}, \bm{t}_3 }(q_{XY}, \hat p_Y |P_X) \label{eq.128}\\
    &= 
    \min_{q_{XY}} \min_{p_{X|Y}}
    J_{{\rm s}, \bm{t}_4 }(q_{XY}, p_{Y|X} |P_X) \label{eq.129}\\
    &=
    \min_{\hat p_Y} \min_{p_{X|Y}}
     \lambda 
    F_{\rm AR, s}^{(-\lambda, \nu)}
    (\hat p_Y, p_{Y|X}|P_X). \label{eq.130}    
\end{align}
\end{proposition}
We have derived Algorithms~\ref{algorithm_GCK1} and~\ref{algorithm_GCK2} from
the double minimization of (\ref{eq.128}) and (\ref{eq.129}).
The property of the double minimization of (\ref{eq.130}) will be discussed 
in Section~\ref{section_sourcecoding_Arimotoalgorithm}.
We will show that the algorithm derived from (\ref{eq.130}) is the same as Arimoto algorithm except that $\rho$ is negative. We then show that Arimoto algorithm works correctly also for negative $\rho$.

\section{Computation of the error exponents in lossy source coding}
In this section, Arimoto's algorithm for computing the error exponents in lossy source coding is discussed and is compared with Algorithms \ref{algorithm_GCK1}, \ref{algorithm_GCK2}, and \ref{algorithm_GCK}. 
As shown in Section~\ref{section:algorithm5}, Csisz\'ar and K\"orner's strong converse is equal to a function similar to Blahut's error exponent, except that the slope parameter $\rho$ takes negative values. Hence, it is natural to expect that the Arimoto algorithm can be used to compute the strong converse exponent in lossy source coding. 
To the best our knowledge, this issue has not been published.   
In Section~\ref{section_sourcecoding_Arimotoalgorithm}, we have 
proved that Csisz\'ar and K\"orner's strong converse exponent
matches with a Blahut-style exponent function with parameter $\rho \in [-1,0]$. 
Thus, the answer to the natural question above is yes.
Note that the strong converse exponent of lossy source coding~\cite{Csiszar-KornerBook} was established after the publication of~\cite{Arimoto1976}, and Arimoto was unable to mention the computation of this exponent.

%
%
%

We will also show that Arimoto's algorithm can be used to compute Ar{\i}kan and Merhav's guessing exponent~\cite{ArikanMerhav1998}.

\subsection{Arimoto algorithm for lossy source coding exponents}
\label{section_sourcecoding_Arimotoalgorithm}

The target of Arimoto's algorithm for lossy source coding
is Blahut's error exponent for lossy source coding, defined by 
\begin{align}
    E_{\rm B}(R, \Delta | P_X)
    =
    \sup_{\rho\geq 0}
    \inf_{\nu \geq 0}
    \left[ 
    \rho R + \rho \nu \Delta - \min_{\hat p_Y} E_{0,\rm s}^{(\rho, \nu)} (\hat p_Y | P_X)
    \right].
\end{align}
The exponent functions satisfy the following property. 
\begin{property}[Theorem~21 and Theorem~22 in \cite{Blahut1974}]
\begin{enumerate}[(a)]
    \item For a fixed $\Delta\geq 0$, $ E_{\rm B}(R, \Delta|P_X)$ is 
    monotone non-decreasing function of $R$.
    For a fixed $R\geq 0$, $ E_{\rm B}(R, \Delta|P_X)$ is 
    monotone non-decreasing $\Delta$. 
    \item $ E_{\rm B}(R, \Delta|P_X)$ is convex in $(R,\Delta)$.
    \item For $R \leq R(\Delta|P_X)$, $E_{\rm B}(R|P_X) = 0$ 
    and for $R > R(\Delta|P_X)$, $E_{\rm B}(R|P_X)$  is strictly positive.
\end{enumerate}
\end{property}

Graphing of the exponent function $E_{\rm B}(R, \Delta | P_X)$ is 
similar to that of $G_{\rm CK}(R, \Delta |P_X) $. 
For the convenience, let us denote the minimum value of $ E_{0,\rm s}^{(\rho, \nu)} (\hat p_Y | P_X) $ by 
\begin{align}
    E_{0,\rm s}^{(\rho, \nu)} (P_X)
    = 
    \min_{\hat p_Y} E_{0,\rm s}^{(\rho, \nu)} (\hat p_Y | P_X). 
    \label{min_E_B}
\end{align}

We define 
\begin{align}
    E_{\rm B}^{(\rho)} (\Delta |P_X)
    =
    \rho \sup_{\nu\geq 0} 
    \left[ 
    \nu \Delta 
    - \frac{1}{\rho} E_{0,\rm s}^{(\rho, \nu)}(P_X)
    \right], \quad \rho> 0.
    \label{L_s}
\end{align}
For $\rho = 0$, $(1/\rho) E_{0,\rm s}^{(\rho,\nu)}(\hat p_Y|P_X)$
is interpreted as 
\begin{align}
    \lim_{\rho \to 0} \frac1{\rho} E_{ 0, s }^{(\rho,\nu)} (\hat p_Y|P_X)
    = -\sum_{x} P_X(x) \log \sum_{y} \hat p_Y(y) {\rm e}^{-\nu d(x,y)}.
    \label{d-tilded_information_0}
\end{align}
Then, we have
\begin{align}
    E_{\rm B}(R,\Delta|P_X)
    = \sup_{\rho \geq 0} \left[
    \rho R + E_{\rm B}^{(\rho)} (\Delta |P_X) 
    \right].
    \label{E_B_Lagrange}
\end{align}
Eq.(\ref{E_B_Lagrange}) shows that $E_{\rm B}(R,\Delta |P_X)$ is obtained by the Legendre-Fenchel transformation (LFT) of $-E_{\rm B}^{(\rho)}(\Delta | P_X)$ as a function of $\rho\geq 0$. Moreover, Eq.(\ref{L_s}) shows that for a fixed $\rho\geq 0$, 
$E_{\rm B}^{(\rho)}(\Delta |P_X)$  is obtained by the the LFT of $(1/\rho) E_{0,\rm s}^{(\rho, \nu)}(P_X) $ as a function of $\nu\geq 0$, multiplied by $\rho$.  
In Fig.~\ref{fig:strong_converse_exponent_source}, 
$R^{(\rho)}(\Delta|P_X)$ is the source-coding generalized cutoff rate 
defined as the $R$-axis intercept of the supporting line of slope $\rho>0$
to the curve $E_{\rm B}(R,\Delta |P_X)$. 

Now we describe the Arimoto's algorithm for computing (\ref{min_E_B}). 
The following theorem in the case of $\rho>0$
was suggested in~\cite{Arimoto1976}, 
although the details of the proof were omitted. 
We give a complete proof for $\rho>0$ as well as $\rho\in [-1,0)$.
The theorem for the case of negative $\rho$ suggests 
that Arimoto algorithm successfully works for computing
the exponent function defined in Definition~\ref{def:GCK_check}.

\begin{theorem}[For positive $\rho$ in \cite{Arimoto1976}]
\label{unwritten_theorem:arimoto}
For any fixed $\rho\in[-1,0)\cup (0,\infty)$, $\nu \geq 0$, 
and any fixed $p_Y$, we have 
\begin{align}
    \min_{ p_{Y|X} } 
    F_{\rm AR, s}^{(\rho, \nu)} (\hat p_Y, p_{Y|X} |P_X) 
&=   \frac{1}{\rho}\log \sum_{x} P_X(x)
  \left\{
  \sum_{y}
  \hat p_Y(y){\rm e}^{- \nu d(x,y)}
  \right\}^{-\rho }\notag 
\\
&= 
\frac1{\rho} E_{0, \textrm{s} }^{(\rho, \nu)} (\hat p_Y | P_X) .
\label{theorem4.eq.1}
\end{align}
The maximum value is achieved if and only if 
\begin{align}
p_{Y|X}(y|x) 
=
\frac{ \hat p_Y(y) {\rm e}^{-\nu d(x,y) } } { \sum_{y'} \hat p_Y(y') {\rm e}^{- \nu d(x,y') } } 
    = 
    p_{Y|X}^* ( \hat p_Y )(y|x).
\end{align}
On the other hand, for any fixed $\rho\in[-1,0)\cup (0,\infty)$, $\nu\geq 0$,
and any fixed $p_{Y|X}$, we have
\begin{align}
   \min_{ \hat p_{Y} } F_{\rm AR}^{(\rho, \nu)} (\hat p_Y, p_{Y|X} |P_X) 
= 
    \frac1{\rho} A_{\rm s}^{(\rho, \nu)}(p_{Y|X} |P_X), 
\label{theorem4.eq.3}
\end{align}
where $A_{\rm s}^{(\rho, \nu)}(p_{Y|X} |P_X)$ is the function
defined in (\ref{definition_As}).
The maximum value is achieved if and only if 
\begin{align}
    \hat p_Y(y)=
    \frac{  \left\{ \displaystyle \sum_{x} P_X(x) {\rm e}^{\rho \nu d(x,y)}  p_{Y|X}^{1+\rho} (y|x)  \right\}^{ 1/(1+\rho) } }
    { \displaystyle \sum_{y'} \left\{ \sum_{x} P_X(x) {\rm e}^{\rho \nu d(x,y')}  p_{Y|X}^{1+\rho} (y'|x)  \right\}^{ 1/(1+\rho)} }
    = 
    \hat p_Y^*( p_{Y|X} )(y)
    . 
\end{align}
\end{theorem}

The proof was omitted in~\cite{Arimoto1976} because it is analogous to the proof for the similar theorem for channel coding.
Since we have extended the range of $\rho$ from $[0,\infty)$ to $[-1,\infty)$, we give the proof of Theorem~\ref{unwritten_theorem:arimoto} to make the paper self-contained. 
To this aim, we give the following lemma.
\begin{lemma} \label{lemma14}
  We have
  \begin{align}
      F_{\rm AR,s}^{(\rho)}(\hat p_Y, p_{Y|X} | P_X)
      &= \frac{1}{\rho} E_{0,\rm s}^{(\rho, \nu)} ( \hat p_Y | P_X) + D_{1+\rho} ( p_{Y|X} || p_{Y|X}^*(\hat p_Y) | p_X^*(\hat p_Y) ) 
      \label{eq.97}\\
      &= \frac{1}{\rho} A_{s}^{(\rho, \nu)} ( p_{Y|X} | P_X ) + D_{1+\rho} ( \hat p_{Y}^* ( p_{X|Y})  || \hat p_{Y})
      \label{eq.98},
  \end{align}
\end{lemma}
where $p_X^*(\hat p_Y)$ is defined by
\begin{align}
    p_X^*(\hat p_Y) (x) = 
    \frac{ P_X(x) \left\{ \sum_{y} \hat p_Y(y) {\rm e}^{-\nu d(x,y) } \right\}^{-\rho} }
    {\sum_{x'} P_X(x') \left\{ \sum_{y} \hat p_Y(y) {\rm e}^{-\nu d(x',y) } \right\}^{-\rho} }. 
    \label{eq:p_X_opt}
\end{align}
See Appendix~\ref{appendix0} for the proof.

\textit{Proof of~Theorem~\ref{unwritten_theorem:arimoto}:} 
From (\ref{eq.97}), for any $\rho\geq -1$, we have 
\begin{align}
    \min_{ p_{Y|X} } F_{\rm AR, s}^{(\rho, \nu)} ( \hat p_Y, p_{Y|X} | P_X) 
    = \frac{1}{\rho} E_{0,\rm s}^{(\rho, \nu)} ( \hat p_Y | P_X ) 
\end{align}
and the minimum value is achieved if and only if $p_{Y|X} = p_{Y|X}^*(\hat p_Y)$ because of
the property of R\'enyi divergence. 
Similarly, from (\ref{eq.98}), we also have 
\begin{align}
    \min_{ \hat p_Y } F_{\rm AR, s}^{(\rho, \nu)} ( \hat p_Y, p_{Y|X} | P_X) 
    = \frac{1}{\rho} A_{\rm s}^{(\rho, \nu)} ( p_{Y|X} | P_X ) 
\end{align}
for any $\rho\geq -1$ and the minimum value is achieved if and only if $ \hat p_{Y} = \hat p_{Y}^*(p_{Y|X})$. 
This completes the proof.
\hfill$\IEEEQED$

From (\ref{theorem4.eq.1}) and (\ref{theorem4.eq.3}) in Theorem~\ref{unwritten_theorem:arimoto},
we have the folloiwng corollaries.
\begin{corollary}
For any fixed $\rho \in [-1,0)\cup(0,\infty)$ and $\nu\geq 0$, we have
\begin{align}
    \min_{ \hat p_Y }  \frac{1}{\rho} E_{0,\rm s}^{(\rho, \nu)}( \hat p_Y |P_X)
    = \min_{ \hat p_Y } \min_{ p_{Y|X} } F_{\rm AR}^{(\rho, \nu)}(\hat p_Y, p_{Y|X}|P_X) 
    = \min_{ p_{Y|X} } \frac{1}{\rho} A_{\rm s}^{(\rho, \nu)}(p_{Y|X}|P_X) . 
\end{align}
\end{corollary}
\begin{corollary}
Define the following function 
\begin{align}
&  \tilde E_{\rm B}(R, \Delta|P_X)
=\max_{\rho \geq 0} \min_{t\geq0}
  \left\{  
  \rho R + \rho \nu \Delta 
  -   \min_{ p_{Y|X} } 
A_{\rm s}^{(\rho, \nu)}( p_{Y|X} |P_X)
  \right\}. 
  \label{eq:alternative_form_source_coding}
\end{align}
Then, for any $R\geq 0$, $\Delta\geq 0$, and $P_X\in \mathcal{X}$, we have 
\begin{align}
    \tilde E_{\rm B}(R, \Delta|P_X) = E_{\rm B} (R, \Delta |P_X) .
\end{align}
\end{corollary}
We call $\tilde E_{\rm B}(R, \Delta|P_X)$ 
as the alternative expression for the Blahut's error exponent.

Algorithm~\ref{alg:Arimoto_algorithm_source_coding} shows Arimoto algorithm.
This algorithm was proposed for computing the Blahut's error exponent, but
we have shown that this algorithm works also for negative $\rho = -\lambda$.
By Theorem~\ref{theorem5}, Csisz\'ar and K\"orner's exponent coincides with
$G_{\rm JO}(R, \Gamma |P_X)$

\begin{algorithm}
\caption{Algorithm for lossy source coding exponents (Arimoto~\cite{Arimoto1976} for positive $\rho$)}
\label{alg:Arimoto_algorithm_source_coding}
\begin{algorithmic}
    \Require A probability distribution of an information source $P_X$, 
    $\rho > -1$, and $\nu \geq 0$.
    A distortion measure $d(x,y)$ is also given. 
    \State 
    Choose initial $\hat p_Y^{[1]}$ such that all components are nonzero.
    \For{$i=1,2,3,\ldots$}
\begin{align}
    p_{Y|X}^{[i]} (y|x) 
    &=
    \frac{ \hat p_Y^{[i]}(y) {\rm e}^{- \nu d(x,y) } } 
         { \sum_{y'} \hat p_Y^{[i]}(y') {\rm e}^{- \nu d(x,y') } } 
         \label{eq.arimoto_algorithm_source_coding_1},\\
    \hat p_Y^{ [i+1] }(y) 
    &=
    \frac{ 
      \left\{ 
        \sum_{x} P_X(x) {\rm e}^{\rho \nu d(x,y)} p_{Y|X}^{[i]} (y|x)^{1+\rho} 
      \right\}^{ 1/(1+\rho) } 
      }
      { 
        \sum_{y'} \left\{ 
        \sum_{x} P_X(x) {\rm e}^{\rho \nu d(x,y')} p_{Y|X}^{[i]} (y'|x)^{1+\rho} 
        \right\}^{ 1/(1+\rho)} 
      }.
      \label{eq.arimoto_algorithm_source_coding_2}
\end{align}
    \EndFor
\end{algorithmic}
\end{algorithm}

It is found that Eq.(\ref{eq.arimoto_algorithm_source_coding_1}) is the same as (\ref{algorithm_8_update_a}) in Algorithm~\ref{algorithm_GCK1} and Eq.(\ref{eq.arimoto_algorithm_source_coding_2}) is the same as (\ref{algorithm_7_update_b}) in Algorithm~\ref{algorithm_GCK2}.
Note that (\ref{algorithm_8_update_b}) in Algorithm~\ref{algorithm_GCK1} does not appear in Algorithm~\ref{alg:Arimoto_algorithm_source_coding} but it is the same as 
(\ref{eq:p_X_opt}) in Lemma~\ref{lemma14}.

For the case of $\rho=0$, 
the update rules (\ref{eq.arimoto_algorithm_source_coding_1}) and (\ref{eq.arimoto_algorithm_source_coding_2}) 
reduces to the rule of the Arimoto-Blahut algorithm for the rate-distortion function.
$F_{\rm AR,s}^{(0, \nu)} (\hat p_Y, p_{Y|X} |P_X) 
$ for $\rho=0$  is interpreted as
\begin{align}
    &\lim_{\rho \to 0 } F_{\rm AR,s}^{(\rho, \nu)} (\hat p_Y, p_{Y|X} |P_X) \notag\\
    &= \sum_{x,y} P_X(x) p_{Y|X}(y|x) \left( \log \frac{p_{Y|X}(y|x)}{\hat p_Y(y)}
    + \nu d(x,y) \right). 
\end{align}

At the end of this subsection, we give a remark regarding the function $(1/\rho) A_s^{(\rho,t)}(\hat p_{Y|X}|P_X)$.
When $\nu =0$, the function is equal to Sibson's mutual information of order $\alpha = 1+\rho$, i.e.,
we have
\begin{align}
    \frac{1}{\rho} A_{\rm s}^{(\rho, 0)}(p_{Y|X}|P_X)
    =
    I^{\rm s}_{1+\rho}( P_X, p_{Y|X} ).
    \label{eq.106}
\end{align}
One can easily check this equality holds. 
Hence $(1/\rho) A_{\rm s}^{(\rho, \nu)} (p_{Y|X} |P_X)$ is considered
as an extension of Sibson's mutual information for lossy source coding.
Note that, for Gallagar's $E_0$-function, $\alpha$ is $1/(1+\rho)$, i.e.,  $ \frac{1}{\rho} E_0^{(\rho,0)}(p_X|W) = I^{\rm s}_{\frac{1}{1+\rho}} ( p_X, W)$.

\subsection{Ar{\i}kan and Merhav's guessing exponent}
In this subsection, we show that Arimoto's algorithm can be used to compute Ar{\i}kan and Merhav's guessing exponent~\cite{ArikanMerhav1998}. 
The optimal error exponent in guessing problem under distortion was established by Ar{\i}kan and Merhav~\cite{ArikanMerhav1998}. 
Their exponent is defined as follows: 
\begin{definition}[\cite{ArikanMerhav1998}]
For $\Delta\geq 0$, $\rho\geq 0$, and $P_X\in \mathcal{P(X)}$, 
$\rho$-th order 
guessing exponent at distortion level $\Delta$ is defined by 
\begin{align}
    E_{\rm AM}^{(\rho)}(\Delta | P_X)
    =\sup_{q_X \in \mathcal{P(X)} } \{ \rho R(\Delta|q_X) - D(q_X||P_X) 
    \}.
    \label{eq.178}
\end{align}
\end{definition}

We have the following lemma.
\begin{lemma}
\label{lemma:E_AM}
For any $\Delta\geq 0$, $\rho\geq 0$, $P_X\in \mathcal{P(X)}$,
we have 
\begin{align}
    E_{\rm AM}^{(\rho)}(\Delta |P_X) = -E_{\rm B}^{(\rho)}(\Delta |P_X). 
\end{align}
\end{lemma}

\textit{Proof:}
The rate distortion function is expressed as~\cite{Blahut1972},\cite[Chapter~8]{Csiszar-KornerBook}
\begin{align}
    R(\Delta|P_X) 
    &= 
    \sup_{\nu\geq 0}
    \min_{\hat p_Y}
    \left[ 
    -\sum_{x} P_X \log \sum_{y} \hat p_Y(y) {\rm e}^{- \nu d(x,y)} -\nu \Delta  
    \right].
    \label{Rate-Distortion-Function-another-expression}
\end{align}
Then, substituting (\ref{Rate-Distortion-Function-another-expression}) into (\ref{eq.178}), we have
\begin{align}
    & E_{\rm AM}^{(\rho)}(\Delta |P_X)\notag\\
    &
    \stackrel{\rm (a)}=
    \sup_{q_X \in \mathcal{P(X)} } 
    \sup_{\nu\geq 0}
    \min_{\hat p_Y}
    \{
    -
    \rho
    \sum_{x}
    q_{X}(x) \log\sum_{y} 
    \hat p_Y(y) \mathrm{e}^{-\nu d(x,y)}
    - \rho \nu \Delta 
    - D(q_X||P_X) 
    \} \notag \\
    &
    \stackrel{\rm (b)}
    =
    \sup_{\nu\geq 0}
    \min_{\hat p_Y}
    \sup_{q_X \in \mathcal{P(X)} } 
    \{
    -
    \sum_{x}
    q_{X}(x) \log
    \frac{q_X(x)}{P_X(x) 
    \left(
        \sum_{y} 
    \hat p_Y(y) \mathrm{e}^{-\nu d(x,y)}
    \right)^{-\rho}
    }
    - \rho \nu \Delta 
    \} \notag \\
    &
    \stackrel{\rm (c)}
    =
    \sup_{\nu\geq 0}
    \min_{\hat p_Y}
    \left\{
    \log 
    \sum_{x}
    P_X(x) 
    \left[
        \sum_{y} 
    \hat p_Y(y) \mathrm{e}^{-\nu d(x,y)}
    \right]^{-\rho}
    - \rho \nu \Delta 
    \right\} \notag \\
    &  \stackrel{\rm (d)}
    =-E_B^{(\rho)}( \Delta |P_X). 
\end{align}
Step (a) follows from (\ref{Rate-Distortion-Function-another-expression}), 
Step (b) holds because the objective function is concave in $q_X$
for a fixed $\hat p_Y$ 
and convex in $\hat p_Y$ for a fixed  $q_X$, 
Step (c) holds because the maximum value of the objective function over $q_X$
is attained if $\displaystyle 
q_X(x) = 
\frac{ P_X(x) 
    \left[
        \sum_{y} 
    \hat p_Y(y) \mathrm{e}^{-\nu d(x,y)}
    \right]^{-\rho}
}{ \sum_{x} P_X(x) \left[
        \sum_{y} 
    \hat p_Y(y) \mathrm{e}^{-\nu d(x,y)}
    \right]^{-\rho} }
$, 
and Step (d) follows from (\ref{L_s}). \hfill$\IEEEQED$

Ar{\i}kan and Merhav showed that for fixed $\Delta\geq 0$, 
$ E_{\rm AM}^{(\rho)}(\Delta | P_X) $ as a function of $\rho$ is the one-sided LFT of Marton's error exponent 
$E_{\rm M}(R, \Delta |P_X)$ as a function of $R$ and its inverse relation that
the one-sided LFT of 
$ E_{\rm AM}^{(\rho)}(\Delta | P_X) $ as a function of $\rho$ 
is the lower convex hull of $E_{\rm M}(R, \Delta |P_X)$  as a function of $R$\cite[Theorem 2]{ArikanMerhav1998}. 
Thus from Lemma~\ref{lemma:E_AM}, we have the following proposition.
\begin{proposition}
Blahut's error exponent is equal to a lower convex full of Marton's error exponent 
that is expressed by
\begin{align}
    E_B(R, \Delta |P_X)
    &= \tilde E_M(R, \Delta |P_X) \notag \\
    &=
    \sup_{\rho \geq 0}
    \min_{q_X}\{
    D(q_X||P_X) + \rho [ R - R(\Delta | q_X) ]
    \}. \label{eq.184}
\end{align}
\end{proposition}

Expression (\ref{eq.184}) is found in~\cite[Eq.(30)]{ArikanMerhav1998}, but 
the explicit relationship with Blahut's exponent is shown here for the first time.
Ar{\i}kan and Merhav described how to calculate $E_{\rm AM}^{(\rho)}(\Delta |P_X)$ using Gallager's formula~\cite[Theorem 9.4.1]{GallagerTEXT} for the rate distortion function, but did not mention its relationship to Arimoto's algorithm. Lemma~\ref{lemma:E_AM} and (\ref{L_s}) ensure that we can employ Arimoto's algorithm for computing Ar{\i}kan and Merhav's guessing exponent. 

\section*{Acknowledgments}
The author thanks Dr. Yuta Sakai for the valuable comments. 
A part of this work was supported by JSPS KAKENHI Grant Number JP19K12156.


\appendices

\section{Proofs of Properties~\ref{property1} and~\ref{property2} }
\label{appendixA}
In this section, we give proofs of Properties~\ref{property1} and~\ref{property2}.

{\it Proof of Property~\ref{property1}:} 
We can decompose 
$    J_{\bm{t}_1}^{(\lambda, \nu)} (q_{XY}, p_X|W) $
as
\begin{align}
    & J_{\bm{t}_1}^{(\lambda, \nu )} (q_{XY}, p_X |W) \notag\\
    & = 
    -(1-\lambda) H( q_{XY} )
    - \lambda H(q_Y) \notag\\
    &\quad - \mathrm{E}_{q_{XY}} [ \log W(Y|X) ]
    -(1-\lambda) \mathrm{E}_{q_{XY}} [ \log p_X(X) ] + \lambda \nu \mathrm{E}_{q_X} [ c(X) ] .
\end{align}
The first term is a joint entropy of $X$ and $Y$ multplied
by $-(1-\lambda)<0$ and thus is a convex function of $q_{XY}$.
The marginal distribution $q_Y$ is a linear function 
of $q_{XY}$ and thus the second term is a convex function of $q_{XY}$.
The third term is a linear function of $q_{XY}$. 
The fourth term is a linear function of $q_{XY}$ for a fixed $p_X$
and a convex function of $p_X$ for a fixed $q_{XY}$ because
$-\log t$ is a convex function of $t$. 
The fifth term is linear in $q_X$. 
This completes the proof.
\hfill$\IEEEQED$

Next, we prove Property~\ref{property2}

\textit{Proof of Property~\ref{property2}:}
We can expand $ J_{\bm{t}_2}^{(\lambda, \nu )}(q_{XY}, \hat p_{X|Y} | W) $ as
\begin{align}
    J_{\bm{t}_2}^{(\lambda, \nu)}(q_{XY}, \hat p_{X|Y} | W)
    =
    D(q_{Y|X} |W |q_X) - \lambda H(q_X) - \lambda \mathrm{E}_{ q_{XY} } [ \log \hat p_{X|Y} (Y|X) ]
    + \lambda \nu \mathrm{E}_{q_X}[c(X)].
\end{align}
Part a) holds because the first, the second and the fourth terms are constant in $\hat p_{X|Y}$,
and the third term is convex (resp. concave) for $\lambda \in [0,1]$ (resp. $\lambda \in [-1,0$) 
because of 
the concavity of $\log$ and linearity of the expectation operation.
Part b) holds because of the convexity (resp. concavity) 
of the second term, while the first, the third and the fourth terms are constant in $q_X$.
Part c) holds because of the convexity of $D(q_{Y|X}||W|q_X)$, which completes the proof.
\hfill$\IEEEQED$

\section{Proof of Theorem 
\ref{theorem_convergence_algorithm4}
}
\label{appendixC}
In this section we give proof of Theorem 
\ref{theorem_convergence_algorithm4}.

\textit{Proof of Theorem \ref{theorem_convergence_algorithm4}:} 
From (\ref{algorithm_4_update_a}), we have
\begin{align}
    \sum_{y'} \hat p_{X|Y}^{[i]} ( x | y') W(y'|x) \mathrm{e}^{-\lambda \nu c(x) }
    &=\frac{\hat p_{X|Y}^{[i]} (x|y)^\lambda W(y|x) \mathrm{e}^{-\lambda \nu c(x) } }{ q_{Y|X}^{[i]} (y|x) } 
    \label{eq.186}
\end{align}
for any $(x,y)\in \mathcal{X\times Y}$. 
From (\ref{algorithm_4_update_b}), 
\begin{align}
\sum_{x'} \left\{    \sum_{y'} \hat p_{X|Y}^{[i]} ( x' | y') W(y'|x') \mathrm{e}^{-\lambda \nu c(x') }
\right\}^{1/\lambda} 
&=
\frac{\left\{    \sum_{y'} \hat p_{X|Y}^{[i]} ( x | y') W(y'|x) \mathrm{e}^{-\lambda \nu c(x) }
\right\}^{1/\lambda}
}{
q_X^{[i]}(x)} 
    \label{eq.187}
\end{align}
holds for all $x\in \mathcal{X}$.
Therefore, we have
\begin{align}
    &A^{(-\lambda, \nu)}( \hat p_{X|Y}^{[i]} | W ) \notag \\
    & \stackrel{\rm (a)}{=} 
    -\lambda \log 
    \sum_{x} 
    \left\{    
    \sum_{y} \hat p_{X|Y}^{[i]} ( x | y) W(y|x) \mathrm{e}^{-\lambda \nu c(x) }
    \right\}^{\frac{1}\lambda} 
    \notag \\
    & \stackrel{\rm (b)}{=} 
    -\lambda \left[
    \log \left\{    
    \sum_{y} \hat p_{X|Y}^{[i]} ( x | y) W(y|x) \mathrm{e}^{-\lambda \nu c(x) }
    \right\}^{\frac{1}\lambda} 
    -\log q_X^{[i]}(x)
    \right] 
    \notag\\
    & \stackrel{\rm (c)}{=} 
    -\log \frac{\hat p_{X|Y}^{[i]} (x|y)^\lambda W(y|x) \mathrm{e}^{-\lambda \nu c(x) } }{ q_{Y|X}^{[i]} (y|x) }
    +\lambda \log q_X^{[i]}(x)
    \notag\\
    & = 
    -\log \frac{ \hat p_{X|Y}^{[i]} (x|y)^\lambda W(y|x) \mathrm{e}^{-\lambda \nu c(x) } }{ q_{Y|X}^{[i]} (y|x) q_X^{[i]}(x)^\lambda }
    \label{eq.162}
\end{align}
for any $(x,y)\in \mathcal{X\times Y}$. 
Step (a) follows from the definition. 
Step (b) follows from (\ref{eq.187}) and
step (c) from (\ref{eq.186}).
Let $q_{XY}^* $ be a joint distribution that attains $ \min_{q_{XY}} \Theta^{(\lambda, \mu)}(q_{XY}|W)$.
Then it follow from Proposition~\ref{proposition_new_alg_B} and Lemma~\ref{lemma5} that 
$ A^{(-\lambda,\nu)} (\hat p_{X|Y}|W)$ is minimized by $ \hat p_{X|Y}^* = q_{X|Y}^*$. 
Then we have
\begin{align}
    0
    & \stackrel{\rm (a)}\leq 
    \Theta^{(\lambda, \lambda \nu)} (q_{XY}^{[i]}|W) - \Theta^{(\lambda, \lambda \nu)} (q_{XY}^*|W) \notag \\ 
    &
    \stackrel{\rm (b)}{\leq}
    A^{(-\lambda,\nu)} (\hat p_{X|Y}^{[i]}|W) - \Theta^{(\lambda, \lambda \nu)}(q_{XY}^*|W) \notag \\
    &
    \stackrel{\rm (c)}{=}
    {\rm E}_{ q_{XY}^*} \left[
    -\log \frac{ \hat p_{X|Y}^{[i]} (X|Y)^\lambda W(Y|X) \mathrm{e}^{-\lambda \nu c(X) }}{ q_{Y|X}^{[i]} (Y|X) q_X^{[i]}(X)^\lambda }
    \right] -  
    {\rm E}_{q_{XY}^*} \left[ \log 
    \frac{q_{Y|X}^*(Y|X)^{1-\lambda} q_Y^*(Y)^\lambda }{W(Y|X) \mathrm{e}^{-\lambda \nu c(X) }}
    \right]
    \notag \\
    &
    {=}
    {\rm E}_{q_{XY}^*} \left[
    \log \frac{ q_{X|Y}^{[i]} (X|Y)^\lambda q_{Y|X}^{[i]}(Y|X)^{1-\lambda} q_Y^{[i]}(Y)^\lambda }
    { \hat p_{X|Y}^{[i]} (X|Y)^\lambda q_{Y|X}^*(Y|X)^{1-\lambda} q_Y^*(Y)^\lambda  }
    \right] 
    \notag \\
    &=
    \lambda
    {\rm E}_{q_{XY}^*} \left[
    \log \frac{q_{X|Y}^{[i]} (X|Y)}{\hat p_{X|Y}^{[i]} (X|Y)}
    \right] 
    -(1-\lambda) D(q_{Y|X}^* || q_{Y|X}^{[i]} | q_X^*)
    -\lambda D(q_Y^* || q_Y^{[i]} )\notag\\
    &
    \stackrel{\rm (d)}\leq
    \lambda
    {\rm E}_{q_{XY}^*} \left[
    \log \frac{ \hat p_{X|Y}^{[i+1]} (X|Y)}{\hat p_{X|Y}^{[i]} (X|Y)}
    \right] \notag\\
    &= \lambda \{ D( q_{X|Y}^* || p_{X|Y}^{[i]} | q_X^* ) -  D( q_{X|Y}^* || p_{X|Y}^{[i+1]} | q_X^* ) \} . 
\end{align}
Step (a) holds because $q_{XY}^*$ minimizes $ \Theta^{(\lambda)} (p_{XY} |W) $. 
Step (b) follows from Proposition~\ref{proposition_new_alg_B}. 
Step (c) follows from (\ref{eq.162}) holds for every $(x,y)\in \mathcal{X\times Y}$. 
Step (d) follows from (\ref{algorithm_4_update_c}) and non-negativity of relative entropy. 

Let $\xi_i = \Theta^{(\lambda, \lambda \nu )} (q_{XY}^{[i]}|W) - \Theta^{(\lambda, \lambda \nu )} (q_{XY}^*|W) $. 
Then
\begin{align}
    0 & \leq \sum_{i=0}^{T-1} \xi_t\notag\\
      & \leq \lambda \{ D(q_{Y|X}^*||p_{Y|X}^{[0]} | q_X^*) -D(q_{Y|X}^*||p_{Y|X}^{[T]} | q_X^*) \} \notag\\
      & \leq \lambda D(q_{Y|X}^*||p_{Y|X}^{[0]} | q_X^*).  \label{eq.137} 
\end{align}
By Proposition~\ref{proposition_new_alg_B},
$\xi_i$ is a monotone decreasing sequence.
Then from (\ref{eq.137}) we have 
$0\leq T \xi_T \leq \lambda D(q_{Y|X}^*||p_{Y|X}^{[0]} | q_X^*)$. Thus 
\begin{align}
    0\leq \xi_T \le \frac{\lambda D(q_{Y|X}^*||p_{Y|X}^{[0]} | q_X^*)}{T} \to 0, \quad T\to \infty
\end{align}
Hence, we have 
\begin{align}
    \lim_{t\to \infty} 
    \Theta^{(\lambda, \lambda\nu)} (q_{XY}^{[i]} | W) 
    =
    \Theta^{(\lambda, \lambda\nu)} (q_{XY}^*|W).
\end{align}
This completes the proof.\hfill$\IEEEQED$

\section{Proofs of Lemmas~\ref{lemma0} and~\ref{lemma14}}
\label{appendix0}
This section gives the proofs of Lemma~\ref{lemma0} and Lemma~\ref{lemma14}. 

\textit{Proof of Lemma~\ref{lemma0}:}
We have 
\begin{align*}
& D_{1+\rho} ( p_{X|Y}^*(p_X) || \hat p_{X|Y} | p_Y^* (p_X) ) \notag\\
& = \frac1{\rho} \log \sum_y p_Y^*(p_X)(y)
\sum_x \tilde p_{X|Y}(p_X)^{1+\rho}(x|y) \hat p_{X|Y}^{-\rho}(x|y) \notag\\
&=
\frac1{\rho} \log \sum_y 
\frac{
\{ \sum_{x} p_X(x) \{ W(y|x) \mathrm{e}^{\rho \nu c(x)} \}^{1/(1+\rho)} \}^{1+\rho}
}{
\sum_{y'}
\{ \sum_{x} p_X(x) \{ W(y'|x) \mathrm{e}^{\rho \nu c(x)}\}^{1/(1+\rho)} \}^{1+\rho}
}
\sum_{x}
\frac{
\{ p_X(x) \{ W(y|x) \mathrm{e}^{\rho \nu c(x)}\}^{1/(1+\rho)} \}^{1+\rho}
}{\{ \sum_{x'} p_X(x) \{ W(y|x') \mathrm{e}^{\rho \nu c(x')}\}^{1/(1+\rho)} \}^{1+\rho}
}
\hat p_{X|Y}^{-\rho}(x|y) \\
&=
\frac1{\rho} \log  
\frac{
1}{
\sum_{y'}
\{ \sum_{x} p_X(x) \{ W(y'|x) \mathrm{e}^{\rho \nu c(x)} \}^{1/(1+\rho)} 
\}^{1+\rho}
}
+\frac{1}{\rho}\log 
\sum_y \sum_{x} p_X^{1+\rho}(x) W(y|x) \mathrm{e}^{\rho \nu c(x)} 
\hat p_{X|Y}^{-\rho}(x|y) \\
&= \frac1\rho E_0^{(\rho, \nu)}(p_X|W) - F_{\rm AR}^{(\rho, \nu)} (p_X, \hat p_{X|Y} |W),
\end{align*}
which proves (\ref{eq.lemma0.1}) and we also have
\begin{align*}
    & D_{1+\rho}(p_X || q_X^*(\hat p_{X|Y} ) ) \\
    & = \frac1{\rho} \sum_{x} p_X^{1+\rho}(x) q_X^*(\hat p_{X|Y}) (x)^{-\rho} \\
    &= \frac1{\rho} \log \sum_{x} p_X^{1+\rho}(x) 
    \frac{  \sum_{y} W(y|x) \mathrm{e}^{ \rho \nu c(x) }\hat p_{X|Y} ^{-\rho} (x|y) }{ 
    \left\{
    \sum_{x'}
    \left[ \sum_{y} W(y|x') \mathrm{e}^{ \rho \nu c(x') }\hat p_{X|Y} ^{-\rho} (x'|y)\right]^{-1/\rho} \right\}^{-\rho}} \\
    &=
    \log \sum_{x'}
    \left[ \sum_{y} W(y|x') \mathrm{e}^{ \rho \nu c(x') } \hat p_{X|Y} ^{-\rho} (x'|y)\right]^{-1/\rho} 
    + \frac1\rho \log \sum_x p_X^{1+\rho} (x)
    \sum_{y} W(y|x) \mathrm{e}^{ \rho \nu c(x) } \hat p_{X|Y}^{-\rho}(x|y)\\
    &= \frac1{\rho} A^{(\rho, \nu)}(\hat p_{X|Y} |W) - F_{\rm AR}^{(\rho, \nu)} (p_X, \hat p_{X|Y} |W),
\end{align*}
which proves (\ref{eq.lemma0.2}). This completes the proof. \hfill$\IEEEQED$

\textit{Proof of Lemma~\ref{lemma14}: }We have 
\begin{align}
    & \frac1\rho E_{0,\rm s}^{(\rho, \nu)} (\hat p_Y | P_X) 
    + D_{1+\rho} ( p_{Y|X} ||p_{Y|X}^*(\hat p_Y)  | p_X^*(\hat p_Y) ) \notag \\  
    & = 
    \frac1{\rho} \log \sum_x P_X(x) \left\{
    \sum_y \hat p_Y(y) {\rm e}^{ -\nu d(x,y)}
    \right\}^{-\rho}\notag\\
    & \quad +
    \frac1{\rho} \log \sum_x \left(
    \frac{P_X(x) \{ \sum_{y} \hat p_Y(y) {\rm e}^{-\nu d(x,y) } \}^{-\rho} }
    { \sum_{x'} P_X(x') \{ \sum_{y} \hat p_Y(y) {\rm e}^{-\nu d(x',y) } \}^{-\rho} }
    \right) 
    \sum_y
    p_{Y|X}^{1+\rho}(y|x) \left( 
    \frac{ \hat p_Y(y) {\rm e}^{ -\nu d(x,y) } }   { \sum_{y'}\hat p_Y(y') {\rm e}^{\rho \nu d(x,y')} } 
    \right)^{-\rho} \notag\\
    &=
    \frac1{\rho} \log \sum_x P_X(x) 
    \left\{ \sum_{y} \hat p_Y(y) {\rm e}^{-\nu d(x,y)}
    \right\}^{-\rho}
        \sum_y
    p_{Y|X}^{1+\rho}(y|x) 
    \frac{ \hat p_Y^{-\rho}(y) {\rm e}^{\rho \nu d(x,y) } }{
    \{ \sum_{y'} \hat p_Y(y') {\rm e}^{-\nu d(x,y')}
    \}^{-\rho}}\notag\\
    &= 
    \frac1{\rho} \log \sum_x P_X(x) 
    \sum_y p_{Y|X}^{1+\rho}(y|x) 
    \hat p_Y^{-\rho}(y) {\rm e}^{\rho \nu d(x,y) }
    = F_{\rm AR,s}^{(\rho, \nu)} ( \hat p_Y, p_{Y|X} |P_X).
\end{align}
This proves that Eq.(\ref{eq.97}) holds. We also have
\begin{align}
    & \frac1\rho A_{\rm s}^{(\rho, \nu)} (p_{Y|X} | P_X) 
    + D_{1+\rho} ( \hat p_Y^* ( p_{Y|X} ) || \hat p_Y)\notag\\
    &=
    \frac{1+\rho}{\rho} \log \sum_{y}\left[
    \sum_x P_X(x) {\rm e}^{\rho \nu d(x,y) } p_{Y|X}^{1+\rho}(y|x)
    \right]^{\frac{1}{1+\rho}} \notag\\
   &\quad +\frac{1}{\rho} \log \sum_{y} \frac{
    \sum_x P_X(x) {\rm e}^{\rho \nu d(x,y) } p_{Y|X}^{1+\rho}(y|x)
   }{
   \left\{
   \sum_{y'}\left[
    \sum_x P_X(x) {\rm e}^{\rho \nu d(x,y') } p_{Y|X}^{1+\rho}(y'|x)
    \right]^{\frac{1}{1+\rho}}
    \right\}^{1+\rho}
   } \hat p_Y^{-\rho}(y)\notag\\
   &=     \frac{1}{\rho} \log \sum_{y}
\sum_x P_X(x) {\rm e}^{\rho \nu d(x,y) } p_{Y|X}^{1+\rho}(y|x)\hat p_Y^{-\rho}(y)
  = F_{\rm AR,s}^{(\rho, \nu)} (\hat p_Y, p_{Y|X} | P_X).
\end{align}
This proves that Eq.(\ref{eq.98}) holds, completing the proof.\hfill$\IEEEQED$

\section{The parameterized objective function 
$ F_{{\rm AR}, \bm{t} }^{(-\lambda, \nu)}(p_{XY}, \hat p_{XY} |W)$ 
and the algorithm derived from it}
\label{appendix:parameterized_Arimoot}

We will derive an algorithm from $\tilde F_{ {\rm JO}, \bm{t}}^{(\lambda, \nu)}
    (q_{XY}, p_{XY}, \hat p_{XY} |W)$ for $\bm{t}\in \mathcal{T}_1 \cap \mathcal{T}_2$. 
This algorithm updates the two joint distributions $p_{
XY} $ and $\hat p_{XY}$ alternately. 
Interestingly, the algorithm updates them even in the case of $\bm{t}=\bm{t}_1  + \bm{t}_2$, although 
in the original Arimoto algorithm, only $p_X$ and $\hat p_{X|Y}$ are updated.

Define
\begin{align}
    F_{{\rm AR}, \bm{t} }^{(-\lambda, \nu)}(p_{XY}, \hat p_{XY} |W)
    =\frac{1}{-\lambda}
    \min_{ q_{XY} } 
    \tilde F_{ {\rm JO}, \bm{t} }^{(\lambda, \nu)} (q_{XY}, p_{XY}, \hat p_{XY} |W). 
\end{align}
The original 
$ F_{{\rm AR}}^{(-\lambda, \nu)}(p_{X}, \hat p_{X|Y} |W)$ is a special case of 
$F_{{\rm AR}, \bm{t} }^{(-\lambda, \nu)}(p_{XY}, \hat p_{XY} |W)
$ with $\bm{t}=\bm{t}_1+\bm{t}_2$.
Then, for any $\bm{t} \in \mathcal{T}$, we have 
\begin{align}
\min_{q_{XY }} \Theta^{(\lambda,\lambda\nu)}(q_{XY}|W)
&=\min_{q_{XY }} 
\min_{p_{XY}} \min_{\hat p_{XY}} 
\tilde F_{ {\rm JO} , \bm{t} } ^{(\lambda, \nu)}
    (q_{XY}, p_{XY}, \hat p_{XY} |W) \notag\\
&= 
\min_{p_{XY}} \min_{\hat p_{XY}} 
-\lambda 
F_{{\rm AR}, \bm{t} }^{(-\lambda, \nu)}(p_{XY}, \hat p_{XY} |W) . 
\label{eq.91}
\end{align}

Hereafter assume $\bm{t} \in \mathcal{T}_1 \cap \mathcal{T}_2 $. 
In this case, the parameterized objective function and $ F_{{\rm AR}, \bm{t} }^{(-\lambda, \nu)}(p_{XY}, \hat p_{XY} |W)$ are given by 
\begin{align}
&F_{{\rm AR}, \bm{t} }^{(-\lambda, \nu)}(p_{XY}, \hat p_{XY} |W) = \frac{ 1+(1-\lambda) (t_2+t_3) } {\lambda} \notag\\
&\cdot \log \sum_{x,y}
\Big[ 
p_X^{ (1-\lambda)(1+t_2) }(x) p_{Y|X} ^{(1-\lambda)t_2 } (y|x) \hat p_Y^{(1-\lambda)t_3 } (y) \hat p_{X|Y}^{(1-\lambda)t_3+\lambda }(x|y) W(y|x) 
    {\rm e}^{-\lambda \nu c(x)}
\Big]^{\frac{1}{ 1+(1-\lambda) (t_2 + t_3) }} .
\label{generalized_objective_function_FAR}
\end{align}
Eq.(\ref{generalized_objective_function_FAR}) reduces to $F_{{\rm AR}}^{(-\lambda, \nu)}(p_{XY}, \hat p_{XY} |W)$ if $t_2=t_3 = 0$. 
From (\ref{generalized_objective_function_FAR}), a family of algorithms with parameter $t_2, t_3\geq 0$ is derived. 
For this objective function, the following lemmas hold.
\begin{lemma}
For a fixed $\lambda \in [0,1]$, $\nu\geq 0$, $\bm{t} \in \mathcal{T}$, and $q_{XY}$, we have
\begin{align}
    \min_{ p_{XY} } 
    \min_{ \hat p_{XY} } 
    \tilde F_{{\rm JO}, \bm{t}}^{(\lambda, \nu)}(q_{XY}, p_{XY}, \hat p_{XY} | W)
    = \Theta^{(\lambda, \lambda \nu)} ( q_{XY} | W).
\end{align}
The minimum is attained if $p_{XY} = q_{XY}$ and $\hat p_{XY} = q_{XY}$. 
\end{lemma}

\begin{lemma}
\label{lemma15}
For a fixed $\lambda \in [0,1]$, $\nu\geq 0$, $\bm{t} \in \mathcal{T}_1 \cap \mathcal{T}_2$, $p_{XY}$ and $\hat p_{XY}$, we have
\begin{align}
    \min_{ q_{XY} } 
    \tilde F_{ {\rm JO}, \bm{t} }^{(\lambda, \nu)}(q_{XY}, p_{XY}, \hat p_{XY} | W)
    = -\lambda F_{{\rm AR}, \bm{t} }^{(-\lambda, \nu)} ( p_{XY}, \hat p_{XY} | W).
\end{align}
The minimum is attained if 
\begin{align}
    q_{XY}(x,y) = \frac{ 
    \left\{
    p_X^{ (1-\lambda)(1+t_2) }(x) p_{Y|X} ^{(1-\lambda)t_2 } (y|x)
    \hat p_Y^{(1-\lambda)t_3 } (y) \hat p_{X|Y}^{(1-\lambda)t_3+\lambda }(x|y) W(y|x) 
    {\rm e}^{-\lambda \nu c(x)}
    \right\}^{\frac{1}{1+(1-\lambda)(t_2+t_3)}}
    }
    {\sum_{x',y'}  
    \left\{
    p_X^{ (1-\lambda)(1+t_2) }(x') p_{Y|X} ^{(1-\lambda)t_2 } (y'|x')
    \hat p_Y^{(1-\lambda)t_3 } (y') \hat p_{X|Y}^{(1-\lambda)t_3+\lambda }(x'|y') W(y'|x') 
    {\rm e}^{-\lambda \nu c(x')}
    \right\}^{\frac{1}{1+(1-\lambda)(t_2+t_3)}} 
    }.
    \label{optimal_q_Lemma17}
\end{align}
\end{lemma}
Eq.(\ref{optimal_q_Lemma17}) reduces to (\ref{optimal_q_Lemma13}) if $t_2 = t_3 = 0$. 
See Appendix \ref{appendixB} for the proof of Lemma~\ref{lemma15}.

Define the following functions.
\begin{align}
    & E_{0, t_2}^{(-\lambda, \nu)} (p_{XY}|W) \notag\\
    & =
    -(1+(1-\lambda) t_2)  \log \sum_y
    \left[ \sum_x p_X(x) p_{Y|X}^{\frac{t_2}{1+t_2}}(y|x) 
    \left\{
    W(y|x)
    {\rm e}^{-\lambda \gamma c(x) }
    \right\}^\frac{1}{ (1-\lambda)(1 + t_2) } 
    \right]^{ \frac{(1-\lambda)(1 + t_2) }{\lambda + (1-\lambda) t_2 } }
\end{align}

\begin{align}
    & A_{ t_3 }^{(-\lambda, \nu)} (\hat p_{XY} |W) \notag\\
    & =
    - (\lambda + (1-\lambda ) t_3 )
    \log \sum_x \left[
    \sum_y \left\{
    W(y|x) {\rm e}^{-\lambda \nu c(x) } 
    \hat p_Y^{(1-\lambda)t_3} (y) 
    \hat p_{X|Y}^{\lambda + (1-\lambda)t_3} (x|y) 
    \right\}^{\frac{1 }{ 1+(1-\lambda)t_3}}
    \right]^{\frac{1+(1-\lambda)t_3}{\lambda+(1-\lambda)t_3}}
\end{align}
When $t_2=0$ and $t_3 = 0$, these functions reduces to 
$E_{0}^{(-\lambda, \nu)}(p_X|W)$ and $A^{(-\lambda, \nu)}(\hat p_{X|Y}|W)$, respectively. 

\begin{lemma}
\label{lemma.parameterized_arimoto1}
For fixed $\lambda\in (0,1)$, $\nu \geq 0$, $\bm{t}\in \mathcal{T}_1 \cap \mathcal{T}_2$, 
and for a fixed $p_{XY} \in \mathcal{P(X\times Y)}$, 
$
-\lambda F_{{\rm AR}, \bm{t} }^{(-\lambda, \nu)} (p_{XY}, \hat p_{XY} |W) 
$ is minimized by
$\hat p_{XY} = ( \hat p_{Y}^*(p_{XY}), \hat p_{X|Y}^*( p_{XY}) ) $, where
\begin{align}
    & \hat p_Y^*(y) 
    = 
    \frac{ 
    \left[
        \sum_x
       p_X(x)
       p_{Y|X}^{\frac{t_2}{1+t_2} }(y|x)
       \left\{ 
       W(y|x)
       {\rm e}^{-\lambda \nu c(x)}
       \right\}^{ \frac{1}{(1-\lambda)(1+t_2)}} 
    \right]^{\frac{(1-\lambda)(1+t_2)}{1+(1-\lambda)t_2}}
    }{
    \sum_{y'}
    \left[
        \sum_x
       p_X(x)
       p_{Y|X}^{\frac{t_2}{1+t_2} }(y'|x)
       \left\{ 
       W(y'|x)
       {\rm e}^{-\lambda \nu c(x)}
       \right\}^{ \frac{1}{(1-\lambda)(1+t_2)}} 
    \right]^{\frac{(1-\lambda)(1+t_2)}{1+(1-\lambda)t_2}}
    }, \label{eq.242}\\
    & \hat p_{X|Y}^*(x|y)
    =
    \frac{
       p_X(x)
       p_{Y|X}^{\frac{t_2}{1+t_2} }(y|x)
       \left\{ 
       W(y'|x)
       {\rm e}^{-\lambda \nu c(x)}
       \right\}^{ \frac{1}{(1-\lambda)(1+t_2)}} 
    }{
    \sum_{x'}
           p_X(x')
       p_{Y|X}^{\frac{t_2}{1+t_2} }(y|x')
       \left\{ 
       W(y|x')
       {\rm e}^{-\lambda \nu c(x')}
       \right\}^{ \frac{1}{(1-\lambda)(1+t_2)}} 
    }\label{eq.243}
\end{align}
and the minimum value is $E_{0, t_2}^{ (-\lambda, \nu) } (p_{XY}|W)$. 
\end{lemma}

\begin{lemma}
\label{lemma.parameterized_arimoto2}
For fixed $\lambda\in (0,1)$, $\bm{t}\in \mathcal{T}_1\cap \mathcal{T}_2$, and 
for a fixed $\hat p_{XY} \in \mathcal{P(X\times Y)}$, 
$-\lambda F_{{\rm AR}, {\bm{t}}}^{(-\lambda, \nu)} (p_{XY}, \hat p_{XY} |W) $ is minimized by
$p_{XY} = ( p_{X}^*(\hat p_{XY}), p_{Y|X}^*( \hat p_{XY}) ) $, where
\begin{align}
    &p_X^*(x) 
    = 
    \frac{ 
    \left[
         \sum_y
       \left\{ 
       \hat p_Y^{ (1-\lambda) t_3}(y)
       \hat p_{X|Y}^{ \lambda + (1-\lambda) t_3 }(x|y)
       W(y|x)       {\rm e}^{ -\lambda \nu c(x) }
       \right\}^{ \frac{1}{ 1 + (1-\lambda) t_3 } } 
    \right]^{\frac{ 1 + (1-\lambda) t_3 }{ \lambda +(1-\lambda) t_3 }}
    }{
    \sum_{x'}
    \left[
        \sum_y
       \left\{ 
       \hat p_Y^{ (1-\lambda) t_3}(y)
       \hat p_{X|Y}^{ \lambda + (1-\lambda) t_3 }(x'|y)
       W(y|x')        {\rm e}^{ -\lambda \nu c(x') }
       \right\}^{ \frac{1}{ 1 + (1-\lambda) t_3 } } 
    \right]^{\frac{ 1 + (1-\lambda) t_3 }{ \lambda +(1-\lambda) t_3 }}
    },\\
    & p_{Y|X}^*(y|x)
    =
    \frac{
       \left\{ 
       \hat p_Y^{(1-\lambda)t_3 }(y)
       \hat p_{X|Y}^{ (1-\lambda)t_3 + \lambda  }(x|y)
       W(y|x)
       \right\}^{ \frac{1}{ 1 + (1-\lambda) t_3 } } 
    }{
    \sum_{y'}
       \left\{ 
       \hat p_Y^{(1-\lambda)t_3 }(y')
       \hat p_{X|Y}^{ (1-\lambda)t_3 + \lambda  }(x|y')
       W(y'|x)
       \right\}^{ \frac{1}{ 1 + (1-\lambda) t_3 } } 
    }
\end{align}
and the minimum value is $ A_{ t_3 }^{ (-\lambda, \nu) } (\hat p_{XY}|W)$. 
\end{lemma}
\begin{IEEEproof}
The outline of the proof of Lemma~\ref{lemma.parameterized_arimoto1} is given here.
Because $(1+(1-\lambda)(t_2+t_2))/\lambda$ is positive for all $\lambda \in (0,1)$ and
the logarithm is a monotone increasing function, $ F_{{\rm AR}, {\bm{t}}}^{(-\lambda, \nu)} (p_{XY}, \hat p_{XY} |W)$
is maximized by $(\hat p_Y, \hat p_{X|Y})$ that maximizes 
\begin{align}
\sum_{x,y}
\Big[ 
p_X^{ (1-\lambda)(1+t_2) }(x) p_{Y|X} ^{(1-\lambda)t_2 } (y|x) \hat p_Y^{(1-\lambda)t_3 } (y) \hat p_{X|Y}^{(1-\lambda)t_3+\lambda }(x|y) W(y|x) 
    {\rm e}^{-\lambda \nu c(x)}
\Big]^{\frac{1}{ 1+(1-\lambda) (t_2 + t_3) }}.
\label{eq.246}
\end{align}
Then, this function is concave in $\hat p_Y$ for fixed $\hat p_{X|Y}$ and $p_{XY}$
and also is concave in $p_{X|Y}$ for fixed $\hat p_Y$ and $p_{XY}$.
Use the method of Lagrange multipliers to find
the $\hat p_Y$ and $\hat p_{X|Y}$ that maximize (\ref{eq.246}). This yields  
(\ref{eq.242}) and (\ref{eq.243}), completing the proof.
Lemma~\ref{lemma.parameterized_arimoto2} is proved in the same way. 
\end{IEEEproof}

An iterative algorithm derived from Lemmas~\ref{lemma.parameterized_arimoto1} and \ref{lemma.parameterized_arimoto2} for computing (\ref{eq.91}) is given in Algorithm~\ref{alg:parameterized_Arimoto}.
Algorithm~\ref{alg:parameterized_Arimoto} is a parameterized algorithm with $t_2,t_3\geq 0$.
We observe that equations in Algorithm~\ref{alg:parameterized_Arimoto} is different from those in
Algorithm~\ref{alg:TSZ_generalized}.

\begin{algorithm}
\caption{Computation of $ \min_{p_{XY}} \min_{\hat p_{XY}} 
F_{{\rm AR}, \bm{t} }^{(-\lambda, \nu)} (p_{XY}, \hat p_{XY} |W)$ }
\label{alg:parameterized_Arimoto}
\begin{algorithmic}
\Require 
The conditional probability of the channel $W$, 
$\lambda \in(0,1)$, $\nu \ge 0$, $t_2\geq 0, t_3 \geq 0$. 
Choose initial joint probability distribution
$ p_{XY}^{[0]} $ such that $ p_{XY}^{[0]}(x,y) = 0 $
if $W(y|x)=0$ and 
$ q_{XY}^{[0]}(x,y) > 0 $
if $W(y|x)>0$.
\For{$i=0,1,2,\ldots$,}
\State 
\begin{align}
    & \hat p_Y^{[i]}(y) 
    = 
    \frac{ 
    \left[
        \sum_x
       p_X^{[i]}(x)
       p_{Y|X}^{[i]}(y|x)^{\frac{t_2}{1+t_2} }
       \left\{ 
       W(y|x)
       {\rm e}^{-\lambda \nu c(x)}
       \right\}^{ \frac{1}{(1-\lambda)(1+t_2)}} 
    \right]^{\frac{(1-\lambda)(1+t_2)}{1+(1-\lambda)t_2}}
    }{
    \sum_{y'}
    \left[
        \sum_x
       p_X^{[i]}(x)
       p_{Y|X}^{[i]}(y'|x)^{\frac{t_2}{1+t_2} }
       \left\{ 
       W(y'|x)
       {\rm e}^{-\lambda \nu c(x)}
       \right\}^{ \frac{1}{(1-\lambda)(1+t_2)}} 
    \right]^{\frac{(1-\lambda)(1+t_2)}{1+(1-\lambda)t_2}}
    },\label{update6a}\\
    & \hat p_{X|Y}^{[i]}(x|y)
    =
    \frac{
       p_X^{[i]}(x)
       p_{Y|X}^{[i]}(y'|x)^{\frac{t_2}{1+t_2} }
       \left\{ 
       W(y'|x)
       {\rm e}^{-\lambda \nu c(x)}
       \right\}^{ \frac{1}{(1-\lambda)(1+t_2)}} 
    }{
    \sum_{x'}
           p_X^{[i]}(x')
       p_{Y|X}^{[i]}(y|x')^{\frac{t_2}{1+t_2} }
       \left\{ 
       W(y|x')
       {\rm e}^{-\lambda \nu c(x')}
       \right\}^{ \frac{1}{(1-\lambda)(1+t_2)}} 
    }, \label{update6b}\\
    & p_X^{[i+1]}(x) 
    = 
    \frac{ 
    \left[
         \sum_y
       \left\{ 
       \hat p_Y^{[i]}(y)^{ (1-\lambda) t_3}
       \hat p_{X|Y}^{[i]}(x|y)^{ \lambda + (1-\lambda) t_3 }
       W(y|x) {\rm e}^{ -\lambda \nu c(x) }
       \right\}^{ \frac{1}{ 1 + (1-\lambda) t_3 } } 
    \right]^{\frac{ 1 + (1-\lambda) t_3 }{ \lambda +(1-\lambda) t_3}}
    }{
    \sum_{x'}
    \left[
        \sum_y
       \left\{ 
       \hat p_Y^{[i]}(y)^{ (1-\lambda) t_3}
       \hat p_{X|Y}^{[i]}(x'|y)^{ \lambda + (1-\lambda) t_3 }
       W(y|x') {\rm e}^{ -\lambda \nu c(x') }
       \right\}^{ \frac{1}{ 1 + (1-\lambda) t_3 } } 
    \right]^{\frac{ 1 + (1-\lambda) t_3 }{ \lambda +(1-\lambda) t_3}}
    }, \label{update6c}\\
    & p_{Y|X}^{[i+1]}(y|x)
    =
    \frac{
       \left\{ 
       \hat p_Y^{[i]}(y)^{(1-\lambda)t_3 }
       \hat p_{X|Y}^{[i]}(x|y)^{ \lambda + (1-\lambda)t_3 }
       W(y|x)
       \right\}^{ \frac{1}{ 1 + (1-\lambda) t_3}} 
    }{
    \sum_{y'}
       \left\{ 
       \hat p_Y^{[i]}(y')^{(1-\lambda)t_3 }
       \hat p_{X|Y}^{[i]}(x|y')^{ \lambda + (1-\lambda)t_3   }
       W(y'|x)
       \right\}^{ \frac{1}{ 1 + (1-\lambda) t_3}} 
    } .
    \label{update6d}
\end{align}

\EndFor 
\end{algorithmic}
\end{algorithm}

\section{Proofs of Lemmas \ref{lemma11}, \ref{lemma12},\ref{lemma20}, \ref{lemma22}, and \ref{lemma15}. }
\label{appendixB}

\begin{IEEEproof}[Proof of Lemma~\ref{lemma11}]
 From Property~\ref{property2} c),
 $ J_{\bm{t}_2}^{(\lambda)} (q_{XY}, \hat p_{X|Y} |W) $
 is convex in $q_{Y|X}$ for a fixed $q_X$ and $\hat p_{X|Y}$
 for $\lambda \in [-1,0)$. Hence, we have
 \begin{align}
    J_{\bm{t}_2}^{(-\rho, \nu)}(q_{XY}, \hat p_{X|Y} | W )
    &= \mathrm{E}_{q} 
    \left[ \log \frac{ q_{Y|X} (Y|X) q_X^{-\rho}(Y) }{W(Y|X) \mathrm{e}^{\rho \nu c(X)} \hat p_{X|Y}^{-\rho}(X|Y)   } \right] \notag\\
    &= \rho H(q_X) + 
    \mathrm{E}_{q_{XY} } \left[ 
    \log \frac{ q_{Y|X} (Y|X) q_X^{-\rho}(Y) }
    {W(Y|X) \mathrm{e}^{\rho \nu c(X)}\hat p_{X|Y}^{-\rho}(X|Y) } 
    \right]\notag\\
    &= \rho H(q_X) + D(q_{Y|X} || q_{Y|X}^*(\hat p_{X|Y} ) | q_X )
    -\sum_{x} q_X(x) \log \sum_y W(y|x) \mathrm{e}^{\rho \nu c(x)}
    \hat p_{X|Y}^{-\rho} (x|y) \notag \\
    &
    \stackrel{\rm (a)}
    \geq
    \rho H(q_X) 
    -\sum_{x} q_X(x) \log \sum_y W(y|x) \mathrm{e}^{\rho \nu c(x)}
    \hat p_{X|Y}^{-\rho} (x|y) \notag \\
    &= -\rho  
    \sum_{x} q_X(x) \log 
    \frac{ q_X(x) }{ 
    \left\{ 
    \sum_y W(y|x) \mathrm{e}^{\rho \nu c(x)}
    \hat p_{X|Y}^{-\rho} (x|y) 
    \right\}^{-1/\rho}
    } \notag\\
    &= \tilde J_{\bm{t}_2}^{(-\rho, \nu)}(q_X, \hat p_{X|Y} |W) . 
\end{align}
In Step (a), equality holds if and only if $q_{Y|X} = q_{Y|X}^*(\hat p_{X|Y})$.
This completes the proof.
\end{IEEEproof}

\begin{IEEEproof}[Proof of Lemma~\ref{lemma12}]
From Property~\ref{property2} b),
 $ J_{\bm{t}_2}^{(-\rho )} (q_{XY}, \hat p_{X|Y} |W) $
 is concave in $q_X$ for a fixed $q_{Y|X}$ and $\hat p_{X|Y}$
 For $\rho\in (0,1]$. In fact, we have
\begin{align*}
&    \max_{q_X} 
    \tilde J_{\bm{t}_2}^{(-\rho)}(q_X, \hat p_{X|Y} |W) \\
&=     \max_{q_X} \left\{ 
-\rho D(q_X || \tilde q_X(\hat p_{X|Y} ) )
    + \rho \log \sum_x \left[
    \sum_y W(y|x) 
    \hat p_{X|Y}^{-\rho} (x|y) 
    \right]^{-1/\rho} \right\}\\
&    = 
    \rho \log \sum_x \left\{
    \sum_y W(y|x) 
    \hat p_{X|Y}^{-\rho} (x|y) 
    \right\}^{-1/\rho}\\ 
&    =A^{(\rho)}(\hat p_{X|Y}|W).
\end{align*}
The maximum is attained by
$q_X=\tilde q_X(\hat p_{X|Y}) $
because of the equality condition of $D(q_X||\tilde q_X(\hat p_{X|Y}))$.
This completes the proof.
\end{IEEEproof}

\begin{IEEEproof}[Proof of Lemma~\ref{lemma20}]
Let the rhs's of (\ref{eq.lemma20.2}) and (\ref{eq.lemma20.3}) be 
$\tilde q_{Y|X}(\hat p_Y)(y|x)$ and $\tilde q_X(\hat p_Y)(x)$.
Then, we have
\begin{align*}
    & J_{{\rm s}, \bm{t}_3 }^{(\lambda, \nu)} (q_{XY}, \hat p_Y |W) 
    = \mathrm{E}_{q_{XY} } \left[ 
    \log \frac{ q_X(X) q_{Y|X}^\lambda(Y|X) }{P_X(X) {\rm e}^{-\lambda\nu d(X,Y)} \hat p_Y^\lambda(Y)}
    \right] \notag \\
    & = D(q_X||P_X) + \lambda \mathrm{E}_{ q_{XY} }
    \left[
    \log 
    \frac{ q_{Y|X}(Y|X) }{ \hat p_Y(Y) \mathrm{e}^{-\nu d(X,Y) } }
    \right] \\
    & = D(q_X||P_X) + \lambda D(q_{Y|X} || \tilde q_{Y|X}(\hat p_Y) | q_X)
    -\lambda \sum_x q_X(x) \log \sum_y
    \hat p_Y(y) {\rm e}^{-\nu d(x,y) }  \\
    & 
    \stackrel{\rm (a)}
    \geq D(q_X||P_X) 
    -\lambda \sum_x q_X(x) \log \sum_y
    \hat p_Y(y) {\rm e}^{-\nu d(x,y) }  \\
    & = \sum_x q_X(x) 
    \log \frac{q_X(x)} {P_X(x) \left\{
    \sum_y
    \hat p_Y(y) {\rm e}^{- \nu d(x,y) }
    \right\}^\lambda } \\
    &= D(q_X|| \hat q_X(\hat p_Y) )
    -\log \sum_{x}P_X(x)\Big\{
    \sum_y
    \hat p_Y(y) {\rm e}^{-\nu d(x,y) }
    \Big\}^\lambda \\
    &
    \stackrel{\rm (b)}
    \geq -\log \sum_{x}P_X(x)\Big\{
    \sum_y
    \hat p_Y(y) {\rm e}^{- \nu d(x,y) }
    \Big\}^\lambda\\
    &= E_{0,\rm s}^{(-\lambda, \nu )} (\hat p_Y||P_X) . 
\end{align*}
Inequalities (a) and (b) follow from the nonnegativity of the conditional and the ordinary relative entropies. 
\end{IEEEproof}

\begin{IEEEproof}[Proof of Lemma~\ref{lemma22}]
Let the rhs's of (\ref{eq.lemma22.2}) and (\ref{eq.lemma22.3}) be $\tilde q_{X|Y}(p_{Y|X})$
and $\tilde q_{Y} (p_{Y|X})$. Then we have
\begin{align*}
    & J_{{\rm s}, \bm{t}_4}^{(\lambda,\nu)}
    (q_{XY}, p_{Y|X} | P_X ) \\
    & =\mathrm{E}_{q_{XY}} \left[
    \log \frac{q_Y^{1-\lambda}(Y) q_{X|Y}(X|Y) 
    }{P_X(X)\mathrm{e}^{-\lambda\nu d(X,Y)} p_{Y|X}^{1-\lambda}(Y|X) } 
    \right]\\
    & = -(1-\lambda) H(q_Y) + 
    \mathrm{E}_{q_{XY}} \left[
    \log \frac{ q_{X|Y}(X|Y) 
    }{P_X(X)\mathrm{e}^{-\lambda\nu d(X,Y)} p_{Y|X}^{1-\lambda}(Y|X) } 
    \right]\\
    & = -(1-\lambda) H(q_Y) + D(q_{X|Y} || \tilde q_{X|Y}(p_{Y|X}) | q_Y )  \\
&\quad  - \sum_{y} q_Y(y) \log \sum_{x} P_X(x) \mathrm{e}^{-\lambda\nu d(x,y)} 
    p_{Y|X}^{1-\lambda}(y|x)  \\
   &\stackrel{\rm (a)}\geq (1-\lambda)
   \sum_y q_Y(y) 
   \log \frac{q_Y(y)}{
   \{
   \sum_x P_X(x) {\rm e}^{-\lambda \nu d(x,y)} p_{Y|X}^{1-\lambda}(y|x) 
   \}^{1/(1-\lambda)}
   }\\
   &=(1-\lambda) D(q_Y || \tilde q_Y(p_{Y|X}))\\
    &\quad - (1-\lambda) \log \sum_{y}
   \left\{
   \sum_x P_X(x) {\rm e}^{-\lambda\nu d(x,Y)} p_{Y|X}^{1-\lambda}(Y|x) 
   \right\}^{1/(1-\lambda)}\\
   &\stackrel{\rm (b)}\geq - (1-\lambda) \log \sum_{y}
   \left\{
   \sum_x P_X(x) {\rm e}^{-\lambda\nu d(x,Y)} p_{Y|X}^{1-\lambda}(Y|x) 
   \right\}^{1/(1-\lambda)}\\
   &=A_{\rm s}^{(-\lambda, \nu )}( p_{Y|X} | P_X ). 
\end{align*}
As before, inequalities (a) and (b) follow
from the nonnegativity of conditional and ordinary relative entropies.
The equality conditions implies that
minimum of $J_{{\rm s}, \bm{t}_4 }^{(\lambda,\nu)}(q_{XY}, p_{Y|X} | P_X )$ is attained if
both $q_Y=\tilde q_Y(p_{Y|X})$ and
$q_{X|Y} = \tilde q_{X|Y}(p_{Y|X})$ are satisfied. 
\end{IEEEproof}

\begin{IEEEproof}[Proof of Lemma~\ref{lemma15}]
Let the right hand side of (\ref{optimal_q_Lemma17}) as $q_{XY}^* ( p_{XY}, \hat p_{XY} ) $. 
For $\bm{t}\in \mathcal{T}_1 \cap \mathcal{T}_2$, 
we have
\begin{align}
    &\tilde F_{{\rm JO}, \bm{t} }^{(\lambda, \nu)} (q_{XY}, p_{XY}, \hat p_{XY} |W) \notag\\
    &=
   \Theta^{(\lambda, \lambda \nu)} ( q_{XY} |W) + (1-\lambda) ( (1+t_2) D(q_X||p_X) 
   + t_2 D(q_{Y|X}  || p_{Y|X} | q_X) + t_3 D(q_Y|\hat p_Y)) \notag\\
   &\quad + 
   ( (1-\lambda) t_3 + \lambda ) D(q_{X|Y} | \hat p_{X|Y} | q_Y) \notag\\
   & =
   {\rm E}_{q_{XY}} \left[
   \frac{ q_{XY}^{1+(1-\lambda)(t_2+t_3)} (X,Y) } 
   { 
   p_X^{(1-\lambda)(1+t_2)}(X) p_{Y|X}^{(1-\lambda)t_2}(Y|X) 
   \hat p_Y^{(1-\lambda)t_3}(Y) \hat p_{X|Y}^{(1-\lambda)t_3+\lambda}(X|Y) 
   W(Y|X) {\rm e} ^{-\lambda \nu c(X)}
   }
   \right]\\
   &
   = \{1+(1-\lambda)(t_2+t_3)\} \Bigg\{ D(q_{XY}|| q_{XY}^*(p_{XY},\hat p_{XY} ) ) 
   \notag\\
   &-\log
   \sum_{x,y}
   \left\{
      p_X^{(1-\lambda)(1+t_2)}(X) p_{Y|X}^{(1-\lambda)t_2}(Y|X) 
      \hat p_Y^{(1-\lambda)t_3}(Y) \hat p_{X|Y}^{(1-\lambda)t_3+\lambda}(X|Y) W(Y|X) {\rm e} ^{-\lambda \nu c(X)}
   \right\}^{\frac{1}{t+(1-\lambda)(t_2+t_3)}} \Bigg\}\\
   &
   \stackrel{\rm (a)} 
   \geq 
   -\{1+(1-\lambda)(t_2+t_3)\}\notag\\
   &\cdot \log
   \sum_{x,y}
   \left\{
      p_X^{(1-\lambda)(1+t_2)}(X) p_{Y|X}^{(1-\lambda)t_2}(Y|X) p_Y^{(1-\lambda)t_3}(Y) p_{X|Y}^{(1-\lambda)t_3+\lambda}(X|Y) W(Y|X) {\rm e} ^{-\lambda \nu c(X)}
   \right\}^{\frac{1}{t+(1-\lambda)(t_2+t_3)}}\notag\\
   &=-\lambda F_{{\rm AR}, \bm{t}} ^{(-\lambda, \nu)} (p_{XY}, \hat p_{XY} | W) .
\end{align} 
In Step (a), equality holds if and only if $q_{XY}=q_{XY}^*( p_{XY}, \hat p_{XY} )$ holds.
This completes the proof.
\end{IEEEproof}

\section{Proof of Lemma~\ref{lemma:tilde_G_CK}}
\label{appendixD}
In this section, we prove Lemma~\ref{lemma:tilde_G_CK}.

{\it Proof of Lemma~\ref{lemma:tilde_G_CK}:} 
Let $q_{XY}^*$ be
a joint distribution that attains
$G(R, \Delta| P)$.
{From its formula, we have 
\begin{align}
R(\Delta | q_X^*) \leq I(q_X^*, q^*_{Y|X}). 
\label{ineq:lm1}
\end{align}
Thus, }
\begin{align*}
G (R, \Delta|P)
&= |I(q_X^*, q_{Y|X}^*) - R|^+ + D(q_X^* || P)\\
&\stackrel{\rm (a)}{\geq} |R(\Delta | q_X^*) - R|^+ + D(q_X^* || P)\\
& \geq \min_{q_{X} \in \mathcal{P(X)}} \{ |R(\Delta | q_X) - R|^+ + D(q_X || P) \}\\
& = G_{\rm CK} (R, \Delta|P).
\end{align*}
{Step (a) follows from (\ref{ineq:lm1}).}
On the other hand, let $\tilde q_{X}^* $
be a distribution that attains
$G_{\rm CK}(R, \Delta| P)$
and let $\tilde q_{Y|X}^* $ be
a conditional distribution that
attains $R(\Delta|\tilde q_X^*)$. 
Then, we have
\begin{align*}
&G_{\rm CK} (R, \Delta|P) 
= |I(\tilde q_X^*, \tilde q_{Y|X}^*) - R|^+ 
+ D(\tilde q_X^* || P) \\
& \geq \min_{
\genfrac{}{}{0pt}{}{q_{XY}: 
}{ {\rm E}_{ q_{XY} } [ d(X,Y)] \leq \Delta}
}
\{ |I(q_X, q_{Y|X}) - R|^+ + D(q_X || P) \} \\
& = G (R, \Delta|P). 
\end{align*}
Thus, we have 
$ G_{\rm CK} (R, \Delta|P) =
G (R, \Delta|P)$, 
which completes the proof.
\hfill\IEEEQED

\section{Proof of property~\ref{pr:G_CK}} 
\label{appendixE}
In this section, we prove Property~\ref{pr:G_CK}. 
By definition, Part a) is obvious.
For the proof of Part b), 
let $q^{(0)}$ and $q^{(1)}$ be 
joint distribution functions that attain 
$\tilde G_{\rm CK}(R_0, \Delta_0|P)$ and $\tilde G_{\rm CK}(R_1, \Delta_1|P)$,
respectively. 
Denote 
\begin{align}
&\Theta(R, q | P) 
\stackrel{\triangle}{=} |I(q_X, q_{Y|X}) - R|^+ + D(q_X || P) \notag \\
&= \max\{ D(q_X || P),
I(q_X, q_{Y|X}) - R + D(q_X || P) \}.\label{Theta_max}
\end{align}
By definition, we have
\begin{align}
\tilde G_{\rm CK}(R_i, \Delta_i |P) = \Theta(R_i, q^{(i)} |P)
\mbox{ for } i=0,1.
\label{optimal_distribution}
\end{align}
For $\alpha_1 = \alpha \in [0,1]$ and 
$\alpha_0 = 1-\alpha$,
we set 
$R_\alpha = \alpha_0 R_0 + \alpha_1 R_1$, 
$\Delta_\alpha = \alpha_0 \Delta_0 + \alpha_1 \Delta_1$, 
and 
$q^{(\alpha)} = \alpha_0 q^{(0)} + \alpha_1 q^{(1)}$.
By linearity of ${\rm E}_{q}[ d(X,Y)]$ 
with respect to $q$,
we have that
\begin{align}
{{\rm E}_{q^{(\alpha)}} [ d(X, Y) ] }
= \sum_{i=0,1} \alpha_i {{\rm E}_{q^{(i)}} [ d(X, Y) ] }
\leq \Delta_{\alpha}. 
\label{constraint_Delta_alpha}
\end{align}
Because $$I(q_X, q_{Y|X}) + D(q_X||P) = 
\sum_{x,y} q_{XY}(x,y)
\log 
\frac{q_{X|Y}(x|y)}{P(x)}$$
is convex with
respect to $q_{XY}$ and $D(q_X||P)$ is convex with
respect to $q_X$, we have
\begin{align}
& I( q_X^{(\alpha)}, q_{Y|X}^{(\alpha)}) + D(q_X^{(\alpha)} || P) \notag\\
& \leq \sum_{i=0,1} \alpha_i  
\left\{
I( q_X^{(i)}, q_{Y|X}^{(i)}) + D(q_X^{(i)} || P)
\right\},  \label{eq.21a}\\
& D(q_X^{(\alpha)} || P) 
\leq \sum_{i=0,1} \alpha_i 
D(q_X^{(i)} || P).  \label{eq.21b}
\end{align}
Therefore, we have the following two chains
of inequalities:
\begin{align}
& I( q_X^{(\alpha)}, q_{Y|X}^{(\alpha)}) + D(q_X^{(\alpha)} || P) 
  - R_{\alpha} \notag \\
& 
\stackrel{\rm (a)}{\leq} \sum_{i=0,1} \alpha_i  
\left\{
I( q_X^{(i)}, q_{Y|X}^{(i)}) + D(q_X^{(i)} || P)
- R_{i}
\right\} \notag \\
& \stackrel{\rm (b)}{\leq} \sum_{i=0,1} \alpha_i 
\Theta(R_i, q^{(i)} | P), \\
& D(q_X^{(\alpha)} || P) 
\stackrel{\rm (c)}{\leq} 
\sum_{i=0,1} \alpha_i 
D(q_X^{(i)} || P) \notag\\
&\stackrel{\rm (d)}{\leq} 
\sum_{i=0,1} \alpha_i 
\Theta(R_i, q^{(i)} | P).
\end{align}
Steps (a) and (c) follow from (\ref{eq.21a}) and (\ref{eq.21b})
and Steps
(b) and (d) follow from the definition
of $\Theta(R_i, q^{(i)} | P)$ for $i=0,1$. 
Then, from (\ref{Theta_max}) we have
\begin{align}
 \Theta(R_\alpha, q^{(\alpha)} | P) 
{\leq} 
\sum_{i=0,1} \alpha_i 
\Theta(R_i, q^{(i)} | P) {.}
\label{eq.25}
\end{align}
Therefore, 
\begin{align*}
&  \tilde G_{\rm CK}(R_\alpha, \Delta_\alpha | P) 
= 
\min_{
\genfrac{}{}{0pt}{}{q \in \mathcal{P(X\times Y)}: 
}{{\rm E}_{q} [d(X,Y)] \leq \Delta_\alpha
}
}
\Theta(R_\alpha, q | P) \\
&
\stackrel{\rm (a)}{\leq} 
\Theta(R_\alpha, q^{(\alpha)} | P) 
\stackrel{\rm (b)}{\leq} 
\sum_{i=0,1} \alpha_i 
\Theta(R_i, q^{(i)} | P) \\
&
\stackrel{\rm (c)}{=} 
\sum_{i=0,1} \alpha_i 
 \tilde G_{\rm CK}(R_i, \Delta_i | P). 
\end{align*}
Step (a) follows from (\ref{constraint_Delta_alpha}),
Step (b) follows from (\ref{eq.25}), and
Step (c) follows from (\ref{optimal_distribution}).

For the proof of Part c), the choice of 
$q_X=P$ gives $\tilde G_{\rm CK}(R,\Delta|P)=0$, 
if $R\geq R(\Delta|P)$. 
If $R<R(\Delta | P)$, 
the choice of $q_X=P$ makes 
the first term of the objective
function strictly positive, while 
any choice of $q \neq P$, 
$D(q || P)$ is strictly positive. 
This completes the proof of Part c).

For the proof of Part d),
let $q^*$ be a joint distribution that attains
$\tilde G_{\rm CK}(R', \Delta |P)$.
Then,
\begin{align*}
 \tilde G_{\rm CK}(R, \Delta |P) 
\leq & 
| I(q_X^*, q_{Y|X}^*) - R |^+ + D(q_X^* || P)\\
\stackrel{\rm (a)}\leq & (R'-R) + 
| I(q_X^*, q_{Y|X}^*) - R' |^+ \\
& + D(q_X^* || P)\\
= &(R'-R) + \tilde G_{\rm CK}(R', \Delta|P). 
\end{align*}
Step (a) follows from $|x|^+ \leq |x-c|^+ + c $ for $c\geq 0$.
This completes the proof. \hfill\IEEEQED

\section{Proof of Lemma~\ref{lemma17} }
\label{appendixF}
In this section, we give the proof of Lemma~\ref{lemma17}.
To this aim, we define the following functions:
\begin{align}
    \tilde G_{\rm CK}^{(\lambda)} ( R, \Delta | P_X)
    &\stackrel{\triangle}{=}
    \min_{ 
    \genfrac{}{}{0pt}{}
    { q_{XY}: }
    { \mathrm{E}_{q_{XY}} [ d(X,Y)] \leq \Delta }
    }
    \{ D(q_X||P_X) + \lambda [ I( q_X, q_{Y|X} ) - R] \}, 
    \label{eq.appendixF.1}\\
    \tilde G_{\rm CK}^{(\lambda,\mu)} ( R, \Delta | P_X)
    &\stackrel{\triangle}{=}
    \min_{ q_{XY} } 
    \big\{ D(q_X||P_X) + \lambda [I(q_X, q_{Y|X} ) -R] \notag\\
    &\hspace{1cm} +\mu [ {\rm E}_{q_{XY}} [d(X,Y)] -\Delta ] \big\}. 
    \label{eq.appendixF.1.1}
\end{align}
We first prove 
\begin{align}
  \tilde G_{\rm CK}(R,\Delta|P_X) 
  = \max_{\lambda \in [0,1]} 
    \tilde G_{\rm CK}^{(\lambda)} (R, \Delta | P_X)
  \label{eq.appendixF.2}
\end{align}
and then prove 
\begin{align}
 \tilde G_{\rm CK}^{(\lambda)} (R, \Delta | P_X) =
 \sup_{ \mu \ge 0 }  
 \tilde G_{\rm CK}^{(\lambda, \mu)}(R, \Delta | P_X ) .
    \label{eq.appendixF.3}
\end{align}
Eqs.(\ref{eq.appendixF.2}) and (\ref{eq.appendixF.3}) imply  
Eq.(\ref{eq.lemma17.3}). 
The function $G_{\rm CK}^{(\lambda)}(R, \Delta|P)$
satisfies the following property: \

\begin{property}
\label{pr:G_CK_lambda}
\

\begin{itemize}
\item[a)] 
$\tilde G_{\rm CK}^{(\lambda)}(R, \Delta|P)$ is a monotone decreasing
function of $R \geq 0$
for a fixed $\Delta \geq 0$ 
and is a monotone decreasing function of $\Delta \geq 0$
for a fixed $R\geq 0$.
\item[b)] $G_{\rm CK}^{(\lambda)}(R, \Delta|P)$ is a convex function of $(R,\Delta)$.
\end{itemize}
\end{property}

\textit{Proof of Property \ref{pr:G_CK_lambda}:}
By definition, Part a) is obvious.
For the proof of Part b), 
choose $R_0, R_1, \Delta_0, \Delta_1\geq 0$ arbitrary. 
Let $q_{XY}^{(0)}$ and $q_{XY}^{(1)}$ be 
joint distribution functions that attain 
$G_{\rm CK}^{(\lambda)}(R_0, \Delta_0|P)$ and 
$G_{\rm CK}^{(\lambda)}(R_1, \Delta_1|P)$,
respectively. 
Denote 
\begin{align}
&\Theta_{\rm s}^{(\lambda)} (q_{XY} | P_X) 
=  D(q_X || P_X) + \lambda I(q_X, q_{Y|X}) .
\label{Theta_lambda}
\end{align}
By definition, we have
\begin{align}
G_{\rm CK}^{(\lambda)} (R_i, \Delta_i |P_X) 
= 
\Theta_{\rm s}^{(\lambda)} ( q_{XY}^{(i)} |P_X) - \lambda R_i
\mbox{ for } i = 0, 1.
\label{optimal_distribution_lambda}
\end{align}
For $\alpha_1 = \alpha \in [0,1]$ and 
$\alpha_0 = 1-\alpha$,
we set 
$R_\alpha = \alpha_0 R_0 + \alpha_1 R_1$, 
$\Delta_\alpha = \alpha_0 \Delta_0 + \alpha_1 \Delta_1$, 
and 
$q_{XY}^{(\alpha)} = \alpha_0 q_{XY}^{(0)} + \alpha_1 q_{XY}^{(1)}$.
By linearity of 
$ 
{ {\rm E}_{q_{XY}}[d(X,Y)] }
$ with respect to $q_{XY}$,
we have that
\begin{align}
{ {\rm E}_{q_{XY}^{(\alpha)} } [ d(X, Y) ] }
= \sum_{i=0,1} \alpha_i 
{ {\rm E}_{q_{XY}^{(i)}} [ d(X, Y) ] }
\leq \Delta_{\alpha} . 
\label{constraint_Delta_alpha_lambda}
\end{align}
We also have
\begin{align}
&\Theta_{\rm s}^{(\lambda)} ( q_{XY} | P_X ) \notag \\
& =  
\lambda {\rm E}_{q_{XY}} [ \log q_{X|Y}(X|Y) ]
-{\rm E}_{q_{XY}} [\log P_X(X)]
-(1-\lambda) H(q_X). \label{eq.appendixF.4}
\end{align}
The second term is a linear function of $q_{XY}$ and
the third term is a convex function of $q_X$
and thus a convex function of $q_{XY}$ because 
$q_X$ is a linear function of $q_{XY}$. 
For the first term, we have 
\begin{align}
    & \sum_{x,y} q_{XY}^{(\alpha)}(x,y) \log q_{X|Y}^{(\alpha)} (x|y) \notag \\
    &= \sum_{x,y} q_{XY}^{(\alpha)}(x,y) \log \frac{q_{XY}^{(\alpha)} (x,y)}{q_Y^{(\alpha)}(y)}\notag \\ 
    &= \sum_{x,y} ( \alpha_0 q_{XY}^{(0)}(x,y)+\alpha_1 q_{XY}^{(1)}(x,y) ) 
    \log \frac{ \alpha_0 q_{XY}^{(0)}(x,y)+\alpha_1 q_{XY}^{(1)}(x,y)  }{ \alpha_0 q_Y^{(0)}(y) + \alpha_1 q_Y^{(1)}(y) }\notag \\ 
    &\stackrel{\rm (a)}\leq 
    \alpha_0 \sum_{x,y} q_{XY}^{(0)}(x,y) \log \frac{ \alpha_0 q_{XY}^{(0)}(x,y) }{\alpha_0 q_{Y}^{(0)}(y)} \notag \\
    &\quad +
    \alpha_1 \sum_{x,y} q_{XY}^{(1)}(x,y) \log \frac{ \alpha_1 q_{XY}^{(1)}(x,y) }{\alpha_1 q_{Y}^{(1)}(y)} \notag \\
    &= \alpha_0 q_{XY}^{(0)}(x,y) \log q_{X|Y}^{(0)} (x|y)
    + \alpha_1 q_{XY}^{(1)}(x,y) \log q_{X|Y}^{(1)} (x|y), 
    \label{eq.appendixF.5}
\end{align}
where step (a) follows from the log-sum inequality. 
Inequality (\ref{eq.appendixF.5}) shows that 
the first term of (\ref{eq.appendixF.4}) is convex with respect to $q_{XY}$
and so is the whole of (\ref{eq.appendixF.4}).
Therefore
\begin{align*}
&  G_{\rm CK}^{(\lambda)} (R_\alpha, \Delta_\alpha | P) 
= 
\min_{
\genfrac{}{}{0pt}{}{
q_{XY} \in \mathcal{P(X\times Y)}: 
}{
{ {\rm E}_{q} [d(X,Y)] }
\leq \Delta_\alpha
}
}
\Theta_{\rm s}^{(\lambda)} (q_{XY} | P) - \lambda R_\alpha\\
&
\stackrel{\rm (a)}{\leq} 
\Theta_{\rm s}^{(\lambda)} (q_{XY}^{(\alpha)} | P)  - \lambda R_\alpha
\stackrel{\rm (b)}{\leq} 
\sum_{i=0,1} \alpha_i \{ 
\Theta_{\rm s}^{(\lambda)} (q_{XY}^{(i)} | P) -\lambda R_i \} \\
&
\stackrel{\rm (c)}{=} 
\sum_{i=0,1} \alpha_i 
 \tilde G_{\rm CK}^{(\lambda)} (R_i, \Delta_i | P). 
\end{align*}
Step (a) follows from (\ref{constraint_Delta_alpha_lambda}),
Step (b) follows from the convexity of $\Theta_{\rm s}^{(\lambda)}(q_{XY}|P_X)$
with respect to $q_{XY}$, and
Step (c) follows from (\ref{optimal_distribution_lambda}).
This completes the proof.\hfill\IEEEQED

This property is used to prove (\ref{eq.appendixF.3}).

{\it Proof of Lemma~\ref{lemma17}}
We first prove (\ref{eq.appendixF.2}).
For any $\lambda \in [0,1]$, 
we have $|x|^+ \geq \lambda x$.
{Let $\hat q_{XY}$ be a joint distribution 
that attains $\tilde G_{\rm CK}(R, \Delta | P)$. }
Then, we have 
\begin{align*}
\tilde G_{\rm CK}(R, \Delta | P) 
= & 
{
D(\hat q_X||P)  + | I(\hat q_X,\hat q_{Y|X} ) - R |^+  
}\\
\geq & 
{
D(\hat q_X||P)  + \lambda [ I(\hat q_X,\hat q_{Y|X} ) - R ] 
}\\
\geq & 
\min_{
\genfrac{}{}{0pt}{}{q_{XY} \in \mathcal{P(X\times Y)}:}{{ {\rm E}_{q_{XY}} [ d(X,Y) ] }
\leq \Delta 
}
}
\big\{
D(q_X||P) + \lambda [ I(q_X, q_{Y|X} ) - R ] 
\big\} \\
= & \tilde G_{\rm CK}^{(\lambda)}(R, \Delta | P) {.}
\end{align*}
Thus, 
$$ 
\tilde G_{\rm CK}(R, \Delta | P) 
\geq 
\max_{0\leq \lambda \leq 1}
\tilde G_{\rm CK}^{(\lambda)}(R, \Delta | P). 
$$
Hence, it is sufficient to show that 
there exists a $\lambda \in [0,1]$
such that 
$ \tilde G_{\rm CK} (R, \Delta | P)
\leq 
\tilde G_{\rm CK}^{(\lambda)}(R, \Delta | P)
$.
From Property \ref{pr:G_CK}, 
there exists a $\lambda \in [0,1]$ such that
for any $R'\geq 0$ we have
\begin{align}
 \tilde G_{\rm CK} (R', \Delta | P)
\geq 
\tilde G_{\rm CK} (R, \Delta | P)
- \lambda (R'-R) {.}
\label{eq:G_CK}
\end{align}
Fix the above $\lambda$. 
Let $q^*$ be a joint distribution
that attains
$ \tilde G_{\rm CK} ^{(\lambda)}(R, \Delta | P)$.
Set $R' = I(q_X^*, q_{Y|X}^*)$.
Then we have
\begin{align}
& \tilde G_{\rm CK}(R, \Delta | P) \notag \\
& \stackrel{\rm (a)}\leq 
 \tilde G_{\rm CK}(R', \Delta | P) + \lambda (R'-R) \notag \\
&=  
\min_{
\genfrac{}{}{0pt}{}{
q_{XY}:}{
{ {\rm E}_{q} [d(X,Y)] }
\leq \Delta 
}
} 
 \big\{ 
 D(q_X||P) + | I(q_X, q_{Y|X}) - R' |^+ \big\} \notag \\
& \hspace{1cm } 
+ \lambda (R'-R) \notag \\
& \leq 
D(q_X^*||P) + | I(q_X^*, q_{Y|X}^*) - R' |^+ + \lambda (R'-R) \notag \\
& \stackrel{\rm (b)}= 
 \lambda [ I(q_X^*, q_{Y|X}^*) - R ] + D(q_X^*||P) \notag \\
&= 
 \tilde G_{\rm CK} ^{(\lambda)}(R, \Delta | P). 
\end{align}
Step (a) follows from
(\ref{eq:G_CK}) and Step (b) comes
from the choice of $R' = I(q_X^*, q_{Y|X}^*)$.
Therefore, there exists a $0 \leq \lambda \leq 1$
such that 
$\tilde G_{\rm CK}^{(\mu)}(R, \Delta | P) = 
\tilde G_{\rm CK}^{(\lambda, \mu)}(R, \Delta | P) $. 

Next, we prove (\ref{eq.appendixF.3}). 
From its formula, it is obvious that
$$
\tilde G_{\rm CK}^{(\lambda)}(R, \Delta|P) 
\geq \max_{\mu \geq 0}
\tilde G_{\rm CK}^{(\lambda, \mu)}(R, \Delta|P). 
$$
Hence, it is sufficient to show that
for any $R\geq 0$ and $\Delta\geq 0$, 
there exists $\mu\geq 0$ such that
\begin{align}
\tilde G_{\rm CK}^{(\lambda)}(R, \Delta|P) 
\leq 
\tilde G_{\rm CK}^{(\lambda, \mu)}(R, \Delta|P). 
\label{ineq_G_CK_mu_lambda}
\end{align}
From Property \ref{pr:G_CK_lambda} part a) and b), $G^{(\lambda)}(R, \Delta|P)$
is a monotone decreasing and convex function of $\Delta \geq 0$ 
{for a 
fixed $R$}. Thus,
there exists $\mu \geq 0$ such that for any $\Delta'\geq 0$,
the following inequality holds:
\begin{align}
  \tilde G_{\rm CK}^{(\lambda)}(R, \Delta'|P) \geq 
  \tilde G_{\rm CK}^{(\lambda)}(R, \Delta|P) 
  - \mu (\Delta' - \Delta) {.}
  \label{convexity_G_CK_lambda}
\end{align}
Fix the above $\mu$. Let $q^*$ be
a joint distribution that attains
$\tilde G_{\rm CK}^{(\mu, \lambda)}(R, \Delta|P)$.
Set $\Delta' = 
{ {\rm E}_{q^*} [ d(X,Y) ] }
$.
Then, we have 
\begin{align*}
& \tilde G_{\rm CK}^{(\lambda)}(R, \Delta|P) 
  \stackrel{\rm (a)}{\leq} 
  \tilde G_{\rm CK}^{(\lambda)}(R, \Delta'|P) 
  - \mu (\Delta - \Delta') \\
& {= 
  \min_{
  \genfrac{}{}{0pt}{}{q_{XY}:}{{ {\rm E}_{q_{XY}} [ d(X,Y) ] }
\leq \Delta'
}
}
\{ D(q_X||P) + 
\lambda [ I(q_X, q_{Y|X}) - R ] \} }\\
&\hspace{5mm} {- \mu (\Delta - \Delta'  ) }
\\
& \stackrel{\rm (b)}\leq 
  \lambda [ D(q_X^*||P) + I(q_X^*, q_{Y|X}^*) - R ] 
  - \mu (\Delta - 
{ {\rm E}_{q_{XY}^*} [ d(X,Y) ] }
) \\
& = \tilde G_{\rm CK}^{(\lambda, \mu)}(R, \Delta|P). 
\end{align*}
Step (a) follows from (\ref{convexity_G_CK_lambda})
and Step (b) follows from the definition of
$\tilde G^{{(\lambda)}}(R, \Delta'|P)$
and the choice of $\Delta' = 
{ {\rm E}_{q^*} [ d(X,Y) ] }
$. 
Thus, for any $\Delta\geq 0$, we have (\ref{ineq_G_CK_mu_lambda}) 
for some $\mu \geq 0$.
This completes the proof. \hfill\IEEEQED

\section{Proof of Lemma~\ref{lemma:two_conditions_lossy}
}
\label{section:update_rule_of_generalized_algorithm_source_coding}

This section gives the proof of Lemma~\ref{lemma:two_conditions_lossy}. 

\begin{IEEEproof}
Substituting (\ref{Theta_s}) and (\ref{def:D_t}) into (\ref{def:J_t_s}), we have
\begin{align}
& J_{{\rm s}, \bm{t}}^{(\lambda, \nu)} ( q_{XY}, p_{XY} | P_X) \notag\\
& = {\rm E}_{ q_{XY} }
\left[
\log 
\frac{q_X^{ (1-\lambda) ( 1 + t_1 ) }(X) q_{Y|X}^{(1-\lambda) t_2 }(Y|X) 
q_Y^{ (1-\lambda) t_3 }(X) q_{X|Y}^{ \lambda + (1-\lambda) t_4 }(X|Y) }{
P(X) {\rm e}^{-\lambda \nu d(X,Y) 
}  
p_X^{ (1-\lambda) t_1 }(X) p_{Y|X}^{ (1-\lambda) t_2 }(Y|X) 
p_Y^{ (1-\lambda) t_3 }(Y) p_{X|Y}^{ (1-\lambda) t_4 }(X|Y)
}
\right]
\end{align}

First assume $ \bm{t} \in \mathcal{T}_3 $, i.e., $t_3 = t_4 + \frac{\lambda}{1-\lambda}$. 
We have 
\begin{align}
& J_{{\rm s}, \bm{t}}^{(\lambda, \nu)} ( q_{XY}, p_{XY} | P_X) \notag\\
& = {\rm E}_{ q_{XY} }
\left[
\log 
\frac{q_X^{ 1 + (1-\lambda) ( t_1 + t_4 ) }(X) q_{Y|X}^{ \lambda + (1-\lambda) (t_2 + t_4) }(Y|X) 
}{
P(X) {\rm e}^{-\lambda \nu d(X,Y) 
}  
p_X^{ (1-\lambda) (t_1+t_4) }(X) p_{Y|X}^{ (1-\lambda) (t_2+t_4) }(Y|X) 
p_Y^{ \lambda }(Y) 
}
\right] \notag \\
& = 
{\rm E}_{ q_{X} }
\left[
\log 
\frac{q_X^{ 1 + (1-\lambda) ( t_1 + t_4 ) }(X) }
{P(X) p_X^{ (1-\lambda) (t_1+t_4) }(X) }
\right]
+
{\rm E}_{ q_{XY} }
\left[
\log 
\frac{
q_{Y|X}^{ \lambda + (1-\lambda) (t_2 + t_4) }(Y|X) 
}{
{\rm e}^{-\lambda \nu d(X,Y) }  
p_{Y|X}^{ (1-\lambda) (t_2+t_4) }(Y|X) 
p_Y^{ \lambda }(Y) 
}
\right]
\end{align}
The first term is independent of $q_{Y|X}$. 
On the second term, we have
\begin{align}
& \min_{ q_{Y|X} }     
{\rm E}_{ q_{XY} }
\left[
\log 
\frac{
q_{Y|X}^{ \lambda + (1-\lambda) (t_2 + t_4) }(Y|X) 
}{
{\rm e}^{-\lambda \nu d(X,Y) }  
p_{Y|X}^{ (1-\lambda) (t_2+t_4) }(Y|X) 
p_Y^{ \lambda }(Y) 
}
\right] \notag\\
& =
( \lambda + (1-\lambda) (t_2 + t_4) )
\min_{q_{Y|X}} 
{\rm E}_{ q_{XY} }
\left[
\log 
\frac{
q_{Y|X}(Y|X) 
}{
\{
{\rm e}^{-\lambda \nu d(X,Y) }  
p_{Y|X}^{ (1-\lambda) (t_2+t_4) }(Y|X) 
p_Y^{ \lambda }(Y) 
\}^{\frac{1}{\lambda + (1-\lambda) (t_2 + t_4) } }
}
\right]  \notag\\
&=
- ( \lambda + (1-\lambda) (t_2 + t_4) ) \sum_{x} q_X(x) 
\log 
\sum_{y}
\{
{\rm e}^{-\lambda \nu d(x,y) }  
p_{Y|X}^{ (1-\lambda) (t_2+t_4) }(y|x) 
p_Y^{ \lambda }(y) 
\}^{\frac{1}{\lambda + (1-\lambda) (t_2 + t_4) } }
\end{align}
The minimum is attained if and only if
\begin{align}
q_{Y|X}(y|x) =    
    \frac1{K_x} 
\{
{\rm e}^{-\lambda \nu d(x,y) }  
p_{Y|X}^{ (1-\lambda) (t_2+t_4) }(y|x) 
p_Y^{ \lambda }(y) 
\}^{\frac{1}{\lambda + (1-\lambda) (t_2 + t_4) } },
\end{align}
where $K_x =
\sum_{y} 
\{
{\rm e}^{-\lambda \nu d(x,y) }  
p_{Y|X}^{ (1-\lambda) (t_2+t_4) }(y|x) 
p_Y^{ \lambda }(y) 
\}^{\frac{1}{\lambda + (1-\lambda) (t_2 + t_4) } }
$. 
Finally, we have
\begin{align}
    & \min_{q_X} \min_{q_{Y|X}} 
    J_{{\rm s}, \bm{t}}^{(\lambda, \nu)} ( q_{XY}, p_{XY} | P_X) \notag\\
    & = \min_{q_X} \Bigg\{ 
    {\rm E}_{ q_{X} }
    \left[
    \log 
    \frac{q_X^{ 1 + (1-\lambda) ( t_1 + t_4 ) }(X) }
    {P(X) p_X^{ (1-\lambda) (t_1+t_4) }(X) }
    \right] \notag\\
    &\quad  - \sum_{x} q_X(x) 
    \log \left[
    \sum_{y}
    \{
    {\rm e}^{-\lambda \nu d(x,y) }  
    p_{Y|X}^{ (1-\lambda) (t_2+t_4) }(y|x) 
    p_Y^{ \lambda }(y) 
    \}^{\frac{1}{\lambda + (1-\lambda) (t_2 + t_4) } }
    \right]^{ \lambda + (1-\lambda) (t_2 + t_4) }
    \Bigg\} \notag \\
    & = \min_{q_X} 
    \sum_{x} q_X(x)
    \log 
    \frac{q_X^{ 1 + (1-\lambda) ( t_1 + t_4 ) }(x) }
    {P(x) p_X^{ (1-\lambda) (t_1+t_4) }(x) \left[
    \sum_{y}
    \{
    {\rm e}^{-\lambda \nu d(x,y) }  
    p_{Y|X}^{ (1-\lambda) (t_2+t_4) }(y|x) 
    p_Y^{ \lambda }(y) 
    \}^{\frac{1}{\lambda + (1-\lambda) (t_2 + t_4) } }
    \right]^{ \lambda + (1-\lambda) (t_2 + t_4) }}
    \notag\\
    &= - \{ 1+ (1-\lambda)(t_1 + t_4) \} \log \sum_{x} 
    \left\{ 
    P_X(x)
    p_X^{ (1-\lambda) (t_1+t_4) } (x) 
    \right\}^{\frac{1}{ 1+(1-\lambda)(t_1+t_4) } } \notag\\
    & \quad \cdot \left[
    \sum_{y}
    \left\{
    {\rm e}^{-\lambda \nu d(x,y)}
    p_{Y|X}^{ (1-\lambda) (t_2+t_4) }(y|x)
    p_Y^{ \lambda }(y)
    \right\}^{ \frac{1}{ \lambda + (1-\lambda) (t_2 + t_4) } }
    \right]^{ \frac{ \lambda + (1-\lambda) (t_2+t_4) }{ 1 + (1-\lambda) (t_1 + t_4) } }
    \label{eq.245}
\end{align}
The minimum is attained if and only if
\begin{align}
    q_X(x) = \frac{1}{L} 
    \left\{ 
    P_X(x)
    p_X^{ (1-\lambda) (t_1+t_4) } (x) 
    \right\}^{\frac{1}{ 1+(1-\lambda)(t_1+t_4) } } \left[
    \sum_{y}
    \left\{
    {\rm e}^{-\lambda \nu d(x,y)}
    p_{Y|X}^{ (1-\lambda) (t_2+t_4) }(y|x)
    p_Y^{ \lambda }(y)
    \right\}^{ \frac{1}{ \lambda + (1-\lambda) (t_2 + t_4) } }
    \right]^{ \frac{ \lambda + (1-\lambda) (t_2+t_4) }{ 1 + (1-\lambda) (t_1 + t_4) } },
\end{align}
where $L$ is a normalization factor given by
$$ L= \sum_{x}   \left\{ 
    P_X(x)
    p_X^{ (1-\lambda) (t_1+t_4) } (x) 
    \right\}^{\frac{1}{ 1+(1-\lambda)(t_1+t_4) } } \left[
    \sum_{y}
    \left\{
    {\rm e}^{-\lambda \nu d(x,y)}
    p_{Y|X}^{ (1-\lambda) (t_2+t_4) }(y|x)
    p_Y^{ \lambda }(y)
    \right\}^{ \frac{1}{ \lambda + (1-\lambda) (t_2 + t_4) } }
    \right]^{ \frac{ \lambda + (1-\lambda) (t_2+t_4) }{ 1 + (1-\lambda) (t_1 + t_4) } }. $$

Next, assume $\bm{t} \in \mathcal{T}_4$, i.e., $t_2 = t_1 + 1$. 
The derivation is analogous to the case of $\bm{t} \in \mathcal{T}_3$
and therefore details are omitted. We describe the results here.
\begin{align}
    & \min_{q_Y} \min_{q_{X|Y}} J_{{\rm s}, \bm{t}}^{(\lambda, \nu)}(q_{XY}, p_{XY} | P_X) \notag\\
    & =
    -(1-\lambda) ( 1 + t_1 + t_3) \log \sum_{y}
    p_Y^{\frac{t_1+t_3}{1+t_1 +t_3 }} (y) \notag\\
    &\quad \cdot \left[ \sum_x
    \left\{
        P_X(x) {\rm e}^{-\lambda \nu d(x,y)}
        p_{Y|X}^{1-\lambda}(y|x) 
        p_{X|Y}^{(1-\lambda)(t_1+t_4)}(x|y)
    \right\}^{\frac{1}{1+(1-\lambda)(t_1+t_4)}}
    \right]^{\frac{1+(1-\lambda)(t_1+t_4)}{(1-\lambda)(1+t_1+t_3)} }
    \label{eq.248}.
\end{align}
The minimum is attained if and only if 
\begin{align}
    q_{X|Y}(x|y) 
    &= \frac{1}{K_y'} 
    \left\{
    P_X(x) {\rm e}^{-\lambda \nu d(x,y)} p_{Y|X}^{1-\lambda}(y|x) 
    p_{X|Y}^{(1-\lambda)(t_1+t_4)}(x|y)
    \right\}^{\frac{1}{1+(1-\lambda)(t_1+t_4)}}, \notag\\\
    q_Y(y) 
    &= \frac{1}{L'}
    p_Y^{\frac{t_1+t_3}{1+t_1+t_3} }(y) 
    \left\{
    \sum_x
    \left[
    P_X(x) {\rm e}^{-\lambda \nu d(x,y)} p_{Y|X}^{1-\lambda}(y|x) 
    p_{X|Y}^{(1-\lambda)(t_1+t_4)}(x|y)
    \right]^{\frac{1}{1+(1-\lambda)(t_1+t_4)} }
    \right\}^{\frac{1+(1-\lambda)(t_1+t_4) }{ (1-\lambda)(1+t_1+t_3) } }, 
\end{align}
where $K'_x$ and $L'$ are the normalization factors defined by
\begin{align}
    K_y' &= \sum_{x}\left\{
    P_X(x) {\rm e}^{-\lambda \nu d(x,y)} p_{Y|X}^{1-\lambda}(y|x) 
    p_{X|Y}^{(1-\lambda)(t_1+t_4)}(x|y)
    \right\}^{\frac{1}{1+(1-\lambda)(t_1+t_4)}} ,
    \notag \\
    L' &= \sum_{y} 
    p_Y^{\frac{t_1+t_3}{1+t_1+t_3} }(y) 
    \left\{
    \sum_x
    \left[
    P_X(x) {\rm e}^{-\lambda \nu d(x,y)} p_{Y|X}^{1-\lambda}(y|x) 
    p_{X|Y}^{(1-\lambda)(t_1+t_4)}(x|y)
    \right]^{\frac{1}{1+(1-\lambda)(t_1+t_4)} }
    \right\}^{\frac{1+(1-\lambda)(t_1+t_4) }{ (1-\lambda)(1+t_1+t_3) } }. 
\end{align}
\end{IEEEproof}

\section{Proof of Theorem{~\ref{theorem.convergence.family.algorithm.source} } }
\label{appendix.convergence.family.algorithm.source}
This section gives the proof of Theorem~\ref{theorem.convergence.family.algorithm.source}.
The outline of the proof is similar to the proof of Theorem 2 in~\cite{Tridenski2020arXiv}.
In order to prove Theorem~\ref{theorem.convergence.family.algorithm.source},
we use the following lemma.
\begin{lemma}
\label{lemma36}
For a given $P_X\in \mathcal{P(X)}$, 
Let $q_{XY}^* $ and $p_{XY}^{[0]}$ 
be arbitrary joint distributions satisfying 
$D(q_{X}^* || P_X ) + D_{\bm{t}} ( q_{XY}^* || p_{XY} ^{[0]} ) <+\infty$.
Fix $\bm{t}\in \mathcal{T}$ arbitrarily. 
Let $q_{XY}^{[0]}$ and $p_{XY}^{[1]}$ be joint distribution obtained by Algorithm~\ref{alg:family_lossy_source_coding}
with parameter $\bm{t}$.
Then, we have 
\begin{align}
    & J_{{\rm s}, \bm{t}}^{(\lambda, \nu)} ( q_{XY}^{[0]}, p_{XY}^{[0]} | P_X )
    -    
    J_{{\rm s}, \bm{t}}^{(\lambda, \nu)} ( q_{XY}^*, q_{XY}^* | P_X )\notag\\
    & \leq
    (1-\lambda) \{
      D_{ \bm{t} } (q_{XY}^* , p_{XY}^{[0]}) 
    - D_{ \bm{t} } (q_{XY}^* , p_{XY}^{[1]}) 
    \}.  \label{eq.254}
\end{align}

\textit{Proof:}
Let $q_{XY}^{(s)} = (1-s) q_{XY}^{[0]} + s q_{XY}^*$ for $s\in [0,1]$
and put
$g(s) = J_{{\rm s}, \bm{t}}^{(\lambda, \nu)} (q_{XY}^{(s)}, p_{XY}^{[0]}|P_X)$.
Because of the definition, 
$J_{{\rm s}, \bm{t}}^{(\lambda, \nu)} (q_{XY}, p_{XY}^{[0]}|P_X)$
is minimized by $q_{XY} = q_{XY}^{[0]}$ for a fixed $p_{XY}^{[0]}$. 
Because $g(s)$ is a continuous function of $s\in [0,1]$, we have
\begin{align}
    \left.\frac{\mathrm{d} g(s)}{\mathrm{d} s}\right\lvert_{s\to 0+} \geq 0.
\end{align}
Then we evaluate the derivative of $g(s)$. To this aim, we evaluate
the derivatives of $ \Theta_{\rm s}^{(\lambda, \lambda \nu)}(q_{XY}^{(s)} |P_X) $ 
and $ D_{\bm{t}}(q_{XY}^{(s)}, p_{XY}^{[0]})$ as functions of $s$.
We have
\begin{align}
    & \left.\frac{\mathrm{d}}{\mathrm{d} s} \Theta_{\rm s}^{(\lambda, \lambda \nu)}(q_{XY}^{(s)} |P_X)\right\lvert_{s\to 0+} \notag \\
    &=
    \left.\frac{\mathrm{d}}{\mathrm{d} s} 
    \left( \sum_{x,y} q_{XY}^{(s)}(x,y) \log 
    \frac{q_X^{ (s) }(x)^{1-\lambda} q_{X|Y} ^{ (s) }(x|y)^{\lambda} }{P_X(x) {\rm e}^{-\mu d(x,y) } }
    \right)
    \right\lvert_{s\to 0+}\notag\\
    &=
    \sum_{x,y} 
    \left.
    \left( \frac{\mathrm{d}}{\mathrm{d} s} 
        q_{XY}^{(s)}(x,y) \right)
    \log 
    \frac{q_X^{ (s) }(x)^{1-\lambda} q_{X|Y} ^{ (s) }(x|y)^{\lambda} }{P_X(x) {\rm e}^{-\mu d(x,y) } }
    \right\lvert_{s\to 0+} 
\notag\\
    &\quad +(1-\lambda)
    \sum_{x,y} 
    \left. 
    q_{XY}^{(s)}(x,y) 
    \frac{1}{q_X^{ (s) }(x)}
    \left( 
    \frac{\mathrm{d} }{\mathrm{d} s}
    q_X^{ (s) }(x)
    \right)
    \right\lvert_{s\to 0+}\notag\\
    &\quad +\lambda
    \left.
    \sum_{x,y} q_{XY}^{(s)}(x,y) 
    \frac{1}{q_{X|Y} ^{ (s) }(x|y)}
    \left( 
    \frac{\mathrm{d}}{\mathrm{d} s} 
    q_{X|Y} ^{ (s) }(x|y)
    \right)
    \right\lvert_{s\to 0+}\notag\\
    &=
    \sum_{x,y} 
    \left(  
        q_{XY}^*(x,y) - q_{XY}^{[0] }(x,y) \right)
    \log 
    \frac{q_X^{ [0] }(x)^{1-\lambda} q_{X|Y} ^{ [0] }(x|y)^{\lambda} }{P_X(x) {\rm e}^{-\mu d(x,y) } }
\notag\\
    &\quad +(1-\lambda)
    \sum_{x,y} 
    q_{XY}^{ [0] }(x,y) 
    \frac{1}{q_X^{ [0] }(x)}
    \left( 
    q^*_{X}(x) - q_{X}^{[0] }(x)
    \right)
    \notag\\
    &\quad +\lambda
    \sum_{x,y} q_{XY}^{ [0] }(x,y) 
    \frac{1}{q_{X|Y} ^{ [0] }(x|y)}
    \frac{q^*_Y(y)}{ q_Y^{[0]} (y) }
    \left(
    q^*_{X|Y} (x|y)
    -
    q_{X|Y} ^{ [0] }(x|y)
    \right)
    \notag\\
    &=
    \sum_{x,y} 
        q_{XY}^*(x,y) 
    \log 
    \frac{q^*_X(x)^{1-\lambda} q^*_{X|Y} (x|y)^{\lambda} }{P_X(x) {\rm e}^{-\mu d(x,y) } }
\notag\\
    &\quad -\sum_{x,y} 
        q_{XY}^*(x,y) 
    \log 
    \frac{q^*_X(x)^{1-\lambda} q^*_{X|Y}(x|y)^{\lambda} }{ q_X^{ [0] }(x)^{1-\lambda} q_{X|Y} ^{ [0] }(x|y)^{\lambda} }
\notag\\
    &\quad - \sum_{x,y} 
        q_{XY}^{[0]}(x,y) 
    \log 
    \frac{q_X^{ [0] }(x)^{1-\lambda} q_{X|Y} ^{ [0] }(x|y)^{\lambda} }{P_X(x) {\rm e}^{-\mu d(x,y) } }
\notag\\
&=
\Theta_{\rm s }^{(\lambda, \lambda\nu)}(q_{XY}^*|P_X)
- \Theta_{\rm s }^{(\lambda, \lambda\nu)}(q_{XY}^{[0]}|P_X)
- (1-\lambda) D(q^*_X||q_X^{[0]}) - \lambda D(q^*_{X|Y} || q_{X|Y}^{[0]} | q^*_Y). 
\end{align}
For evaluating the derivative of 
$ D_{ \bm{ t } }(q_{XY}^{(s)}, p_{XY}^{[0]}) $, we have to evaluate the derivative
of four types of divergences. Only the derivative of $D(q_{Y|X}^{(s)}|| p_{Y|X}^{[0]} | q_X^{(s)})$ is shown here. \begin{align}
&  \left.\frac{\rm d}{{\rm d}s} D(q_{Y|X}^{(s)} || p_{Y|X} ^{[0]} | q_{X}^{(s)} ) \right\lvert_{s\to 0+} \notag \\
&= \left.\frac{\rm d}{{\rm d}s} {\rm E}_{q_{XY}^{(s)}} \left[ \log 
\frac{q_{Y|X}^{(s)}(Y|X) }{ p_{Y|X} ^{[0]}(Y|X) }  
\right]
 \right\rvert_{s\to 0+} \notag \\
&=
\left.
\sum_{x,y} \left(
\frac{\rm d}{{\rm d}s} 
q_{XY}^{(s)}(x,y)
\right)
\log \frac{q_{Y|X}^{(s)}(y|x) }{ p_{Y|X} ^{[0]}(y|x) }  
\right\rvert_{s\to 0+} \notag\\
&\quad +
\left.
\sum_{x,y} 
q_{XY}^{(s)}(x,y)
\left(
\frac{\rm d}{{\rm d}s} 
\log q_{Y|X}^{(s)}(y|x)
\right)
 \right\lvert_{s\to 0+} \notag\\
&=
\left.
\sum_{x,y} (
q_{XY}^*(x,y) - q_{XY}^{[0]}(x,y) 
)
\log \frac{q_{Y|X}^{(s)}(y|x) }{ p_{Y|X} ^{[0]}(y|x) }  
\right\rvert_{s\to 0+} \notag\\
&\quad +
\left.
\sum_{x,y} 
q_{XY}^{(s)}(x,y)
\frac{1}{q_{Y|X}^{(s)}(y|x)
}
\left(
\frac{\rm d}{{\rm d}s} 
q_{Y|X}^{(s)}(y|x)
\right)
 \right\lvert_{s\to 0+} \notag\\
&=
\sum_{x,y} (
q_{XY}^*(x,y) - q_{XY}^{[0]}(x,y) 
)
\log \frac{q_{Y|X}^{[0]}(y|x) }{ p_{Y|X} ^{[0]}(y|x) }  
\notag\\
&\quad +
\sum_{x,y} 
q_{XY}^{[0]}(x,y)
\frac{1}{q_{Y|X}^{[0]}(y|x)
}
\left(
q^*_{Y|X}(y|x) - q_{Y|X}^{[0]}(y|x) 
\right)
\notag\\
&=
\sum_{x,y} (
q_{XY}^*(x,y) - q_{XY}^{[0]}(x,y) 
)
\log \frac{q_{Y|X}^{[0]}(y|x) }{ p_{Y|X} ^{[0]}(y|x) }  
\notag\\
& 
=
D(q^*_{Y|X} || p_{Y|X}^{[0]} | q^*_X)
- 
D(q^*_{Y|X} || q_{Y|X}^{[0]} | q^*_X)
-
D(q_{Y|X}^{[0]} || p_{Y|X}^{[0]} | q^*_X).
\end{align}
Other terms are derived in a similar way.
Thus, we have 
\begin{align}
    & \left.\frac{\mathrm{d} }{\mathrm{d} s} 
    D_{ \bm{ t } }(q_{XY}^{(s)}, p_{XY}^{[0]}) 
    \right\lvert_{s\to 0+} \notag\\
    & =
    D_{ \bm{ t } }(q_{XY}^*, p_{XY}^{[0]}) 
    -
    D_{ \bm{ t } }(q_{XY}^{[0]}, p_{XY}^{[0]}) 
    -
    D_{ \bm{ t } }(q_{XY}^*, q_{XY}^{[0]}) .
\end{align}  
Then we have
\begin{align}
    0 & \le \left.\frac{\mathrm{d} g(s)}{\mathrm{d} s} 
    \right\lvert_{s\to 0+}  \notag \\
    &=
      \Theta_{\rm s }^{(\lambda, \lambda\nu)}(q_{XY}^* |P_X)
    - \Theta_{\rm s }^{(\lambda, \lambda\nu)}(q_{XY}^{[0]} |P_X)
    - (1-\lambda) D(q^*_X||q_X^{[0]}) - \lambda D(q^*_{X|Y} || q_{X|Y}^{[0]} | q^*_Y)\notag\\
    & \quad +(1-\lambda) \{
    D_{ \bm{ t } }(q_{XY}^*, p_{XY}^{[0]}) 
    -
    D_{ \bm{ t } }(q_{XY}^{[0]}, p_{XY}^{[0]}) 
    -
    D_{ \bm{ t } }(q_{XY}^*, q_{XY}^{[0]}) \} \notag\\
    & = 
    J_{{\rm s}, \bm{t}}^{(\lambda, \nu)} ( q_{XY}^*, p_{XY}^{[0]}  | P) 
    -
    J_{{\rm s}, \bm{t}}^{(\lambda, \nu)} ( q_{XY}^{[0]}, p_{XY}^{[0]}  | P) 
    - (1-\lambda) D(q^*_X||q_X^{[0]}) - \lambda D(q^*_{X|Y} || q_{X|Y}^{[0]} | q^*_Y)
    -(1-\lambda) D_{ \bm{ t } }(q_{XY}^*, q_{XY}^{[0]}) \notag\\
    & 
    \stackrel{\rm (a)}
    \leq  
    J_{{\rm s}, \bm{t}}^{(\lambda, \nu)} ( q_{XY}^* , p_{XY}^{[0]} | P)
    -
    J_{{\rm s}, \bm{t}}^{(\lambda, \nu)} ( q_{XY}^{[0]}, p_{XY}^{[0]} | P) 
    -(1-\lambda) D_{ \bm{ t } }(q_{XY}^*, q_{XY}^{[0]}) \notag\\
    &\stackrel{\rm (b)}
    =
    J_{{\rm s}, \bm{t}}^{(\lambda, \nu)} ( q_{XY}^* ,  q_{XY}^* | P) 
    + (1-\lambda) D_{\bm{t}} ( q_{XY}^* ,  p_{XY}^{[0]} ) 
    -
    J_{{\rm s}, \bm{t}}^{(\lambda, \nu)} ( q_{XY}^{[0]}, p_{XY}^{[0]} | P) 
    -(1-\lambda) D_{ \bm{ t } }(q_{XY}^*, q_{XY}^{[0]}) .
    \label{eq.257}
\end{align}
Step (a) holds because the fourth and fifth terms are non-positive.
Step (b) follow from the equation $ 
J_{{\rm s}, \bm{t}}^{(\lambda, \nu)} ( q_{XY}^* , p_{XY}^{[0]} | P)
=
J_{{\rm s}, \bm{t}}^{(\lambda, \nu)} ( q_{XY}^* ,  q_{XY}^* | P) 
+ (1-\lambda) D_{\bm{t}} ( q_{XY}^* ,  p_{XY}^{[0]} )
$ holds. 
From (\ref{update2_Algorithm_Family_Source}), we have $p_{XY}^{[1]} = q_{XY}^{[0]}$. 
Substituting this into (\ref{eq.257}), we obtain (\ref{eq.254}).
This completes the proof.

\end{lemma}
    
{\it Proof of Theorem~\ref{theorem.convergence.family.algorithm.source}:}
From (\ref{def:J_t_s}) and (\ref{def:hat_J_t_s}), we have 
\begin{align}
    & \min_{ p_{XY} } \hat J_{{\rm s}, \bm{t}}^{(\lambda, \nu)} ( p_{XY} | P_X)
    =
    \min_{ p_{XY} } \min_{ q_{XY} } J_{{\rm s}, \bm{t}}^{(\lambda, \nu)} ( q_{XY}, p_{XY} | P_X) \\
    & =
    \min_{ q_{XY} } J_{{\rm s}, \bm{t}}^{(\lambda, \nu)} ( q_{XY}, q_{XY} | P_X)
    =
    \min_{ q_{XY} } \Theta_{\bm{t}, \rm s}^{(\lambda, \nu)} ( q_{XY} | P_X)
\end{align}
Let $q_{XY}^*$ be the optimal distribution that minimizes $\Theta_{\rm s}^{(\lambda, \lambda \nu)}(q_{XY}|P_X) $.
Then, from Lemma~\ref{lemma36}, for every $i=0,1,2$, we have
\begin{align}
    0
    &\stackrel{\rm (a)}
    \leq 
    J_{{\rm s}, \bm{t}}^{(\lambda, \nu)} ( q_{XY}^{[i]}, p_{XY}^{[i]} | P_X) 
    -
    J_{{\rm s}, \bm{t}}^{(\lambda, \nu)} ( q_{XY}^* ,  q_{XY}^* | P_X) \notag \\
    &\leq 
    (1-\lambda) \{ 
    D_{ \bm{ t } }(q_{XY}^*, p_{XY}^{[i]}) 
    -
    D_{\bm{t}} ( q_{XY}^* ,  p_{XY}^{[i + 1]} ) 
    \}.
\end{align}
Step (a) holds because $q_{XY}^*$ minimizes $ J_{{\rm s}, \bm{t}}^{(\lambda, \nu)} ( q_{XY},  q_{XY} | P_X) $. 
Let $\xi_i = J_{{\rm s}, \bm{t}}^{(\lambda, \nu)} ( q_{XY}^{[i]}, p_{XY}^{[i]} | P_X) 
    -
J_{{\rm s}, \bm{t}}^{(\lambda, \nu)} ( q_{XY}^* ,  q_{XY}^* | P_X) $. Then,
\begin{align}
    0 
    &\leq
    \sum_{i=0}^{T-1} \xi_i \notag\\
    &\leq 
    (1-\lambda) \{ 
    D_{ \bm{ t } }(q_{XY}^*, p_{XY}^{[0]}) 
    -
    D_{\bm{t}} ( q_{XY}^* ,  p_{XY}^{[T]} ) 
    \} \notag \\
    &\leq
    (1-\lambda) 
    D_{ \bm{ t } }(q_{XY}^*, p_{XY}^{[0]}) 
    \label{eq.260}
\end{align}
By proposition~\ref{proposition_Algorithm_Family_Source}, $\xi_t$ is a monotone decreasing sequence.
Then from (\ref{eq.260}) we have
$0\leq T\xi_T \leq (1-\lambda)     D_{ \bm{ t } }(q_{XY}^*, p_{XY}^{[0]}) $.
Thus, 
\begin{align}
    0\leq \xi_T \leq \frac{(1-\lambda) 
    D_{ \bm{ t } }(q_{XY}^*, p_{XY}^{[0]})}{T}
    \to 0, \quad T\to \infty. 
\end{align}
Hence, we have 
\begin{align}
    \lim_{i\to \infty }
    J_{{\rm s}, \bm{t}}^{(\lambda, \nu)} ( q_{XY}^{[i]}, p_{XY}^{[i]} | P_X) 
    =
    J_{{\rm s}, \bm{t}}^{(\lambda, \nu)} ( q_{XY}^{*}, q_{XY}^{*} | P_X) 
\end{align}
This completes the proof. \hfill\IEEEQED

\section{Proof of Proposition~\ref{limit_of_A_s} }
\label{appendixH}
The limit of $(1/\rho) A_{\rm s}^{(\rho, \nu)} (\hat p_{Y|X} | P_X) $
as $\rho$ approaches to $0$ is evaluated as follows. 

\textit{Proof of Proposition~\ref{limit_of_A_s}: }
We have
\begin{align*}
& \lim_{\rho \to 0} \frac1 \rho A_{\rm s}^{(\rho,\nu)} ( \hat p_{Y|X} | P_X) \\
& \stackrel{\rm (a)}{=} 
\lim_{\rho \to 0} \frac{\partial}{\partial \rho} A_{\rm s}^{(\rho,\nu)} ( \hat p_{Y|X} | P_X) \\
& = 
\lim_{\rho \to 0} \frac{\partial}{\partial \rho} \left\{
(1+\rho) \log \sum_y \left[
\sum_{x} P_X(x) {\rm e}^{\rho \nu d(x,y)} \hat p_{Y|X}^{1+\rho}(y|x)
\right]^{1/(1+\rho)} \right\}
\\
& = 
\lim_{\rho \to 0} 
\log \sum_y \left[
\sum_{x} P_X(x) {\rm e}^{\rho \nu d(x,y)} \hat p_{Y|X}^{1+\rho}(y|x)
\right]^{1/(1+\rho)} \\
&\quad + 
\lim_{\rho \to 0} \frac{\partial}{\partial \rho} 
\log \sum_y \left[
\sum_{x} P_X(x) {\rm e}^{\rho \nu d(x,y)} \hat p_{Y|X}^{1+\rho}(y|x)
\right]^{1/(1+\rho)} 
\\
& = 
0+  
\lim_{\rho \to 0}  
\frac{ \frac{\partial}{\partial \rho} \sum_y \left[
\sum_{x} P_X(x) {\rm e}^{\rho \nu d(x,y)} \hat p_{Y|X}^{1+\rho}(y|x)
\right]^{1/(1+\rho)} 
}{ \sum_y \left[
\sum_{x} P_X(x) {\rm e}^{\rho \nu d(x,y)} \hat p_{Y|X}^{1+\rho}(y|x)
\right]^{1/(1+\rho)} 
}
\\
& = 
\lim_{\rho \to 0}  
\frac{\partial}{\partial \rho} \sum_y \left[
\sum_{x} P_X(x) {\rm e}^{\rho \nu d(x,y)} \hat p_{Y|X}^{1+\rho}(y|x)
\right]^{1/(1+\rho)} 
\\
&\stackrel{(b)}=
\lim_{\rho \to 0}  
\sum_y 
\left[
\sum_{x} P_X(x) {\rm e}^{\rho \nu d(x,y)} \hat p_{Y|X}^{1+\rho}(y|x)
\right]^{1/(1+\rho)}\\
&\phantom{\lim_{\rho \to 0}  
\sum_y 
} \cdot \left(
\frac{-1}{(1+\rho)^2} 
\log  \left\{
\sum_{x} P_X(x) {\rm e}^{\rho \nu d(x,y)} \hat p_{Y|X}^{1+\rho}(y|x)
\right\}
+
\frac{1}{1+\rho} 
\frac{ \frac{\partial}{\partial \rho} 
\sum_{x} P_X(x) {\rm e}^{\rho \nu d(x,y)} \hat p_{Y|X}^{1+\rho}(y|x)
}{\sum_{x} P_X(x) {\rm e}^{\rho \nu d(x,y)} \hat p_{Y|X}^{1+\rho}(y|x)}
\right)
\\
& = 
\lim_{\rho \to 0}  
\sum_y 
\left( 
\sum_{x} P_X(x) \hat p_{Y|X}(y|x)
\right)
\left(
-\log  
\sum_{x} P_X(x) \hat p_{Y|X}(y|x)
+
\frac{ \frac{\partial}{\partial \rho} 
\sum_{x} P_X(x) {\rm e}^{\rho \nu d(x,y)} \hat p_{Y|X}^{1+\rho}(y|x)
}{ \sum_{x} P_X(x) \hat p_{Y|X}(y|x)}
\right)
\\
& = 
\lim_{\rho \to 0}  
\sum_y 
\left( 
\sum_{x} P_X(x) \hat p_{Y|X}(y|x)
\right)
\Bigg(
-\log  
\sum_{x} P_X(x) \hat p_{Y|X}(y|x) \\
& \hspace{3cm}+
\frac{ 
\sum_{x} t d(x,y) P_X(x) {\rm e}^{\rho \nu d(x,y)} \hat p_{Y|X}^{1+\rho}(y|x)
+
\sum_{x} P_X(x) {\rm e}^{\rho \nu d(x,y)} \hat p_{Y|X}^{1+\rho}(y|x) \log \hat p_{Y|X}^{1+\rho}(y|x)
}{ \sum_{x} P_X(x) \hat p_{Y|X}(y|x)}
\Bigg)\\
& = 
\sum_y 
\bigg\{
-\sum_{x} P_X(x) \hat p_{Y|X}(y|x)
\log  
\sum_{x'} P_X(x') \hat p_{Y|X}(y|x') \\
& \phantom{=\sum_y \bigg\{ } \, \,+
\sum_{x} t d(x,y) P_X(x) \hat p_{Y|X}(y|x)
+
\sum_{x} P_X(x)  \hat p_{Y|X}(y|x) \log \hat p_{Y|X}(y|x)
\bigg\}
\\
&= I(P_X, \hat p_{Y|X} ) + \nu \mathrm{E}_{q_{XY} } [ d(X,Y)].
\end{align*} 
Step (a) follows from the L'Hôpital's rule.
Step (b) holds because
we have
$
\left( f(x)^{g(x)} \right)^{\prime}
=
f(x)^{g(x)} \left(
g'(x) \log f(x) + g(x) \frac{f'(x)}{f(x)}
\right)
$. \hfill$\IEEEQED$


\section{Proofs of Lemmas~\ref{lemma_GCK_minimization1}, \ref{lemma_GCK_minimization2}, and~\ref{lemma:F_s_tilde} }
\label{appendixG}
In this section, we give proofs of 
Lemmas~\ref{lemma_GCK_minimization1}, 
\ref{lemma_GCK_minimization2}, and~\ref{lemma:F_s_tilde}.

{\it Proof of Lemmma~\ref{lemma_GCK_minimization1}:} We have
\begin{align*}
    & F_{\rm JO, s}^{(\lambda, \nu)}(q_{XY}, \hat q_{XY} |P_X) \\
    &= 
    {\rm E}_{q_{XY} }\left[ \log 
    \frac{ \hat q_X^{1-\lambda}(X) \hat q_{X|Y}^{ \lambda }(X|Y)}{ P_X(X) {\rm e}^{-\lambda \nu d(X,Y) } } \right]
    +D( q_{XY}|| \hat q_{XY} ) \\
    &= 
    {\rm E}_{q_{XY} }\left[ \log 
    \frac{ \hat q_X^{1-\lambda}(X) \hat q_{X|Y}^{ \lambda }(X|Y) q_{XY}(X,Y) }{ P_X(X) {\rm e}^{- \lambda \nu d(X,Y) } \hat q_{XY}(X,Y)} \right] \\
    &= 
    {\rm E}_{q_{XY} }\left[ \log 
    \frac{ q_{XY}(X,Y) }{ P_X(X) {\rm e}^{-\lambda \nu d(X,Y) } \hat q_{Y|X}^{1-\lambda} (Y|X) \hat q_Y^\lambda(Y)} \right] \\
    &=\Theta_{\rm s}^{(\lambda, \lambda \nu )}(q_{XY}|P_X)
    +     {\rm E}_{q_{XY} }\left[ \log 
    \frac{ q_{Y|X}^{1-\lambda}(Y|X) q_Y^\lambda(Y)}{ \hat q_{Y|X}^{1-\lambda} (Y|X) \hat q_Y^\lambda(Y)} \right] \\
    &=\Theta_{\rm s}^{(\lambda, \lambda \nu)}(q_{XY}|P_X)
    +(1-\lambda)D(q_{Y|X}|\hat q_{Y|X}|q_X)+\lambda D(q_Y|\hat q_Y).
\end{align*}
Hence, by non-negativity of divergence we have
$$ 
F_{\rm JO, s}^{(\lambda, \nu)}(q_{XY}, \hat q_{XY} |P_X)  \geq 
\Theta_{\rm s}^{(\lambda, \lambda \nu)}(q_{XY}|P_X),
$$
where equality holds if $\hat q_{XY} = q_{XY}$.
This completes the proof.\hfill$\IEEEQED$

{\it Proof of Lemma~\ref{lemma_GCK_minimization2}:} We have
\begin{align*}
    & F_{\rm JO, s}^{(\lambda, \nu)}(q_{XY}, \hat q_{XY} |P_X) \\
    &= 
    {\rm E}_{q_{XY} }\left[ \log 
    \frac{ q_{XY}(X,Y) }{ P_X(X) {\rm e}^{-\lambda \nu d(X,Y) } \hat q_Y^\lambda(Y) \hat q_{Y|X}^{1-\lambda} (Y|X) }\right] \\
    &=
    {\rm E}_{q_{XY} }\left[ \log 
    \frac{ q_{XY}(X,Y) }{ \tilde q_{XY}(\hat q_{XY}) (X,Y)  } \right] 
    -\log \sum_{x,y} 
    P_X(x) {\rm e}^{-\lambda \nu d(x,y) } \hat q_Y^\lambda(y) \hat q_{Y|X}^{1-\lambda} (y|x). 
\end{align*}
Hence, by non-negativity of relative entropy, we have
$$
F_{\rm JO, s}^{(\lambda, \nu)}(q_{XY}, \hat q_{XY} |P_X) 
\geq 
-\log \sum_{x,y} 
    P_X(x) {\rm e}^{-\lambda \nu d(x,y) } \hat q_Y^\lambda(y) \hat q_{Y|X}^{1-\lambda} (y|x),
$$
where equality holds if and only if
$ q_{XY}= \tilde q_{XY}(\hat q_{XY}) $ holds.
This completes the proof.\hfill$\IEEEQED$

Next we give the proof of Lemma~\ref{lemma:F_s_tilde}.
The proof is almost the same as the proof of Lemma~\ref{lemma_GCK_minimization2}.

\begin{IEEEproof}[Proof of Lemma~\ref{lemma:F_s_tilde}]
We have  
\begin{align}
    &
    \tilde F_{\rm s}^{(\lambda, \nu)} ( q_{XY}, \hat p_{Y}, p_{Y|X}|P_X)\notag\\
    &=
    \Theta_{\rm s}^{(\lambda, \lambda \nu)}(q_{XY} |P_X) + \lambda D(q_Y||\hat p_{Y} )
     + (1-\lambda) D(q_{Y|X} || p_{Y|X} | q_X) \notag\\
     &=
    {\rm E}_{q_{XY} }\left[ \log 
    \frac{ q_{XY}(X,Y) }{ P_X(X) {\rm e}^{-\lambda \nu d(X,Y) } \hat p_Y^\lambda(Y) p_{Y|X}^{1-\lambda} (Y|X) }\right] \\
    &=
    {\rm E}_{q_{XY} }\left[ \log 
    \frac{ q_{XY}(X,Y) }{ \tilde q_{XY}( \hat p_{Y}, p_{Y|X} ) (X,Y)  } \right] 
    -\log \sum_{x,y} 
    P_X(x) {\rm e}^{-\lambda \nu d(x,y) } \hat p_Y^\lambda(y) p_{Y|X}^{1-\lambda} (y|x),
\end{align}
where
\begin{align}
    \tilde q_{XY}( \hat p_{Y}, p_{Y|X} ) (x,y)
    =
    \frac{
    P_X(x) {\rm e}^{-\lambda \nu d(x,y) } \hat p_Y^\lambda(y) p_{Y|X}^{1-\lambda} (y|x)
    }{
    \sum_{x',y'} 
    P_X(x') {\rm e}^{-\lambda \nu d(x',y') } \hat p_Y^\lambda(y') p_{Y|X}^{1-\lambda} (y'|x')
    }.
\end{align}
Hence, by non-negativity of relative entropy, we have 
\begin{align}
    \tilde F_{\rm s}^{(\lambda,\nu)} ( q_{XY}, \hat p_{Y}, p_{Y|X}|P_X)
    \geq 
    -\log \sum_{x,y} 
    P_X(x) {\rm e}^{-\lambda \nu d(x,y) } \hat p_Y^\lambda(y) p_{Y|X}^{1-\lambda} (y|x).
\end{align}
Equality holds if and only if
$q_{XY} = \tilde q_{XY}( \hat p_{Y}, p_{Y|X} )$, 
which completes the proof.

\end{IEEEproof}

\end{document}